\documentclass[aps,prx,superscriptaddress,twocolumn]{revtex4-2}

\usepackage{comment}
\usepackage[utf8]{inputenc}
\usepackage{mathtools}
\usepackage{amssymb}
\usepackage{amsmath}
\usepackage{amsfonts}
\usepackage{bm}
\usepackage{bbold}
\usepackage{upgreek}
\usepackage{xfrac}
\usepackage{soul}
\usepackage{xcolor}
\usepackage{derivative}
\usepackage{physics}
\usepackage{comment}
\usepackage{dsfont}
\usepackage{graphicx,lipsum}
\usepackage{float}

\usepackage[colorlinks=true,citecolor=blue]{hyperref}
\hypersetup{colorlinks=true,citecolor=blue,linkcolor=blue,urlcolor=blue}
\usepackage{url}

\DeclarePairedDelimiter{\ceil}{\lceil}{\rceil}

\usepackage{txfonts}
\DeclareSymbolFont{matha}{OML}{txmi}{m}{it}
\DeclareMathSymbol{\varv}{\mathord}{matha}{118}
\usepackage{bbm}

 \makeatletter
\def\@fnsymbol#1{\ensuremath{\ifcase#1\or \dagger\or \ddagger\or
   \mathsection\or \mathparagraph\or \|\or **\or \dagger\dagger
   \or \ddagger\ddagger \else\@ctrerr\fi}}
    \makeatother

\DeclareFontEncoding{LS1}{}{}
\DeclareFontSubstitution{LS1}{stix}{m}{n}

\usepackage{setspace}
\begin{document}

\title{Quantum algorithm for solving nonlinear differential equations based on physics-informed effective Hamiltonians}

\author{Hsin-Yu Wu}
\affiliation{Department of Physics and Astronomy, University of Exeter, Stocker Road, Exeter EX4 4QL, United Kingdom}
\affiliation{Pasqal, 7 Rue Léonard de Vinci, 91300 Massy, France}
\author{Annie E. Paine}
\affiliation{Pasqal, 7 Rue Léonard de Vinci, 91300 Massy, France}
\author{Evan Philip}
\affiliation{Pasqal, 7 Rue Léonard de Vinci, 91300 Massy, France}
\author{Antonio A. Gentile}
\affiliation{Pasqal, 7 Rue Léonard de Vinci, 91300 Massy, France}
\author{Oleksandr Kyriienko}
\affiliation{School of Mathematical and Physical Sciences, University of Sheffield, Sheffield S10 2TN, United Kingdom}

\date{\today}

\begin{abstract}
We propose a distinct approach to solving linear and nonlinear differential equations (DEs) on quantum computers by encoding the problem into ground states of effective Hamiltonian operators. Our algorithm relies on constructing such operators in the Chebyshev space, where an effective Hamiltonian is a sum of global differential and data constraints. Once the effective Hamiltonian is formed, solutions of differential equations can be obtained using the ground state preparation techniques (e.g. imaginary-time evolution and quantum singular value transformation), bypassing variational search. Unlike approaches based on discrete grids, the algorithm enables evaluation of solutions beyond fixed grid points and implements constraints in the physics-informed way. Our proposal inherits the best traits from quantum machine learning-based DE solving (compact basis representation, automatic differentiation, nonlinearity) and quantum linear algebra-based approaches (fine-grid encoding, provable speed-up for state preparation), offering a robust strategy for quantum scientific computing in the early fault-tolerant era.
\end{abstract}

\maketitle

\section*{Introduction}

Solving differential equations (DEs) is a crucial task in many areas of science and engineering, ranging from simulating chemical reactions~\cite{Gavalas1968} to computational fluid dynamics~\cite{anderson1995computational,cfd_textbook} and financial analysis~\cite{Oksendal2003}. As closed-form solutions are often unavailable for practical tasks, many problems require using numerical methods~\cite{Gear1981}, which can rely on mesh-based approaches (finite-difference and finite-element solvers~\cite{LeVeque, Jin}) or spectral methods~\cite{hussaini1983spectral}. The former often rely on discretization techniques and iterative~\cite{Saad2003} or multigrid methods~\cite{wesseling1992introduction}. The latter use a finite set of orthogonal basis functions (e.g. Fourier, Chebyshev or Legendre polynomials) to provide a global, non-discretized approximation \cite{Boyd2013,trefethen2019approximation}. For stiff, nonlinear, and multiscale problems such as turbulence \cite{StanfordTurbulenceRev,BeverleyTurbulence,Schumacher2009,Schmidt2018} applying direct numerical methods is often prohibitive at scale when requiring fine meshes or very large basis sets. This calls for alternative methods for addressing computationally challenging DE-based problems.  

Data-driven approaches paired with machine learning (ML) offer one distinct way to address scientific and engineering problems. Taking fluid dynamics as an example, experiments for high–Reynolds number turbulent flows~\cite{ZAGAROLA_SMITS_1998,BeverleyTurbulence} provide insights that cannot be readily obtained in direct numerical simulations \cite{StanfordTurbulenceRev,Eivazi_2024}. Enforced by differential constraints embedded into ML models, neural network-based data processing can enable efficient description of fluid flows \cite{Brunton2020rev,Duraisamy_annurev}, model discovery~\cite{Rudy2017,Both2021,Chen2021}, and feature extraction~\cite{Strofer2019,Obayashi2021,UNNIKRISHNAN2023}. Examples of machine learning architectures in this domain include physics-informed neural networks (PINNs)~\cite{Raissi2019,Raissi2020Science,cai2021physics_heat,Eivazi2022}, Fourier neural operators~\cite{li2020fourier}, Navier-Stokes flow nets~\cite{Jin2021}, symmetry-aware approaches like physics-informed dynamic mode decomposition~\cite{Baddoo2023}, rotation-equivariant graph neural networks~\cite{Lino2022}, and many more~\cite{toscano2024pinnspikans}. Such physics-informed methods to solving scientific problems have led to the emergence of the scientific machine learning (SciML) ecosystem~\cite{rackauckas2020universal,Takamoto2024}. To move forward, SciML needs further breakthroughs in learning and solving DEs.

Quantum computing (QC) can offer new possibilities for solving differential equations and addressing machine learning problems \cite{Biamonte2017,algorithms_survey,Au-Yeung_2024}. Starting from the HHL algorithm \cite{harrow2009quantum}, various methods were proposed for solving DEs based on linear equation solvers~\cite{morales2025,Cao_2013,Montanaro2016,Xin2020,MartinSanz2023}, including those relying on linear combinations of unitaries \cite{childs2017quantum,Childs2012,Berry2017,jennings2023efficient,Gribling2024}, quantum signal processing \cite{Lin2020optimalpolynomial,Linden2022quantumvsclassical,Krovi2023improvedquantum}, adiabatic inversion \cite{subacsi2019quantum,Costa2022,DongLin2022} and eigenstate filtering \cite{Lin2020optimalpolynomial,costa2023improving}. Other suggested methods include quantum reservoir computing~\cite{Pfeffer2022,Pfeffer2023}, Schrödingerization~\cite{jin2022quantum,jin2022quantumEXT,jin2023analog,jin2023quantum,hu2024quantum}, quantum iterative solvers \cite{williams2024iterative,quantum_iterative_solver,time_marching_multigrid,jin2023quantum,quantum_multigrid}, Fourier transform-based solvers \cite{LiuCirak2024}, lattice Boltzmann methods \cite{Mezzacapo2015,Succi2015,TODOROVA2020,Steijl2023,schalkers2022efficient,schalkers2023importance,Budinski2022,chrit2023fully,zamora2024efficient}, and smooth particle hydrodynamics \cite{AUYEUNG2024sph,auyeung2025sph}. Nonlinear differential equations were mostly approached with Carleman linearization \cite{JPLiu2021,Krovi2023improvedquantum,costa2023improving,tanaka2023carleman,ingelmann2023quantum,wu2024quantum,gonzalezconde2024,Sanavio2024,sanavio2024carleman}, use of quantum nonlinear processing \cite{lubasch2020variational,jaksch2022variational}, or Chebyshev-based models \cite{Paine2023}. While aiming to speed up the process of DE solving, the aforementioned protocols typically do not operate with data constraints and extra effort is needed for bridging them with data-driven SciML.

Quantum physics-informed approaches emerged by combining quantum machine learning tools with differential constraints and data \cite{kyriienko2021solving,Paine2021,paine2023quantum,Markidis2022qpinns}. They are used for various problems and applications including equation discovery \cite{heim2021quantum}, generative modeling \cite{Kyriienko2024protocols,kasture2022protocols,wu2024quantum,delejarza2025QCPM}, nonlinear flows \cite{Jaderberg2024} and weather forecasting \cite{Jaderberg2024weather}. Recently, quantum DE solvers were shown to serve as a source of quantum data that can be studied with an emerging quantum SciML framework~\cite{Williams2024readout}. However, as the framework relies on a variational procedure for preparing solutions, quantum SciML often depends on the model trainability and the need of bespoke models that avoid vanishing gradients \cite{McClean2018,Larocca2025}.
\begin{figure*}[t]
\begin{center}
\includegraphics[width=1.0\linewidth]{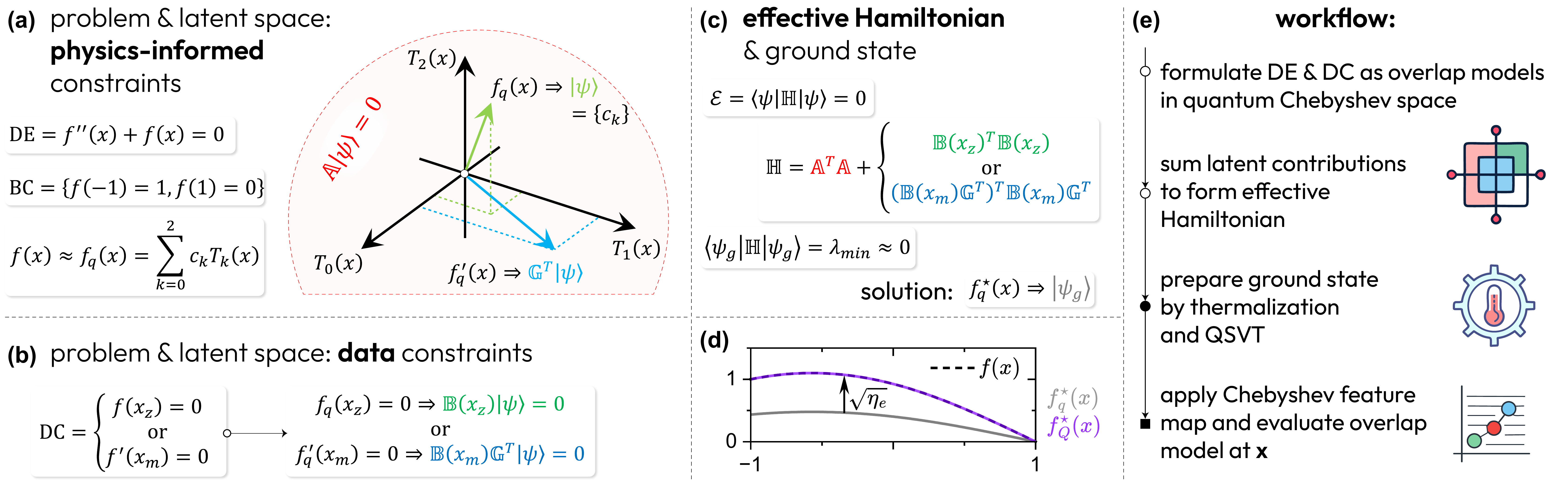}
\end{center}
\caption{\textbf{Concept of the effective Hamiltonian approach for solving differential equations.} \textbf{(a)} Physics-informed differential constraints for the overlap model $f_q(x)$, and visualization of a corresponding null vector in the latent space of Chebyshev polynomials. \textbf{(b)} Regularization of the model via invariant data constraint (DC) in the latent representation. \textbf{(c)} Effective energy as an expectation value of an effective Hamiltonian $\mathbb{H}$ being a sum of physics-informed and data-driven latent contributions. \textbf{(d)} DE solution plotted as an overlap between the latent state vector and $|\psi_g \rangle$ being the ground state $\mathbb{H}$. The scale is adjusted by coefficient $\sqrt{\eta_e}$ based on available data. \textbf{(e)} Summary of the workflow.}
\label{fig:LE}
\end{figure*}

In this work, we propose a physics-informed quantum differential equation solver that avoids variational training yet encodes differential and data constraints that enable SciML-like workflow~\cite{qk_patent}. Our approach relies on formulating an effective Hamiltonian $\mathbb{H}$ that embeds relevant DE-based problems in a latent space, with quantum Chebyshev feature space \cite{Williams2023} being the most suitable option. The quantum solutions are then obtained by preparing low-energy states for $\mathbb{H}$, as performed with quantum imaginary-time evolution~\cite{Motta}, algorithmic thermalization~\cite{chen2023quantum}, eigenstate filtering \cite{Lin2020nearoptimalground}, and generally quantum singular value transformations \cite{Gilyen, Martyn}.

The paper is structured as follows. First, we introduce the core idea. Second, we present a technical background for formulating DE-based effective Hamiltonians in the quantum Chebyshev basis, and describe the methodology of performing the overlap measurements. Next, we demonstrate the use of the effective Hamiltonian algorithm for various examples that include linear, nonlinear and partial DEs. We present the quantum circuits for relevant subroutines. Finally, in the discussion section we reflect on potential improvements and future steps for developing quantum SciML.


\section*{The algorithm}

\subsection*{Core idea}

Let us start by imagining an ordinary differential equation for some univariate function $f(x)$ that contains derivatives of this function, the function itself, and other $x$-dependent terms. We say that the differential equation is solved when all conditions are satisfied and the differential equation loss is equal to zero, $\mathcal{D}[f]=0$. Let us then formulate a quantum model for solving the problem, which shall obey the constraints and represent the solution $f(x)$. This can be written as an overlap-based model $f_q(x) = \langle \tau(x)|\psi\rangle$ between some quantum state $|\psi\rangle$ and a quantum state $|\tau(x)\rangle$ that labels independent basis functions of $x$. The amplitudes of $|\psi\rangle$ store the weights of each basis function. They must be set such that all constraints in our differential equation are satisfied and the loss approaches zero, $\mathcal{D}[f_q]=0$. Since the weights are defined by the choice of $|\psi\rangle$ for the given basis, to solve the differential equation this quantum state needs to match the constrains. We can imagine this as selecting a vector representing a state in a coordinate system specified by the basis set. 

Our main idea is to formulate a set of constraints for $|\psi\rangle$ in the form of operators. For instance, applying differentiation to $f_q(x)$ is equivalent to applying a (generally non-unitary) operator onto the quantum state, $df_q(x)/dx = \langle \tau(x)|\mathbb{G}^{T}|\psi\rangle$, where we call $\mathbb{G}$ a derivative operator for the selected basis \cite{Paine2023}. We can define other constrains in an operator form and sum them up to get an operator $\mathbb{A}$ for which $|\psi\rangle$ is a null vector. If written in a positive semi-definite form, such operator can be seen as an effective Hamiltonian with $|\psi\rangle$ being its ground state associated with zero energy. Our goal is to describe the procedure how such Hamiltonians can be formed, and how their ground states can be prepared for solving different DEs.

To visualize the process, let us look at one simple case and consider a second-order ordinary differential equation, 
\begin{equation}
\label{eq:DE-first-example}
\frac{d^2f(x)}{dx^2}+f(x)=0.
\end{equation}
The overlap model $f_q(x)$ can be formed on a set of weighted Chebyshev polynomials of the first kind, $\{T_{k}(x)\}$ \cite{Williams2023}. For brevity, we truncate $f_q(x)$ to be a polynomial of degree 2 with coefficients $\{c_k\}_{k=0}^2$. We build $f_q(x)$ by selecting coefficients that satisfy the DE, stored by the corresponding quantum state $|\psi\rangle$. The corresponding decomposition is shown in Fig.~\ref{fig:LE}(a) and explained below.

The second derivative of $f_q(x)$ can be represented in the latent space by the repeated application of generators, $d^2f_q(x)/dx^2 = \langle \tau(x)|{\mathbb{G}^T}^2|\psi\rangle$ Summing the differential constraints as physics-informed contributions, we obtain the latent space representation of the exemplary DE as $\mathcal{D}[f]=0 \mapsto \langle \tau(x)|\mathbb{A}|\psi\rangle = 0$ and $\mathbb{A}|\psi\rangle = 0 \cdot |\psi\rangle$, where $\mathbb{A} = {\mathbb{G}^T}^2 + \mathds{1}$. Here, $\mathds{1}$ is the identity matrix. 

Next, we need to impose boundary conditions ($\mathcal{BC}$) to pin a unique solution for the DE. In this work, we do so by effectively supplying a data constraint ($\mathcal{DC}$) that is invariant under scaling operations for reasons that we will elicit in the following. 
A natural choice is to impose the function value or its derivative being zero, i.e. $f(x_z) =0$ or $f'(x_m) =0$, as depicted in Fig.~\ref{fig:LE}(b). The corresponding latent representations of such constraints are $\mathbb{B}(x_z)$ and $\mathbb{B}(x_m)\mathbb{G}^T$, respectively, where $\mathbb{B}(x)$ is a data transformation operator dependent on $x$. These operators are real matrices designed to regularize $|\psi\rangle$.

For each latent space constrain we form a Gram matrix by pre-multiplying individual operators by their transpose. When summed together they form an \emph{effective Hamiltonian} operator $\mathbb{H}$. Note that $\mathbb{H}$ is a real symmetric positive-definite matrix with distinct eigenvalues and mutually orthonormal eigenvectors. For a given pair $\{ |\psi\rangle, \mathbb{H} \}$ we can associate an energy $\mathcal{E}$, being an expectation value $\mathcal{E} = \expval{\mathbb{H}}{\psi}$ [see Fig.~\ref{fig:LE}(c)]. This is the key step of our approach, as it translates the requirement to satisfy DE-based constraints into finding the zero-energy eigenstate of $\mathbb{H}$. 
In other words, we recast the original DE into an eigenvalue problem $\mathbb{H} |\psi_g \rangle = \lambda_{min} |\psi_g \rangle$, where we are interested in the lowest-energy eigenstate $|\psi_g \rangle $ associated with the minimum eigenvalue $\lambda_{min} \approx 0$. By substituting $|\psi_g \rangle$ into the quantum model $f_q(x)$, we get (non-scaled) overlap ($f_q^{\star}(x)$). We can easily evaluate this prefactor once there is access to a single data constraint, $\sqrt{\eta_e} = f(x_s)/f_q^{\star}(x_s)$, for $ x_s \notin x_z$. The process is as shown in Fig.~\ref{fig:LE}(d). The necessity to rescale $f_q^{\star}(x)$ in the problem space to provide the full quantum model solution $f_Q^{\star}(x) = \sqrt{\eta_e} f_q^{\star}(x)$ justifies the choice of invariant $\mathcal{DC}$ made previously. 

The workflow is summarized in Fig.~\ref{fig:LE}(e). We note that designing models of the maximal degree $N_{\mathrm{Cheb}}$ requires preparing the ground state $|\psi_g\rangle$ of $\mathbb{H}$ on just $n = \lceil \log_2(N_{\mathrm{Cheb}}) \rceil$ qubits. Many functions can be comfortably represented by limited-degree Chebyshev expansions \cite{trefethen2019approximation}. The developed models can be mapped into real-space over an extended register \cite{delejarza2025QCPM}, providing fine-grid solutions. Finally, they can be used for learning on models in the latent spaces \cite{Williams2024readout}.


\subsection*{Technical details}
Next, we describe technical details of the algorithm. The starting point corresponds to the choice of the latent space, which we choose a space of Chebyshev polynomials of the first kind. Since Chebyshev polynomials $T_{k}(x)$ satisfy a discrete orthogonality condition and form a complete orthogonal basis in the domain $x \in [-1,1]$, any arbitrary function that is smooth and continuous on the same interval can be expressed as $f(x) = \sum_{k=0}^{\infty} c_k T_k(x)$~\cite{trefethen2019approximation}. Similarly to this Chebyshev expansion, a quantum model $f_Q(x)$ can be built as a scaled state overlap for $n$-qubit states (assuming truncated series of maximal degree $2^n$). This quantum model is represented by the Hermitian inner product of an $x$-dependent Chebyshev basis state $\langle\tau(x)|_n$ and a quantum state $|\psi \rangle_n$ composed of $2^n$ unknown coefficients to be determined, given by
\begin{align}
\label{eq:quantum model_n}
    f_Q(x) = \sqrt{\eta} f_q(x) = \sqrt{\eta}  \langle \tau(x)|_n |\psi \rangle_n, \quad \text{s.t.}\; \langle \psi |_n | \psi \rangle_n = 1,
\end{align}
where $\sqrt{\eta}$ denotes a scaling factor ($\eta >0$). Here, $f_q(x)$ represents the non-scaled state overlap, and a purely real-value quantum Chebyshev state reads \cite{Williams2023} 
\begin{align}
\label{eq:tau(x)}
    |\tau(x)\rangle_n = \frac{1}{2^{n/2}}T_0(x)|\mathrm{\o}\rangle + \frac{1}{2^{(n-1)/2}} \sum_{k=1}^{2^n-1} T_k(x)|k\rangle,
\end{align}
where $|\mathrm{\o}\rangle \equiv |0\rangle^{\otimes n}$ denotes a $n$-qubit product state with all qubits being in zero state. We note that the states in Eq.~\eqref{eq:tau(x)} are orthonormal when evaluated at the Chebyshev nodes $\bar{x}_j \coloneqq \cos \left[ \pi(j+1/2)/2^n \right] \in (-1,1)$, defined as the roots or zeros of $T_{2^n}(x)$, and these Chebyshev states $\{|\tau(\bar{x}_j)\rangle\}_{j=0}^{2^n-1}$ form an orthonormal basis set. The normalized Chebyshev state defined as $|\tilde{\tau}(x)\rangle_n = |\tau(x)\rangle_n / \norm{\,|\tau(x)\rangle_n\,}$ can be prepared by applying to a $(n+1)$-qubit zero product state an $x$-dependent Chebyshev feature map $\hat{\mathcal{U}}_{\tau}(x)$ acting on one ancilla qubit and $n$ system qubits, followed by post-selecting the $|0\rangle$ state in the ancilla qubit~\cite{Williams2023}. For completeness, we describe and visualize the corresponding circuit in Fig.~\ref{fig:ChebFM} of Supplementary Materials (SM). In the following derivations, the normalization term $\norm{\,|\tau(x)\rangle_n\,}$ is discarded, since it can be implicitly incorporated into the scaling factor.

Analyzing Eq.~\eqref{eq:quantum model_n}, we observe that $|\psi \rangle_n$ does not explicitly depend on $x$. To differentiate the function $f_Q(x)$, one needs to take a derivative of the quantum Chebyshev feature map yielding the differentiation matrix~\cite{Trefethen2000,Paine2023}, $\mathbb{G}_n$. The full derivative reads as
\begin{align}
\label{eq:quantum model derivative}
    f'_Q(x) = \frac{df_Q(x)}{dx} = \sqrt{\eta}  \langle \tau(x)|_n \mathbb{G}_n^T |\psi \rangle_n,
\end{align}
where $\mathbb{G}_n^T$ is a constant upper triangular matrix with subscript $n$ denoting the number of qubits and superscript $T$ denoting transpose (see details in section A of SM). In this work, non-unitary matrices are symbolized using double-struck capital letter (also called blackboard bold) throughout the paper. The $m^{th}$-order derivative of the quantum model is immediately expressible by matrix-multiplying $\mathbb{G}_n^T$ by itself $m$ times. 

In addition, using the product-to-sum identity of Chebyshev polynomials~\cite{Trefethen2019,Paine2023}, one can readily prove that 
\begin{align}
\label{eq:Mx}
    x^p \langle \tau(x)|_n = \langle \tau(x)|_{n+1} \mathbb{M}_{x^p},
\end{align}
with $\mathbb{M}_{x^p}$ being an $x$-independent matrix and the integer $p \in [0,2^n]$. For $p=0$, we have 
\begin{align}
\label{eq:M1}
\langle \tau(x)|_n = \langle \tau(x)|_{n+1} \mathbb{M}_1,
\end{align}
Similarly,
\begin{align}
\label{eq:Nx}
    x \langle \tau(x)|_n \otimes \langle \tau(x)|_n = \langle \tau(x)|_{n+1} \mathbb{N}_x,
\end{align}
and
\begin{align}
\label{eq:N1}
\langle \tau(x)|_n \otimes \langle \tau(x)|_n = \langle \tau(x)|_{n+1} \mathbb{N}_1,
\end{align}
These non-square mapping operators are unique constant matrices (see section A of SM for more details) and allow for the representation of the $n$-qubit quantum state on a higher-dimensional Chebyshev basis. Importantly, both $\langle \tau(x)|_n$ and $\langle \tau(x)|_{n+1}$ are the Chebyshev states with linearly independent basis sets, but $x^p \langle \tau(x)|_n$ and $(\langle \tau(x)|_n \otimes \langle \tau(x)|_n)$ are not.

Next, we need to deal with the boundary/initial conditions. Typical boundary conditions, $\mathcal{BC} = \{f(x_0),f'(x_1)\}$, are imposed at different boundaries (e.g. $x_0=-1$ and $x_1=1$) to narrow down the solution space and guarantee a unique solution to the DE. Unlike boundary conditions, data constraints can be anywhere in the domain and play a crucial role in guiding the model toward the desired outcome.

There are two types of data constraints required to solve DEs with this approach. The first one is the \emph{invariant} constraint, $\mathcal{DC_I} \in \{f(x_z)=0\}$ or $\mathcal{DC_I} \in \{f'(x_m)=0\}$, which remains unchanged under scaling. 
$\mathcal{DC_I}$ is used to enforce physics-informed restrictions on the null space of the latent DE in order to obtain a state overlap solution $f_q^{\star}(x)$. Also $\mathcal{DC_I}$ must be formulated in the form of the aforementioned quantum model in one of the following ways:
\begin{align}
\label{eq:bc1}
f(x_z)=0  \rightarrow \sqrt{\eta}  \langle \tau(x)|_n \mathbb{B}_n(x_z) |\psi \rangle_n =0, 
\end{align}
\begin{align}
\label{eq:bc2}    
f'(x_m)=0 \rightarrow \sqrt{\eta}  \langle \tau(x)|_n \mathbb{B}_n(x_m) \mathbb{G}_n^T |\psi \rangle_n =0, 
\end{align}
where $\mathbb{B}_n(x_z)$ and $\mathbb{B}_n(x_m)$ are obtained from $\langle \tau(x_z)|_n =\langle \tau(x)|_n \mathbb{B}_n(x_z)$ and $\langle \tau(x_m)|_n =\langle \tau(x)|_n \mathbb{B}_n(x_m)$, respectively. Specifically, $\mathbb{B}_n(x) = \sqrt{2^n} |0\rangle^{\otimes n} \langle\tau(x)|_n$ is an $x$-dependent rank-one matrix with nonzero entries in the first row (see section A of SM for details). This can be implemented efficiently on quantum computers \cite{rattew2022preparing}.

The other required constraint is the \emph{regular} constraint, $\mathcal{DC_R} \in \{f(x_s)\neq0, \forall x_s \notin x_z\} $ or $\mathcal{DC_R} \in \{f'(x_s)\neq0, \forall x_s \notin x_m\}$. 
This type of constraint is used to determine the scaling factor after the state overlap solution $f_q^{\star}(x)$ is obtained. The property of nonzero function and derivative values leads to the following identities:
\begin{align}
\label{eq:identity1}
f(x_s) = y_s  \rightarrow   \frac{f_Q(x_s)}{y_s} = 1 = \sqrt{\eta}  \langle \tau(x)|_n \mathbb{D}_n^{(0)}(x_s)  |\psi \rangle_n ,
\end{align}
\begin{align}
\label{eq:identity2}    
f'(x_s) = t_s \rightarrow \frac{f'_Q(x_s)}{t_s} = 1 = \sqrt{\eta} \langle \tau(x)|_n \mathbb{D}_n^{(1)}(x_s) |\psi \rangle_n,
\end{align}
%
where $\mathbb{D}_n^{(0)}(x_s) = \mathbb{B}_n(x_s)/f(x_s)$ and $\mathbb{D}_n^{(1)}(x_s) = \mathbb{B}_n(x_s)\mathbb{G}_n^T/f'(x_s)$ are rank-one matrices dependent on $x$. With Eqs.~\eqref{eq:identity1},~\eqref{eq:identity2} and~\eqref{eq:Mx}, the independent variable $x$ to the $p^{th}$ power can be expressed in terms of the quantum model encoded with a $\mathcal{DC_R}$ as
\begin{align}
\label{eq:x^p1}
x^p  = \sqrt{\eta} \langle \tau(x)|_{n+1} \mathbb{M}_{x^p} \mathbb{D}_n^{(k)}(x_s) |\psi \rangle_n,
\end{align}
where $k\in\{0,1\}$. For an arbitrary function $r(x)$ present in DE that can be represented by a Maclaurin series expansion, one can replace this function with a quantum model using Eq.~\eqref{eq:x^p1}, leading to
%
\begin{align}
\label{eq:r(x)}
    r(x) = \sum_{p=0}^{\infty}c_p x^p \approx \sqrt{\eta} \langle \tau(x)|_{n+1} \left(  \sum_{p=0}^{\bar{p}\leq2^n}c_p\mathbb{M}_{x^p}\right) \mathbb{D}_n^{(k)}(x_s) |\psi \rangle_n.
\end{align}
Note that $r(x)$ in Eq.~\eqref{eq:r(x)} needs a regular constraint $f(x_s)\neq 0$ for $k=0$ or $f'(x_s)\neq 0$ for $k=1$ encoded in the expression. For simplicity, in the following we focus on the $k=0$ case and only use the first $(\bar{p}+1)$ terms to approximate the function, where $\bar{p}\leq2^n$. Eqs.~\eqref{eq:Nx},~\eqref{eq:N1},~\eqref{eq:identity1} and~\eqref{eq:x^p1},~\eqref{eq:r(x)} will be extensively used in particular for the cases of nonlinear and inhomogeneous DEs.

\subsection*{Quantum subroutines}

So far, we have illustrated in detail the steps involved in Fig.~\ref{fig:LE}(a,b), showing how to represent the DE and DC in the Chebyshev space. Examples for constructing the corresponding effective Hamiltonians, formed by the summation and transposition of the operators shown in Fig.~\ref{fig:LE}(c), are presented in the following section. Here, we provide an overview of quantum subroutines that implement the effective Hamiltonian algorithm in terms of quantum circuits.
\begin{figure}[t!]
\begin{center}
\includegraphics[width=1.0\linewidth]{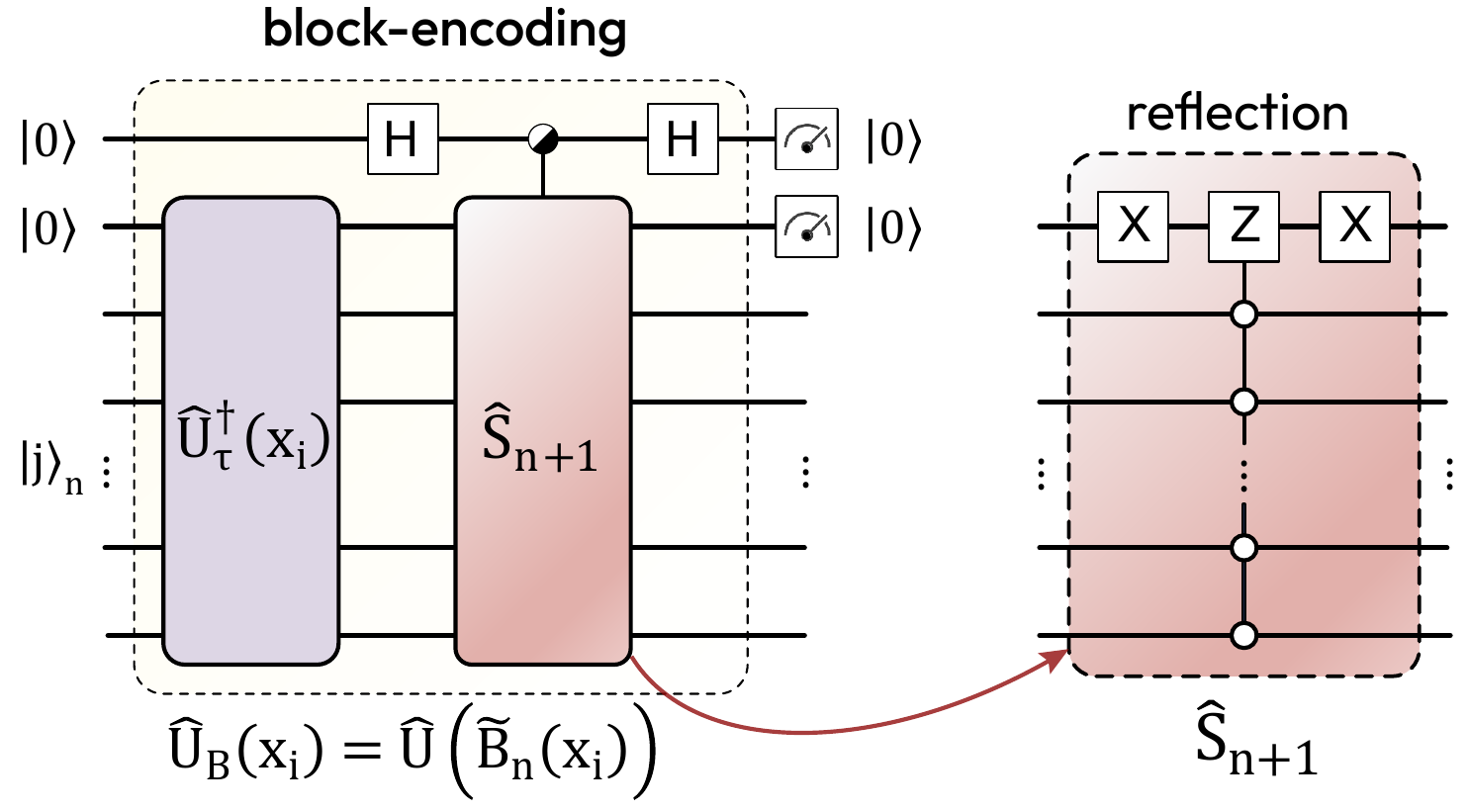}
\end{center}
\caption{
\textbf{Data (invariant) constraint circuit.}
Circuit implementing a block-encoding $\hat{\text{U}}(\tilde{\mathbb{B}}_n(x_i))$ of the data constraint $\tilde{\mathbb{B}}_n(x_i)$ via simple linear combinations of unitaries. Here, $\hat{\mathcal{U}}_{\tau}^\dagger(x_i)$ is an inverse Chebyshev feature map with $x_i \in \{x_z,x_m\}$ (see \cite{Williams2023} and SM, Fig.~\ref{fig:ChebFM}). $\hat{\text{S}}_{n+1} = 2 |0\rangle^{\otimes (n+1)} \langle0|^{\otimes (n+1)} -\mathds{1}_{n+1}$ is a Grover-style operator for performing a reflection about the all zeros state. Note that the global phase gate contributing the value of $-1$ in $\hat{\text{S}}_{n+1}$ circuit matters for reflections implemented as parts of a larger circuit (i.e. in the case of nonlinear DE constraints), but here is omitted for simplicity. The black-and-white circle on ancilla means that the unitary can be controlled by either 0 or 1.
}
\label{fig:U_B}
\end{figure}

To run the proposed algorithm, we need to perform ground state preparation (GSP) for effective Hamiltonians composed for each problem. This can be done with various approaches, including quantum imaginary-time evolution (based on QSVT \cite{Low2019,Gilyen} or LCU \cite{Childs2012}) and quantum thermalization \cite{chen2023quantum}, with their pros and cons in terms of depth and width \cite{algorithms_survey}. Also, near-optimal ground state preparation can be achieved with eigenstate filtering techniques for cases with bounded gaps \cite{Lin2020nearoptimalground}. In our simulations, we adopt a quantum imaginary-time evolution (QITE) (see Sec.~\ref{sec:QITE} of Methods), and implement indefinite parity QSVT sequences that run $\exp(-\tau \mathbb{H})$, confirming that corresponding GSP leads to high-quality solutions. We comment on the question of gap and GSP time in the Discussion section. 

Note that our algorithm can rely on any efficient approach suitable for the task, and ultimately the choice depends on the number of queries to $\mathbb{H}$ and the time required for GSP. However, no matter what we choose, there is a need to block-encode effective Hamiltonians. Here, we follow the general strategies outlined in Ref.~\cite{Low2019}, noting that block-encodings require implementing linear combinations of unitaries or oracles for loading matrix elements. Efficient approaches to block-encoding were studied recently in Refs.~\cite{Daan_FABLE,Daan2022,Sunderhauf2024blockencoding,lapworth2025precondition}. Although exact implementations vary from problem to problem, we stress that, in general, the required $\mathbb{H}$ contains a small number of terms that can be block-encoded separately and efficiently collected with LCU. Our goal is to show that each term can be block-encoded, specifically involving: 1) derivatives as a major part of physics-informed constraints; 2) data constraints that bias correct solutions; 3) multiplication circuits. We design circuits for block-encodings and provide details in Methods (Sec.~\ref{sec:Dn(xs)andGnT}), as well as visualizing operators at the beginning of Supplemental Materials. Below we provide a short version for these procedures.

Specifically, let us show how to block-encode some of the data constraints. Notice that $\mathbb{B}_n(x)$ is a structured and sparse matrix dependent on $n$ and $x \in \{x_z,x_m\}$. In particular, $\mathbb{B}_n(x) = \sqrt{2^n} |0\rangle^{\otimes n} \langle\tau(x)|_n$ can be interpreted as taking the top-left $2^n \times 2^n$ block of $\sqrt{2^n} \Bigl( |0\rangle^{\otimes (n+1)} \langle0|^{\otimes (n+1)} \Bigr) \hat{\mathcal{U}}_{\tau}^\dagger(x)$. The zero-state projector follows a simple LCU decomposition written as $|0\rangle^{\otimes (n+1)} \langle0|^{\otimes (n+1)} = (\hat{\text{S}}_{n+1} + \mathds{1}_{n+1})/2$, where the unitary operator $\hat{\text{S}}_{n+1} = 2 |0\rangle^{\otimes (n+1)} \langle0|^{\otimes (n+1)} -\mathds{1}_{n+1} $ implements the reflection about the all zeros state~\cite{grover1996algorithms}. We also note that the circuit for $\hat{\text{S}}_{n+1}$ is not unique and can be optimized or recompiled for any quantum computing architecture. 
By applying the Chebyshev feature map $\hat{\mathcal{U}}_{\tau}^\dagger(x_i)$ \cite{Williams2023} (also see SM) at point $x_i$ to the LCU decomposition of projector, we complete the unitary circuit  $\hat{\text{U}}(\tilde{\mathbb{B}}_n(x))$ simply denoted as $\hat{\text{U}}_{\text{B}}(x)$. This unitary is schematically illustrated in Fig.~\ref{fig:U_B}. It requires $n+2$ qubits, runs with high probability, and embeds a scaled matrix $\tilde{\mathbb{B}}_n(x) = \mathbb{B}_n(x)/\sqrt{2^{n+1}}~\forall x \in \{x_z,x_m\}$. 

Similar approach holds for block-encoding $\mathbb{G}_n^T$ matrix exclusively dependent on $n$. The nature of this upper triangular matrix allows us to systematically construct its quantum circuit based on LCU operations and block encoding strategies proposed by Camps et al.~\cite{Daan_FABLE,Daan2022}. While the procedure is detailed in Methods (Sec.~\ref{sec:Dn(xs)andGnT}), here would like to highlight that Chebyshev differentiation operators can be implemented efficiently in the quantum circuit form. This follows from the structure of $\mathbb{G}_n$ that can be described in a sequence form. In this case, oracles need to store only a small number of base terms and the order of terms, making it qualitatively similar to matrices with degenerate terms \cite{Sunderhauf2024blockencoding}.

Finally, once operators are block-encoded and QSVT sequences are run, there is a turn to read-out solution. While generally it is a problem \cite{Biamonte2017,algorithms_survey,Williams2024readout}, our models based on feature maps enable function evaluation at specified points $x_i$. This follows from measuring the state overlap $f_q(x_i)$ [Eq.~\eqref{eq:quantum model_n}]. In general, such overlap measurement can be performed through the Hadamard test~\cite{Mitarai2019,Paine2023} between the Chebyshev basis state $\langle 0_a| \langle \tau(x)|_n = \langle 0_a \mathrm{\o} |\hat{\mathcal{U}}_{\tau}^\dagger (x)$ and the prepared ground state $|0_a\rangle |\psi_g \rangle_n$. 
However, such test relies on the use of global controlled-unitary operations via an extra ancillary qubit. This increases complexity, and for early fault-tolerant devices alternative strategies can be adopted.

A potentially better option that avoids ancillas and controls is the interferometric measurement protocol~\cite{Kyriienko,Stefano}, which involves both the ground and the combined states (see Sec.~\ref{sec:M_ODE},~\ref{sec:M_PDE} and~\ref{sec:M_NDE} of Methods) and adopted for the experiments carried out in this work. Specifically, in Methods, we show that the real part of $f_Q(x)$ can be retrieved from the interferometric measurement such that a single overlap measurement can be obtained from two separate evaluations of the expectation value of the observable $\hat{\mathcal{O}} = |0_a \mathrm{\o}\rangle \langle 0_a \mathrm{\o}|$ using the same circuit architecture. This corresponds to measuring the overlap probability $P(x) = |\langle \tau(x)|_n |\psi_g \rangle_n |^2 = |\langle 0_a| \langle \tau(x)|_n |0_a\rangle |\psi_g \rangle_n |^2$.  The corresponding quantum circuit is schematically illustrated in Fig.~\ref{fig:CDEs}(a). It must be emphasized that we are not interested in the absolute value square of the state overlaps, but the state overlaps themselves. Note that we have $f_Q(x) \equiv \text{Re}[f_Q(x)]$ as the Chebyshev basis state $\langle \tau(x)|_n$ is by construction real only and the solution state $|\psi_g\rangle_n$ is also real because the effective Hamiltonian is a real Hermitian operator. Therefore, no additional experiments are required to extract the imaginary part.
Where needed, the final anchoring/rescaling of $f_q^{\star}(x)$ can be done in classical post-processing as already discussed in Fig.~\ref{fig:LE}(d).


\section*{Results}

\subsection*{I. Ordinary differential equations with constant coefficients}

Let us now apply the developed algorithm to solve exemplary differential equation and test its operation. We first consider a second‐order ordinary differential equation (ODE) with constant coefficients $\mathcal{C}=\{a, b, c\} \in \mathbb{R}$,
\begin{equation}
\label{eq:CDE}
\begin{aligned}
  a\frac{d^2f(x)}{dx^2} + b\frac{df(x)}{dx} + cf(x) = 0. 
\end{aligned}
\end{equation}
The corresponding quantum model for Eq.~\eqref{eq:CDE} reads $ \langle \tau(x)|_n \sqrt{\eta} \mathbb{A} |\psi \rangle_n = 0$, where $\mathbb{A} = (a{\mathbb{G}_n^T}^2+b\mathbb{G}_n^T+c\mathds{1}_n )$ is an $x$-independent matrix. Because $ \langle \tau(x)|_n$ is the Chebyshev state with a linearly independent basis, we can get rid of the $x$-dependent part and keep intact its latent representation, that is, $\sqrt{\eta} \mathbb{A} |\psi \rangle_n = 0$. This is equivalent to the condition of $\eta \langle \psi |_n \mathbb{A}^T \mathbb{A} |\psi \rangle_n = 0$ by multiplying its transpose from the left. In the same manner, according to Eqs.~\eqref{eq:bc1} and~\eqref{eq:bc2}, the latent representations of both types of invariant constraint, $\mathcal{DC_I} = \{f(x_z)=0\}$ and $\mathcal{DC_I} = \{f'(x_m)=0\}$, are $\eta \langle \psi|_n \mathbb{B}_n^T(x_z) \mathbb{B}_n(x_z) |\psi \rangle_n = 0 $ and $\eta \langle \psi |_n \mathbb{G}_n \mathbb{B}_n^T(x_m) \mathbb{B}_n(x_m) \mathbb{G}_n^T |\psi \rangle_n = 0$, respectively. For the sake of succinctness, from now on, the Gram operator $\mathcal{T}(\cdot) = (\cdot)^{T}(\cdot)$ is used to concisely represent the operands in the formula.
\begin{figure}[t!]
\begin{center}
\includegraphics[width=1.0\linewidth]{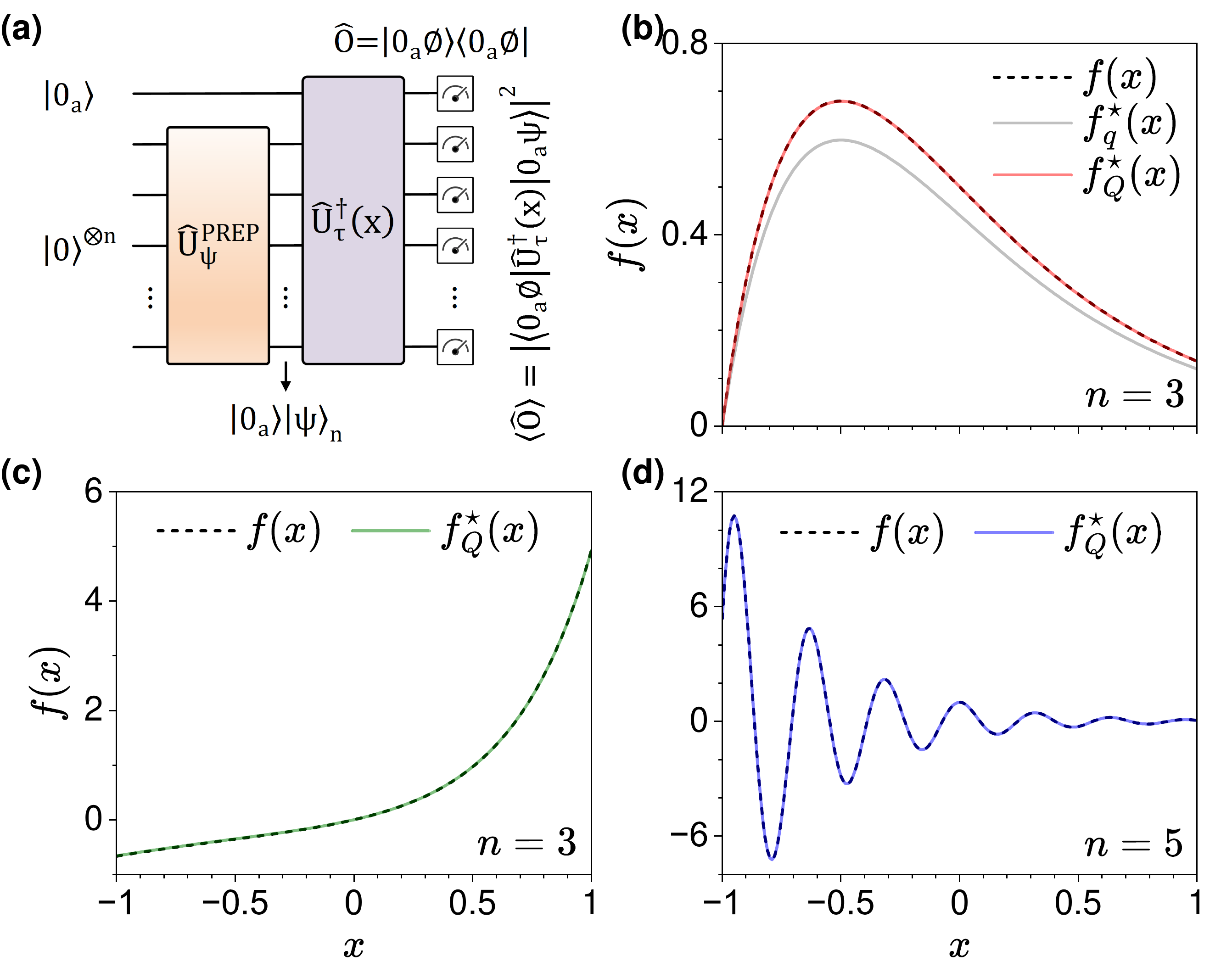}
\end{center}
\caption{
\textbf{Results of solving second-order DEs with constant coefficients.} \textbf{(a)} Quantum circuit for evaluating the state overlaps, $ \langle 0_a \tau(x) | 0_a \psi \rangle$, through the interferometric measurement where a single overlap measurement is obtained from two separate measurements of overlap probabilities [Eq.~\eqref{eq:Refx}]. The circuit is initialized with a zero-product state and a state preparation circuit is responsible to prepare the desired state of a problem of interest, followed by an inverse Chebyshev feature map $\hat{\mathcal{U}}_{\tau}^\dagger (x)$ to obtain the overlap probabilities in the Chebyshev basis. In all cases, Eq.~\eqref{eq:CDE} with coefficients $\mathcal{C} = \{a,b,c\}$ and boundary conditions $\mathcal{BC} = \{f(0),f'(0)\}$ are used to determine a unique analytical solution $f(x)$. 
Plots of $f(x)$ and the evaluated quantum model $f_Q^{\star}(x)$ for
\textbf{(b)} $\mathcal{C} = \{1,4,4\}, \mathcal{BC} = \{0.5,-0.5\}$ and $\mathcal{DC_I} = \{f(x_z=-1)\}, \eta_{e} = 1.29$.
\textbf{(c)} $\mathcal{C} = \{1,-2,-3\}, \mathcal{BC} = \{0,1\}$ and $\mathcal{DC_I} = \{f(x_z=0)\}, \eta_{e} = 32.47$.
\textbf{(d)} $\mathcal{C} = \{1,5,400\}, \mathcal{BC} = \{1,0\}$ and $\mathcal{DC_I} = \{f'(x_m=0)\}, \eta_{e} = 451.71$.
In each panel, $n$ and $\eta_{e}$ represent the number of qubits and the scaling factor of the quantum model, respectively. 
}
\label{fig:CDEs}
\end{figure}
Next, we write a total energy function $\mathcal{E}$ as a sum of individual latent contributions given by
\begin{equation}
\label{eq:Heff_LDE}
\begin{aligned}
    \mathcal{E} =\eta \langle \psi|_n \mathbb{H} |\psi \rangle_n = 0,   
\end{aligned}
\end{equation}
where $\mathbb{H} = \mathcal{T}(\mathbb{A}) + \mathcal{T}(\mathbb{B}_n(x_z))$ is the effective Hamiltonian operator composed of the latent representations of an $x$-independent DE and an $x$-dependent $\mathcal{DC_I}$. Mathematically, $\mathcal{E}$ is formulated as a quadratic function. Note that the second term in $\mathbb{H}$ can be replaced with $\mathcal{T}(\mathbb{B}_n(x_m) \mathbb{G}_n^T)$ in the case that the solution to DE only has zero slopes without zero crossings (see Fig.~\ref{fig:CDESF1}b). 

To showcase the use of the proposed algorithm, three possible types of solution to Eq.~\eqref{eq:CDE} are sequentially investigated. With real repeated roots for given $\mathcal{C}$ and $\mathcal{BC}$ (see the caption of Fig.~\ref{fig:CDEs}(b) for details), this ODE has an analytical solution $f(x)=0.5 \left( \text{exp}(-2x) + x\, \text{exp}(-2x) \right)$, which represents a log-normal-like function (black dashed curve). For the supplied $\mathcal{DC_I}$, the lowest-energy eigenstate of $\mathbb{H}=\mathcal{T}(\mathbb{A})+\mathcal{T}(\mathbb{B}_n(-1))$ denoted as $|\psi_g \rangle_n$ is prepared, and the resulting state overlap solution $f_q^{\star}(x) = \langle \tau(x)|_n |\psi_g \rangle_n$ (gray solid curve) is directly proportional to the exact solution. 
In addition, the exact value of the scaling factor can be determined by $\sqrt{\eta_e}= f(x_s)/f_q^{\star}(x_s)$ via $\mathcal{DC_R} = \{f(x_s) \neq 0\}$ for an arbitrary $ x_s \notin x_z$. The quantum model is evaluated by the quantum circuit shown in Fig.~\ref{fig:CDEs}(a). 
The resulting solution $f_Q^{\star}(x)= \sqrt{\eta_e} f_q^{\star}(x)$, displayed as a red solid curve in Fig.~\ref{fig:CDEs}(b), closely follows the analytical solution. 

Figs.~\ref{fig:CDEs}(c,d) show other exemplary DEs with real distinct and complex pairs of roots, whose analytical solutions are $f(x)= \frac{1}{4} \left(\text{exp}(3x) - \text{exp}(-x) \right)$ and $f(x)=\text{exp}(-5x/2) \left( \text{cos}(15\sqrt{7}x/2) + \sqrt{7}/21 \, \text{sin}(15\sqrt{7}x/2) \right)$, respectively. The evaluated quantum models coincide with their respective exact solutions, as expected. We also present an extended analysis in Supplementary Fig.~\ref{fig:CDESF1} for other cases of ODEs with constant coefficients.

\subsection*{II. Ordinary differential equations with variable coefficients}

We apply the proposed algorithm to address the general Legendre differential equation with $0 \leq m\leq l \in \mathbb{N}$ ,
\begin{equation}
\label{eq:VDE}
\begin{aligned}
  (1-x^2)\frac{d^2f(x)}{dx^2} -2x\frac{df(x)}{dx}+  \left( l(l+1)-\frac{m^2}{1-x^2} \right) f(x) = 0. 
\end{aligned}
\end{equation}
The complete solution to Eq.~\eqref{eq:VDE} is $c_1 P^m_l(x)+ c_2 Q^m_l(x)$, where $P^m_l(x)$ and $Q^m_l(x)$ are two linearly independent functions called associated Legendre functions of the first and second kind, respectively. They are frequently used when solving Laplace's equation and scattering problem in spherical coordinates ~\cite{HYWu2023}. 
\begin{figure}[th]
\begin{center}
\includegraphics[width=1.0\linewidth]{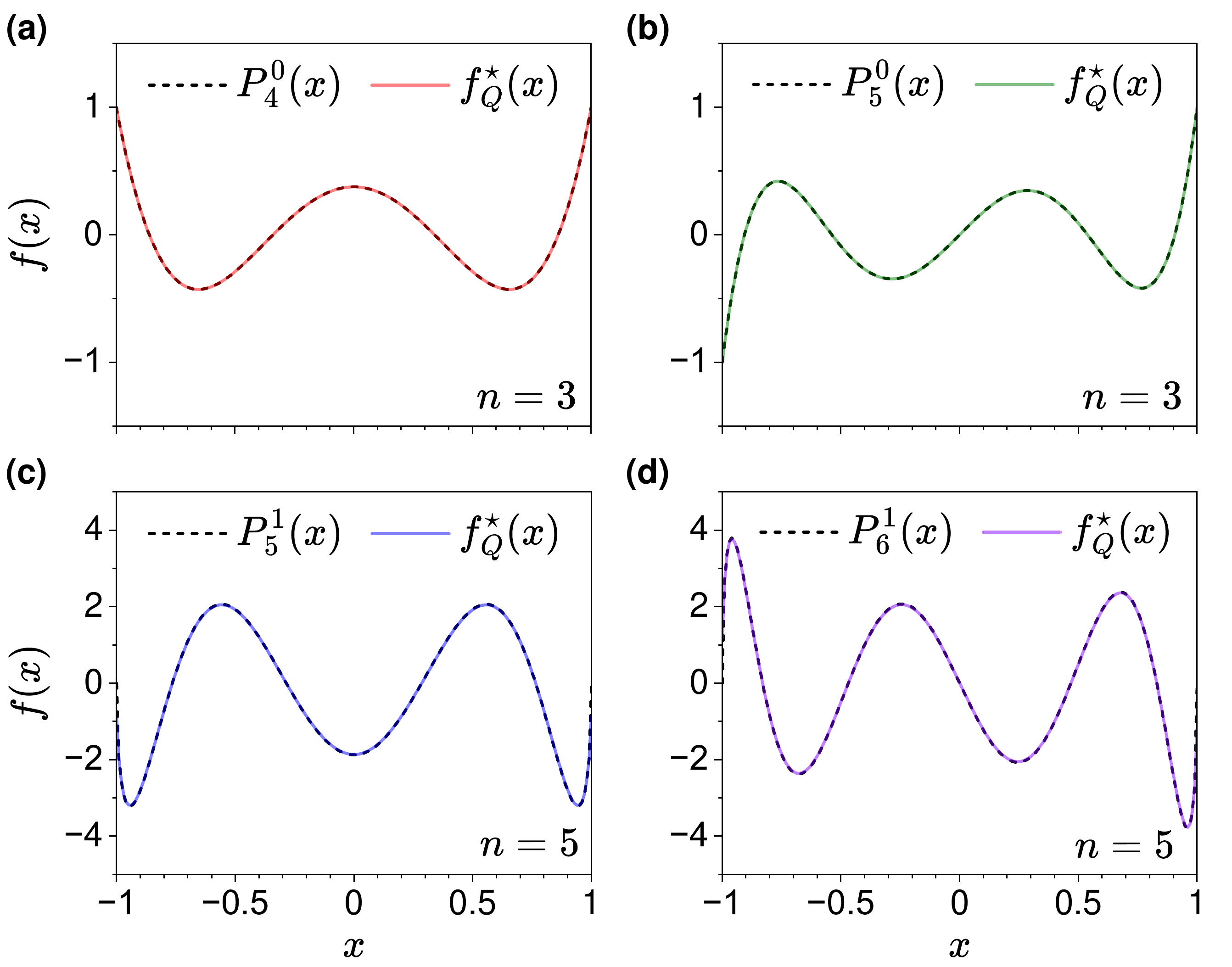}
\end{center}
\caption{
\textbf{Results of solving Legendre’s and associated Legendre’s differential equations.}
For $m=0$, Eq.~\eqref{eq:VDE} with $l$-dependent boundary conditions $\mathcal{BC} = \{f(-1)=(-1)^l,f(1)=1\}$ is used to determine a unique analytical solution $f(x) = P^0_l(x)$. Plots of $P^0_l(x)$ and $f_Q^{\star}(x)$ for \textbf{(a)} $l=4$ and $\mathcal{DC_I} = \{f'(x_m=0)\}$, $\eta_{e}=1.75$ and \textbf{(b)} $l=5$ and $\mathcal{DC_I} = \{f(x_z=0)\}$, $\eta_{e}=1.49$. For $m=1$, Eq.~\eqref{eq:VDE} with fixed boundary conditions $\mathcal{BC} = \{f(-1)=0,f(1)=0\}$ is used to determine a unique analytical solution $f(x) = P^1_l(x)$. Plots of $P^1_l(x)$ and $f_Q^{\star}(x)$ for \textbf{(c)} $l=5$ and $\mathcal{DC_I} = \{f'(x_m=0)\}, \eta_{e}=108.51$ and \textbf{(d)} $l=6$ and $\mathcal{DC_I} = \{f(x_z=0)\}, \eta_{e}=135.38$. In each panel, $n$ and $\eta_{e}$ represent the number of qubits and the scaling factor of the quantum model, respectively.
}
\label{fig:VDEs}
\end{figure}
After substituting Eq.~\eqref{eq:quantum model_n} into Eq.~\eqref{eq:VDE}, we note that multiplication of $x^p$ with the $n$-qubit Chebyshev basis state implies that the leading power of $x$ is raised from $2^n-1$ to $2^n-1+p$ in the Chebyshev polynomials, meaning that $n+1$ qubits is necessary to ensure that each DE term shares the same linearly independent basis set [see Eqs.~\eqref{eq:Mx} and~\eqref{eq:M1}]. This key step enables us to remove the $x$-dependent part and keep intact the latent representation as previously described. Therefore, the quantum model in Eq.~\eqref{eq:quantum model_n} can be written in an alternative form via Eq.~\eqref{eq:M1} as
\begin{align}
\label{eq:quantum model_np1}
    f_Q(x) = \sqrt{\eta} f_q(x) &= \sqrt{\eta} \langle \tau(x)|_{n+1} \mathbb{M}_1 |\psi \rangle_n,  \quad \text{s.t.}\; \langle \psi |_n | \psi \rangle_n = 1.
\end{align}
Likewise, $\mathcal{DC_I}$ must be expressed in terms of $(n+1)$-qubit Chebyshev basis state, and thus Eqs.~\eqref{eq:bc1} and~\eqref{eq:bc2} are modified as follows:
\begin{equation}
\label{eq:bc1n+1}
\begin{aligned}
 f(x_z) = 0 \rightarrow \sqrt{\eta} \langle \tau(x)|_{n+1} \mathbb{B}_{n+1}(x_z) \mathbb{M}_1 |\psi \rangle_n =0, 
\end{aligned}
\end{equation}
\begin{equation}
\label{eq:bc2n+1} 
\begin{aligned}   
f'(x_m) = 0 \rightarrow \sqrt{\eta} \langle \tau(x)|_{n+1} \mathbb{B}_{n+1}(x_m) \mathbb{M}_1 \mathbb{G}_n^T |\psi \rangle_n =0, 
\end{aligned}
\end{equation}
where $\mathbb{B}_{n+1}(x_z)$ and $\mathbb{B}_{n+1}(x_m)$ are obtained from $\langle \tau(x_z)|_{n+1} =\langle \tau(x)|_{n+1} \mathbb{B}_{n+1}(x_z)$ and $\langle \tau(x_m)|_{n+1} =\langle \tau(x)|_{n+1} \mathbb{B}_{n+1}(x_m)$, respectively. 
Specifically, $\mathbb{B}_{n+1}(x) = \sqrt{2^{n+1}} |0\rangle^{\otimes (n+1)} \langle\tau(x)|_{n+1}$ is an $x$-dependent rank-one matrix with nonzero entries in the first row (see SM for details). For the azimuthally symmetric case ($m=0$), the quantum model of Eq.~\eqref{eq:VDE} is $ \langle \tau(x)|_{n+1} \sqrt{\eta} \mathbb{A} |\psi \rangle_n = 0$, where $\mathbb{A} = (\mathbb{M}_1-\mathbb{M}_{x^2}) {\mathbb{G}_n^T}^2-2\mathbb{M}_{x}\mathbb{G}_n^T+l(l+1)\mathbb{M}_1 $ is an $x$-independent matrix. With $l$-dependent $\mathcal{BC}$ [see caption in Fig.~\ref{fig:VDEs}], this ODE has an analytical solution $f(x)=P^0_l(x)$, the Legendre polynomials, as shown in the black dashed curves. $\mathcal{DC_I} = \{f'(x_m=0)\}$ and $\{f(x_z=0)\}$ are applied to the cases of even and odd integers of $l$, respectively. Thus we have the corresponding effective Hamiltonian operators
$\mathbb{H}_e=\mathcal{T}(\mathbb{A})+\mathcal{T}(\mathbb{B}_{n+1}(0)\mathbb{M}_1\mathbb{G}_n^T)$ and
$\mathbb{H}_o=\mathcal{T}(\mathbb{A})+\mathcal{T}(\mathbb{B}_{n+1}(0)\mathbb{M}_1)$. The total energy function $\mathcal{E}$ is the same as Eq.~\eqref{eq:Heff_LDE}.

In the case of $m \neq 0$, after multiplying both sides of Eq.~\eqref{eq:VDE} by $(1-x^2)$, we obtain the latent governing equation $\mathbb{A} = (\mathbb{M}_1-2\mathbb{M}_{x^2}+\mathbb{M}_{x^4}) {\mathbb{G}_n^T}^2 + 2 (\mathbb{M}_{x^3}-\mathbb{M}_{x})\mathbb{G}_n^T - l(l+1)\mathbb{M}_{x^2} +  \left( l(l+1)-m^2  \right) \mathbb{M}_1$. For $m=1$ with fixed $\mathcal{BC}$, this DE has an analytical solution $f(x)=P^1_l(x)$, the associated Legendre polynomials, as shown in the black dashed curves. Given the same $\mathcal{DC_I}$ applied to the case of integer $l$ of opposite parity, we have
$\mathbb{H}_o=\mathcal{T}(\mathbb{A})+\mathcal{T}(\mathbb{B}_{n+1}(0)\mathbb{M}_1\mathbb{G}_n^T)$ and
$\mathbb{H}_e=\mathcal{T}(\mathbb{A})+\mathcal{T}(\mathbb{B}_{n+1}(0)\mathbb{M}_1)$. 
Following the same procedures as previously described, the ground states $|\psi_g \rangle_n$ for each case of effective Hamiltonian operators are prepared. 
We then use the same quantum circuit [Fig.~\ref{fig:CDEs}(a)] to evaluate quantum models [Eq.~\eqref{eq:quantum model_n}] using Eq.~\eqref{eq:Refx}. As shown in Fig.~\ref{fig:VDEs}, the evaluated quantum models (colored solid curves) closely follow the target functions for $m=0$ and exhibit best-fit behavior even though the singular points occur at $x=-1$ and $x=1$ for $m=1$. Please also refer to Supplementary Figs.~\ref{fig:VDESF2} and~\ref{fig:VDESF3} for other cases of ODEs with variable coefficients.

\subsection*{III. Inhomogeneous ordinary differential equations}

\begin{figure}[t!]
\begin{center}
\includegraphics[width=1.0\linewidth]{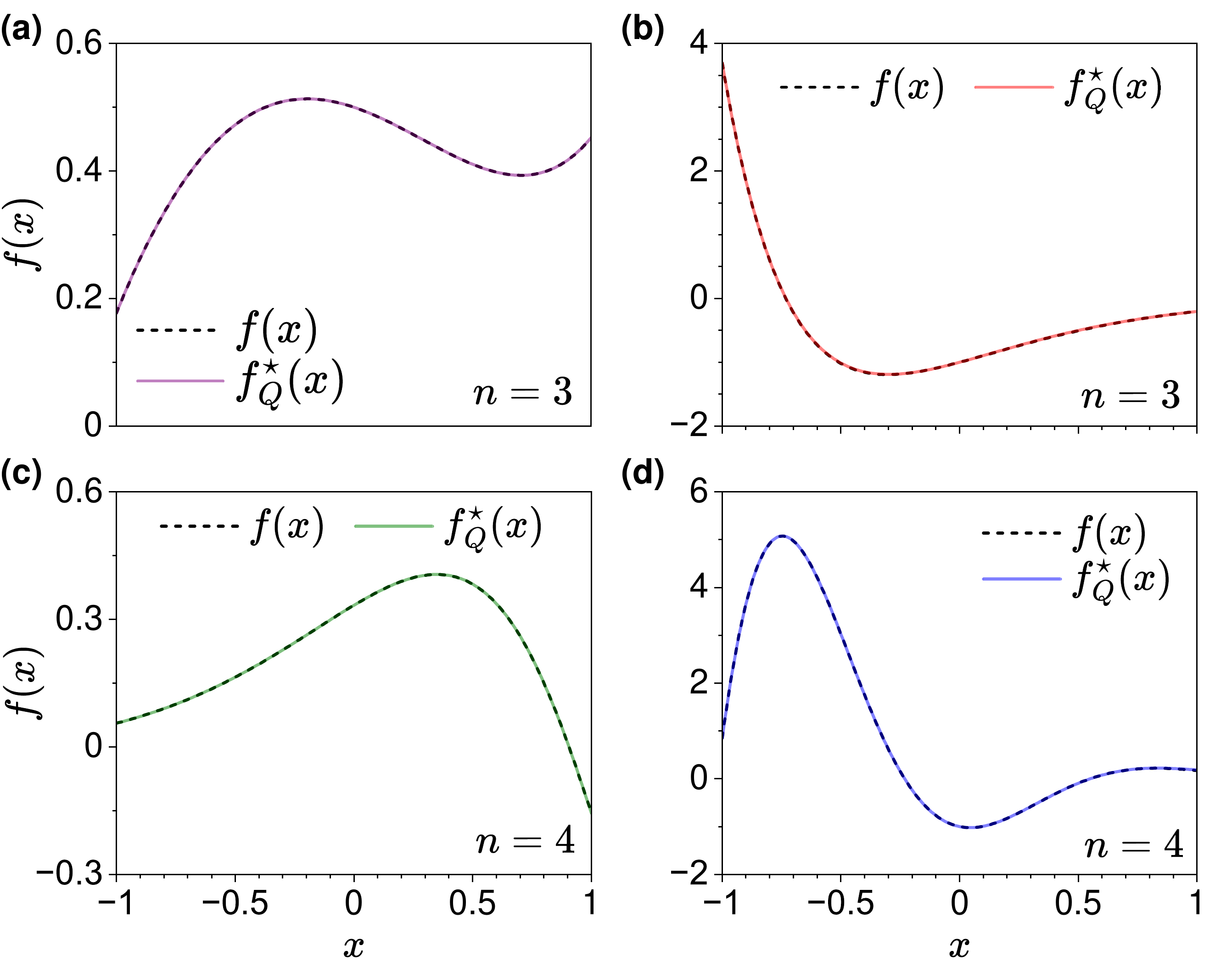}
\end{center}
\caption{
\textbf{Results of solving second-order inhomogeneous DEs.} In all cases, Eq.~\eqref{eq:IDE} with coefficients and a source function denoted as $\mathcal{C} = \{o(x),p(x),q(x);r(x)\}$ and boundary conditions $\mathcal{BC} = \{f(0),f'(0)\}$ are used to determine a unique analytical solution $f(x)$. 
Plots of $f(x)$ and $f_Q^{\star}(x)$ for 
\textbf{(a)} $\mathcal{C} = \{(x-1),-x,1;(x-1)^2\}, \mathcal{BC} = \{1/2,-1/8\}$ and $\mathcal{DC_I} = \{f'(x_m=-0.2)\}, \eta_{e} = 1.35$.
\textbf{(b)} $\mathcal{C} = \{1,4,4;e^{-2x}\}, \mathcal{BC} = \{-1,1\}$ and $\mathcal{DC_I} = \{f(x_z=-0.73)\}, \eta_{e} = 14.74$.
\textbf{(c)} $\mathcal{C} = \{1,-5,6;xe^{2x}\}, \mathcal{BC} = \{1/3,1/3\}$ and $\mathcal{DC_I} = \{f(x_z=0.91)\}, \eta_{e} = 0.82$.
\textbf{(d)} $\mathcal{C} = \{1,4,20;xe^{-2x}\text{sin}(4x)\}, \mathcal{BC} = \{-1,-1\}$ and $\mathcal{DC_I} = \{f'(x_m=-0.75)\}, \eta_{e} = 77.51$.
In each panel, $n$ and $\eta_{e}$ represent the number of qubits and the scaling factor of the quantum model, respectively. 
}
\label{fig:IDEs}
\end{figure}

We now examine a second‐order inhomogeneous differential equation, with coefficients and a source function on either side of the equal sign denoted as $\mathcal{C}=\{o(x), p(x), q(x);r(x)\} \in \mathbb{R} $, given by
\begin{equation}
\label{eq:IDE}
\begin{aligned}
  o(x) \frac{d^2f(x)}{dx^2} + p(x) \frac{df(x)}{dx} + q(x) f(x) = r(x). 
\end{aligned}
\end{equation}
For given $\mathcal{C}$ and $ \mathcal{BC}$ (see caption in Fig.~\ref{fig:IDEs} for details), this DE has an analytical solution $f(x)=1.5\text{exp}(x)-0.125x(8x+13)-1$. We follow the same procedures as described above and use Eq.~\eqref{eq:x^p1} to find the latent representation of $r(x)$.

The quantum model of Eq.~\eqref{eq:IDE} is $ \langle \tau(x)|_{n+1} \sqrt{\eta} \mathbb{A} |\psi \rangle_n = 0$, where 
$\mathbb{A} = (\mathbb{M}_x-\mathbb{M}_1){\mathbb{G}_n^T}^2-\mathbb{M}_x\mathbb{G}_n^T+\mathbb{M}_1 - (\mathbb{M}_{x^2}-2\mathbb{M}_{x}+\mathbb{M}_1)\mathbb{D}_n^{(0)}(x_s) $ is a data-encoded matrix given with an arbitrary regular constraint $f(x_s)\neq 0$. With the supplied $\mathcal{DC_I}$, the lowest-energy eigenstate of $\mathbb{H}=\mathcal{T}(\mathbb{A})+\mathcal{T}(\mathbb{B}_{n+1}(-0.2)\mathbb{M}_1\mathbb{G}_n^T)$ is prepared and then the quantum model [Eq.~\eqref{eq:quantum model_n}] is evaluated by the quantum circuit [Fig.~\ref{fig:CDEs}(a)]. 
The scaling factor can be determined through $f(x_s)$, the same as the one appearing in $\mathbb{D}_n^{(0)}(x_s)$. The evaluated quantum model agrees with the exact solution, as shown in Fig.~\ref{fig:IDEs}(a). For complicated source functions, the Maclaurin expansion of $r(x)$ is applied and the first $\bar{p}+1$ terms are taken for approximation [Eq.~\eqref{eq:r(x)}].

Next, let us return to Eq.~\eqref{eq:CDE} with $r(x)$ on the right-hand side. The corresponding latent governing equation is $\mathbb{A} = \mathbb{M}_1 ( a{\mathbb{G}_n^T}^2+b\mathbb{G}_n^T+c\mathds{1}_n )- \left(  \sum_{p=0}^{\bar{p}}c_p\mathbb{M}_{x^p}\right)\mathbb{D}_n^{(0)}(x_s)$. Again, we consider three exemplary DEs with real repeated, real distinct, and complex pairs of roots [see caption in Figs.~\ref{fig:IDEs}(b,c,d)], whose analytical solutions are $f(x)=\frac{1}{2}\text{exp}(-2x)(x^2-2x-2)$, $f(x)=\frac{1}{6}\left( 4 \text{exp}(3x)-\text{exp}(2x)(3x^2+6x+2) \right)$ and $f(x)=\frac{1}{64}\text{exp}(-2x) \left( -4(x^2+16) \text{cos}(4x) + (x-48) \text{sin}(4x) \right)$, respectively. Given $\mathcal{DC_I}$, the effective
Hamiltonian operators are $\mathbb{H}=\mathcal{T}(\mathbb{A})+\mathcal{T}(\mathbb{B}_{n+1}(x_z)\mathbb{M}_1)$ for the first two cases and $\mathbb{H}=\mathcal{T}(\mathbb{A})+\mathcal{T}(\mathbb{B}_{n+1}(x_m)\mathbb{M}_1\mathbb{G}_n^T)$ for the last. The resulting quantum models evaluated with $\bar{p}=3,5,7$ reveal excellent agreement with the exact solutions, as shown in Figs.~\ref{fig:IDEs}(b,c,d), respectively. Understanding how to represent any function approximated by the Maclaurin series expansion as a quantum model is the key to successfully addressing inhomogeneous DEs.

\subsection*{IV. Partial differential equations}

We proceed to apply the proposed algorithms from univariate to multivariate cases. For simplicity, we first demonstrate how to tackle a partial differential equation (PDE) involving partial derivatives of a dependent variable $f(x,y)$ with respect to two independent variables $x$ and $y$. The unknown function can be expressed as $f(x,y) = \sum_{k=0}^{2^n-1}\sum_{l=0}^{2^n-1}c_{k,l}T_k(x)T_l(y)$. 
\begin{figure*}[t!]
\begin{center}
\includegraphics[width=1.0\linewidth]{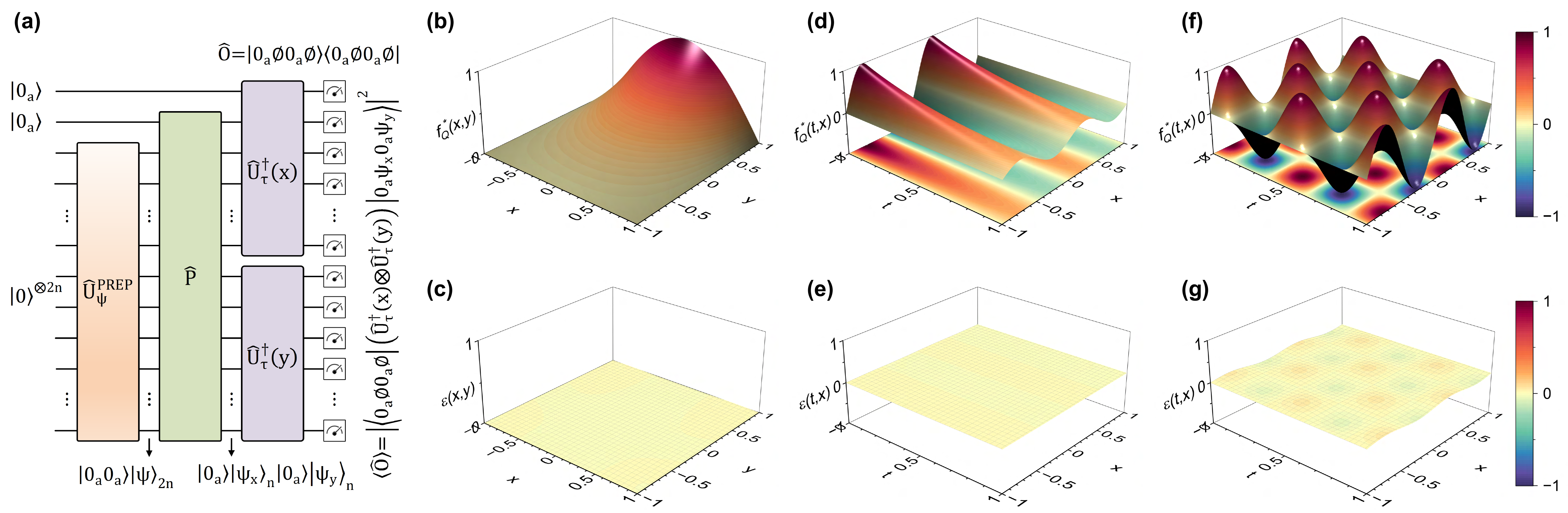}
\end{center}
\caption{
\textbf{Results of solving partial differential equations.}
\textbf{(a)} Quantum circuit for evaluating the two-dimensional state overlaps, $ \langle 0_a \tau(x) 0_a \tau(y) | 0_a \psi_x 0_a \psi_y \rangle$, assuming the the interferometric measurement approach [see Eq.~\eqref{eq:Refxy}]. 
Here, $\hat{\text{P}}$ is a permutation circuit that reshuffles states in a specific order. Parallel inverse Chebyshev feature maps are used to read out overlap probabilities in the Chebyshev basis. \textbf{(b,d,f)} Colormap surface plots of effective Hamiltonian-based solutions for Laplace's equation [Eq.~\eqref{eq:Laplce} with $n=3$ and $\eta_{e}=5.21559$, shown in panel \textbf{(b)}]; heat equation [Eq.~\eqref{eq:Heat} with $n=4$, $\eta_{e}=603.863$, shown in panel \textbf{(d)}]; and wave equation [Eq.~\eqref{eq:Wave} with $n=5$ and $\eta_{e}=216.469$, shown in panel \textbf{(f)}]. \textbf{(c,e,g)} Corresponding error plots $\varepsilon$ are shown at the bottom for each plot, defined as the difference between the approximated and true function values. All plots share the same color bar on the right. Please also refer to the Sec.~\ref{sec:AEHW} of Methods where the permutation circuit is removed.
}
\label{fig:PDEs}
\end{figure*}
In analogy with the two-dimensional Chebyshev expansion, we build a quantum model $f_Q(x,y)$ with parallel registers of equal width $n_x=n_y=n$ to encode two independent variables [Fig.~\ref{fig:PDEs}(a)]. This is given by a scaled two-dimensional state overlap,
\begin{equation}
\label{eq:2D quantum model}
\begin{aligned}
f_Q(x,y)= \sqrt{\eta} f_q(x,y)= \sqrt{\eta} \tau(x,y)|_{2n} | \psi \rangle_{2n}, \ \text{s.t.} \; \langle \psi |_{2n} | \psi \rangle_{2n} = 1,
\end{aligned}
\end{equation}
where $\langle \tau(x,y)|_{2n}$ is a shorthand notation for the two-dimensional linearly-independent Chebyshev basis state $\langle \tau(x)|_n \otimes \langle \tau(y)|_n$ and $f_q(x,y)$ represents the two-dimensional state overlap. We first consider Laplace's equation~\cite{laplaceeq1,laplaceeq2}, an elliptic PDE arises in many applications (such as electrostatics and fluid dynamics) to describe steady-state phenomena. The equation reads
\begin{equation}
\label{eq:Laplce}
\begin{aligned}
\pdv[2]{f(x,y)}{x} + \pdv[2]{f(x,y)}{y} = 0, \\
\mathcal{BC} = \{ f(\pm1,y)=0, \, f(x,-1)=0, \\ f(x,1)=\text{sin}(\pi(x+1)/2)\},
\end{aligned}
\end{equation}
and has an analytical solution $f(x,y) = \text{csch}(\pi) \, \text{cosh}(\pi x/2) \, \text{sinh}(\pi (y+1)/2)$. The quantum model of this PDE is $\langle \tau(x,y)|_{2n} \sqrt{\eta} \mathbb{A} |\psi \rangle_{2n} = 0$, where $\mathbb{A} = ( {\mathbb{G}_n^T}^2 \oplus {\mathbb{G}_n^T}^2 ) $ is an $x$-independent matrix with the symbol $\oplus$ denoting the Kronecker sum operation. Once again, $\mathcal{DC_I}$ must be formulated in the form of the above-mentioned quantum model. Dirichlet and Neumann invariant constraints, respectively, take the form
\begin{equation}
\label{eq:bc1xy}
\begin{aligned}
f(x_z,y) = 0 \rightarrow
\sqrt{\eta} \langle \tau(x,y)|_{2n} \left( \mathbb{B}_n(x_z) \otimes \mathds{1}_n \right) |\psi \rangle_{2n} = 0, \\ 
f(x,y_z) = 0 \rightarrow
\sqrt{\eta} \langle \tau(x,y)|_{2n} \left( \mathds{1}_n \otimes \mathbb{B}_n(y_z)  \right) |\psi \rangle_{2n} = 0, 
\end{aligned}
\end{equation}
\begin{equation}
\label{eq:bc2xy}
\begin{aligned}
\pdv{f(x_m, y)}{x} = 0 \rightarrow
\sqrt{\eta} \langle \tau(x,y)|_{2n} \left( \mathbb{B}_n(x_m) \mathbb{G}_n^T \otimes \mathds{1}_n \right) |\psi \rangle_{2n} = 0, \\ 
\pdv{f(x, y_m)}{y} = 0 \rightarrow
\sqrt{\eta} \langle \tau(x,y)|_{2n} \left( \mathds{1}_n \otimes \mathbb{B}_n(y_m) \mathbb{G}_n^T \right) |\psi \rangle_{2n} = 0, 
\end{aligned}
\end{equation}
where $y_z$ ($y_m$) refers to the $y$ coordinate of a data point at which $f(x,y_z) = 0$ ($ \partial_y{f(x,y_m)} = 0)$. 
Similarly to Eqs.~\eqref{eq:bc1} and~\eqref{eq:bc2}, $\mathbb{B}_n(y_z)$ and $\mathbb{B}_n(y_m)$ are obtained from $\langle \tau(y_z)|_n =\langle \tau(y)|_n \mathbb{B}_n(y_z)$ and $\langle \tau(y_m)|_n =\langle \tau(y)|_n \mathbb{B}_n(y_m)$, respectively. They are $y$-dependent rank-one matrices with few nonzero entries. In contrast to 2D collocation points in PINNs, the regions of minimum variation of the function value and/or the first derivative are chosen as invariant constraints, $\mathcal{DC_I} = \{ f(\pm 1,y)=0, \, f(x, -1)=0\}$. Therefore, the effective Hamiltonian operator $\mathbb{H} = \mathcal{T}( \mathbb{A} ) + \mathcal{T}( \mathbb{B}_n(\pm 1) \otimes \mathds{1}_n ) + \mathcal{T}( \mathds{1}_n \otimes \mathbb{B}_n(-1))$ is obtained and the total energy function $\mathcal{E}$ for PDE is written as
\begin{equation}
\label{eq:Heff_PDE}
\begin{aligned}
    \mathcal{E} = \eta \langle \psi|_{2n} \mathbb{H} |\psi \rangle_{2n} = 0.  \end{aligned}
\end{equation}
The minimum $\mathcal{E}$ can be found when $|\psi\rangle_{2n}$ is the lowest-energy eigenstate ($|\psi_g \rangle_{2n}$) of $\mathbb{H}$. The exact value of the scaling factor is obtained via $\sqrt{\eta_e} = f(x_s,y_s)/f_q^{\star}(x_s,y_s)$ using an arbitrary $\mathcal{DC_R} \in \{f(x_s,y_s)\neq 0\}$. The quantum model is evaluated with the quantum circuit schematically illustrated in Fig.~\ref{fig:PDEs}(a). 
The resulting quantum model solution is displayed as a colormap surface plot in Fig.~\ref{fig:PDEs}(b), being consistent with the exact solution, a hyperbolic sine and cosine function, as demonstrated in the difference plot in Fig.~\ref{fig:PDEs}(c).

We now move on to deal with time-dependent PDEs in one spatial dimension. The dependent variable is a function of time $t$ and $x$. Two representative examples, the heat and wave equations~\cite{heateq1,heateq2,heateq3,waveeq1,waveeq2}, are examined as these parabolic and hyperbolic PDEs are widely used to describe many physical and engineering problems such as diffusion processes/heat flow and vibrating systems/wave motion, respectively. These equations correspond to
\begin{equation}
\label{eq:Heat}
\begin{aligned}
\pdv{f(t,x)}{t} - k \pdv[2]{f(t,x)}{x} &= 0, \\ 
k=\frac{1}{25}, \, \mathcal{IBC} = \{ f(0,x) = \text{sin}(2\pi x), & \, f(t,\pm1)=0\},
\end{aligned}
\end{equation}
\begin{equation}
\label{eq:Wave}
\begin{aligned}
\pdv[2]{f(t,x)}{t} - c^2\pdv[2]{f(t,x)}{x} &= 0, \\
c=2, \, \mathcal{IBC} = \{ f(0,x)=\text{sin}(2\pi x), \, \pdv{f(0,x)}{t} &=0, \, f(t,\pm1)=0\},
\end{aligned}
\end{equation}
where $\mathcal{IBC}$ is the set of initial and boundary conditions to specify the distribution at the initial time and domain boundaries to pin down a unique solution. Eqs.~\eqref{eq:Heat} and~\eqref{eq:Wave} have unique analytical solutions $f(t,x) = \text{exp}(-4 \pi^2 t/25) \, \text{sin}(2 \pi x)$ and $f(t,x) = \text{cos}(4 \pi t) \, \text{sin}(2 \pi x)$, respectively.

The corresponding quantum model is obtained from Eq.~\eqref{eq:2D quantum model} and reads as $f_Q(t,x)= \sqrt{\eta} \langle \tau(t,x)|_{2n} |\psi \rangle_{2n}$, with $\tau(t,x)|_{2n}$ being the two-dimensional linearly-independent Chebyshev basis state $\langle \tau(t)|_n \otimes \langle \tau(x)|_n$. Both PDEs have the same form of the quantum model, $ \langle \tau(t,x)|_{2n} \sqrt{\eta} \mathbb{A} |\psi \rangle_{2n} = 0$, with their respective $\mathbb{A} = \left( \mathbb{G}_n^T \otimes \mathds{1}_n - k \mathds{1}_n \otimes {\mathbb{G}_n^T}^2 \right) $ and $\mathbb{A} = \left( {\mathbb{G}_n^{T}}^2 \otimes \mathds{1}_n - c^2 \mathds{1}_n \otimes {\mathbb{G}_n^T}^2 \right)$. The independent variables $x$ and $y$ in Eqs.~\eqref{eq:bc1xy} and~\eqref{eq:bc2xy} are replaced by $t$ and $x$, respectively. Given $\mathcal{DC_I} = \{f(t,\pm 1)=0, \, \pdv{f(t, \pm 3/4)}{x}=0 \}$ and $\mathcal{DC_I} = \{ f(1/8,x)=0, \, f(7/8,x)=0, \, f(t,\pm1)=0 \}$, we ultimately have $\mathbb{H} = \mathcal{T}( \mathbb{A} ) + \mathcal{T}( \mathds{1}_n \otimes \mathbb{B}_n(\pm 1) ) + \mathcal{T}( \mathds{1}_n \otimes \mathbb{B}_n(\pm 3/4) \mathbb{G}_n^T )$ for Eq.~\eqref{eq:Heat} and $\mathbb{H} = \mathcal{T} ( \mathbb{A} ) + \mathcal{T}( \mathbb{B}_n(1/8) \otimes \mathds{1}_n ) + \mathcal{T}( \mathbb{B}_n(7/8) \otimes \mathds{1}_n ) + \mathcal{T}( \mathds{1}_n \otimes \mathbb{B}_n(\pm 1) )$ for Eq.~\eqref{eq:Wave}. The
quantum models are evaluated following the
same procedures previously described and the results are shown as colormap surface plots in Figs.~\ref{fig:PDEs}(c,d), respectively. Overall, the evaluated quantum models closely follow mixed decaying and oscillating (heat) solution and highly oscillatory (wave) solution, with some minor error at local maxima and minima due to finite basis set size, as shown in Figs.~\ref{fig:PDEs}(e,g).

\subsection*{V. Nonlinear differential equations}

Finally, we apply the algorithms developed to address nonlinear differential equations (NDEs). We first consider a second-order NDE with quadratic nonlinearity on the first derivative, whose analytical solution is an even function, $1-x^2/8$, written as
\begin{equation}
\label{eq:NDE1}
\begin{aligned}
4\frac{d^2f(x)}{dx^2} + 2\left( \frac{df(x)}{dx} \right)^2 + f(x) = 0, \\ \mathcal{BC} = \{ f(0) = 1, f'(0) = 0 \},
\end{aligned}
\end{equation}
We solve the problem following the same workflow as in the linear cases described previously. After substituting Eqs.~\eqref{eq:quantum model_n} and \eqref{eq:quantum model derivative} into Eq.~\eqref{eq:NDE1}, we notice that the quantum state is encoded twice as $|\psi \rangle_{n} \otimes |\psi \rangle_{n}$ in the nonlinear term (contributed by the first derivative squared) whereas the quantum state for the linear term is only encoded once. To maintain the same dimension of Hilbert space, we intentionally multiply the linear terms with the identity (Eq.~\eqref{eq:identity1}) such that $|\psi \rangle_{n} \otimes |\psi \rangle_{n}$ appears in all linear terms at the expense of the Hilbert space doubling. This is expected when dealing with nonlinear problems. Accordingly, the quantum model [Eq.~\eqref{eq:quantum model_n}] and invariant constraints [Eqs.~\eqref{eq:bc1}, \eqref{eq:bc2}] after the introduction of the identity read as
\begin{equation}
\label{eq:nonlinear_quantum model}
\begin{aligned}
     f_Q(x) = \eta f_q(x) &= \eta \bigl(\langle \tau(x)|_n \otimes \langle \tau(x)|_n\bigr) \Bigl(\mathbb{D}_n^{(0)}(x_s) \otimes \mathds{1}_n \Bigr) |\Psi \rangle_{2n}, \\
     &= \eta \langle \tau(x)|_{n+1} \mathbb{N}_1 \Bigl(\mathbb{D}_n^{(0)}(x_s) \otimes \mathds{1}_n \Bigr) |\Psi \rangle_{2n}, \\&
     \text{s.t.} \; \langle \Psi|_{2n}  |\Psi \rangle_{2n} = 1,
\end{aligned}
\end{equation}
\begin{equation}
\label{eq:nonlinear_bc1n+1}
\begin{aligned}
&f(x_z) = 0  \rightarrow \\ 
\eta \langle \tau(x)|_{n+1} \mathbb{N}_1 & \Bigl( \mathbb{D}_n^{(0)}(x_s) \otimes \mathbb{B}_n (x_z) \Bigr) |\Psi \rangle_{2n} = 0,
\end{aligned}
\end{equation}
\begin{equation}
\label{eq:nonlinear_bc2n+1} 
\begin{aligned}   
&f'(x_m) = 0 \rightarrow \\ 
\eta \langle \tau(x)|_{n+1} \mathbb{N}_1 & \Bigl( \mathbb{D}_n^{(0)}(x_s) \otimes \mathbb{B}_{n}(x_m) \mathbb{G}_n^T  \Bigr) |\Psi \rangle_{2n} = 0,
\end{aligned}
\end{equation}
where $|\Psi \rangle_{2n}$ is shorthand notation of the tensor product of two identical quantum states $|\psi \rangle_n \otimes |\psi \rangle_n$ and Eq.~\eqref{eq:N1} is employed to create a linearly independent basis set. Note that the quantum model [Eq.~\eqref{eq:nonlinear_quantum model}] contains $\mathbb{D}_n^{(0)}(x_s)$ 
accompanied with $f(x_s)$, implying that a quantum circuit capable of uploading a regular constraint is necessary for solving NDEs (see Sec.~\ref{sec:Dn(xs)andGnT} of Methods).
\begin{figure}[t!]
\begin{center}
\includegraphics[width=1.0\linewidth]{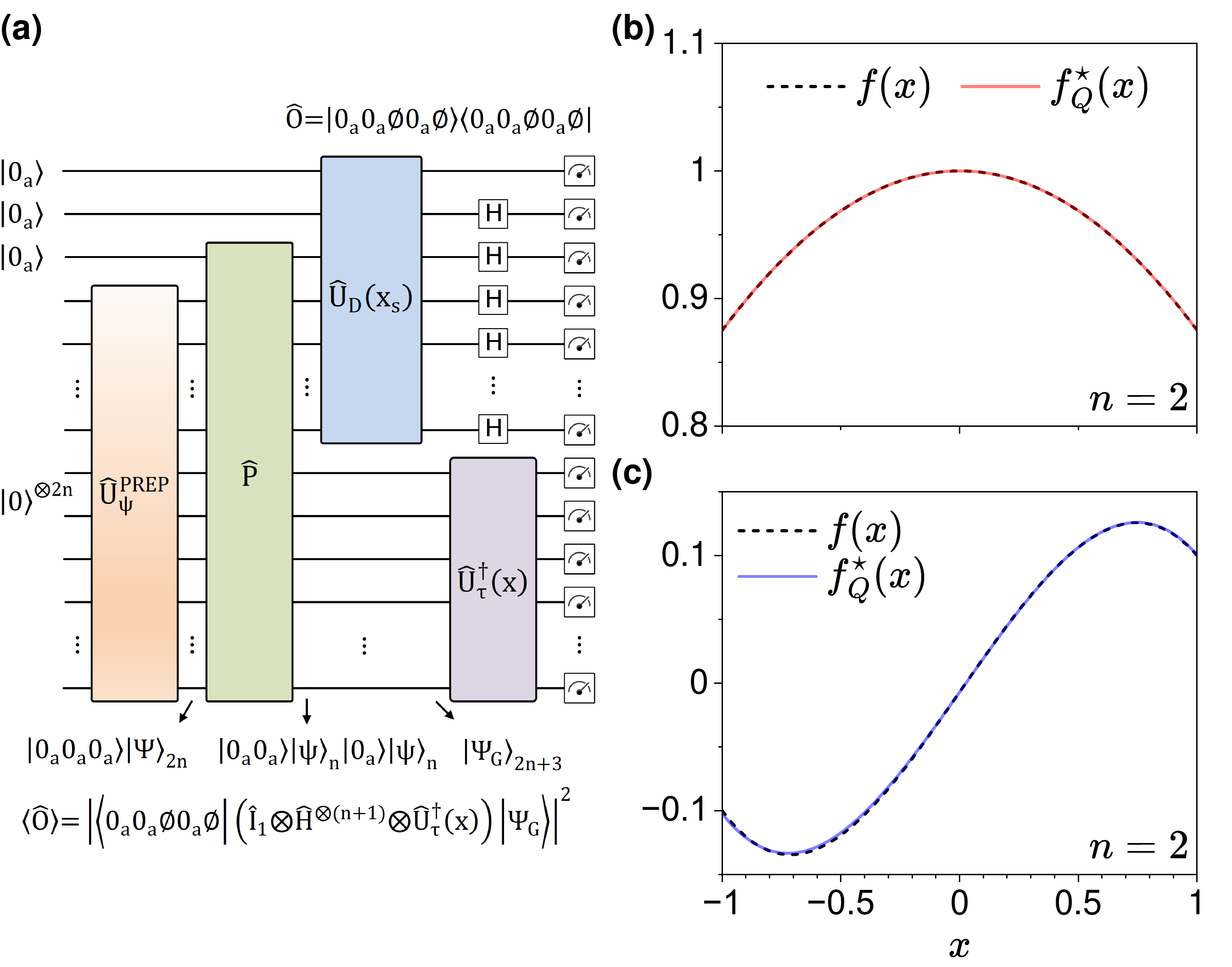}
\end{center}
\caption{
\textbf{Results of solving nonlinear differential equations.}
\textbf{(a)} Quantum circuit for evaluating state overlaps $ \Bigl(\langle0_a| \langle+_a| \langle+|^{\otimes n} \langle0_a| \langle\tau(x)| \Bigr)|\Psi_G \rangle$ 
through the interferometric measurement, where a single overlap measurement is obtained from two separate measurements of overlap probabilities [Eq.~\eqref{eq:RefxNDE}]. The circuit is initialized with a zero-product state, followed by a state preparation, a permutation and a data constraint $\hat{\text{U}}_{\text{D}}(x_s)$ circuit (see Methods). Finally, a layer
of Hadamards and an inverse Chebyshev feature map are applied on the first and second $(n+1)$-qubit registers, respectively, to obtain the overlap probabilities in the Hadamard-Chebyshev basis. 
Plots of $f(x)$ and $f_Q^{\star}(x)$ for \textbf{(b)} Eq.~\eqref{eq:NDE1} and $\mathcal{DC_I} = \{f'(x_m=0)\}$, $\eta_e$ = 3.75367. \textbf{(c)} Eq.~\eqref{eq:NDE2} and $\mathcal{DC_I} = \{f(x_z=0.02615)\}$, $\eta_e$ = 0.07984. In each panel, $n$ and $\eta_e$ represent the number of qubits and the scaling factor of the quantum model, respectively. Please also refer to the Sec.~\ref{sec:AEHW} of Methods for an alternatives without the permutation.
}
\label{fig:NDEs}
\end{figure}

In particular, it should be highlighted that Eq.~\eqref{eq:nonlinear_quantum model} is essentially identical to Eq.~\eqref{eq:quantum model_n} and both equations are applied to the linear and nonlinear terms of NDE, respectively. The quantum model of this NDE is thus $ \langle \tau(x)|_{n+1} \eta \mathbb{A} |\Psi \rangle_{2n} = 0$, where $\mathbb{A} = \mathbb{N}_1 \Bigl[ \mathbb{D}_n^{(0)}(x_s) \otimes \Bigl(4{\mathbb{G}_n^T}^2 + \mathds{1}_n \Bigr) +2(\mathbb{G}_n^T \otimes \mathbb{G}_n^T) \Bigr]$ is a data-encoded matrix with $\mathbb{D}_n^{(0)}(x_s)$ associated with the linear terms. Given $\mathcal{DC_I} = \{f'(x_m=0)\}$, the effective Hamiltonian operator is $\mathbb{H}=\mathcal{T}(\mathbb{A})+ \mathcal{T} \Bigl(\mathbb{N}_1 \Bigl( \mathbb{D}_n^{(0)}(x_s) \otimes \mathbb{B}_{n}(x_m) \mathbb{G}_n^T \Bigr) \Bigr)$. The total energy function $\mathcal{E}$ for NDE is expressed as
\begin{equation}
\label{eq:Heff_NDE}
\begin{aligned}
    \mathcal{E} = \eta^2 \langle \Psi|_{2n} \mathbb{H} |\Psi \rangle_{2n}= 0.    
\end{aligned}
\end{equation}
Similar to linear DEs, solving the second-order NDEs requires one invariant constraint to obtain the state overlap solution and one regular constraint to determine the scaling factor. The regular constraint is selected as $\mathcal{DC_R} = \{f(x_{s}=0) = 1\}$, the same as the one appearing in $\mathbb{H}$.

The consequence of the doubled Hilbert space is that $\mathbb{H}$ possesses at least $2^{2n-1}$ degenerate states associated with zero eigenvalue $\lambda_{\mathrm{deg}} = 0$. Most importantly, the task of solving the selected NDE is converted to that of seeking a degenerate state ($|\Psi_g \rangle_{2n} = |\psi_g \rangle_n \otimes |\psi_g \rangle_n $) that closely matches the analytical solution, which can be verified by substituting the chosen ground state into Eq.~\eqref{eq:nonlinear_quantum model} and determining the scaling factor with the given $\mathcal{DC_R}$. On the other hand, the quantum model in Eq.~\eqref{eq:nonlinear_quantum model} indicates that the same independent variable $x$ is encoded to both registers. In this case, based on the identical quantum model in Eq.~\eqref{eq:quantum model_n}, one can simply replace the inverse Chebyshev feature map in the first register with a layer of Hadamards acting on each qubit and then measure the overlap in the Hadamard-Chebyshev basis, as schematically illustrated in Fig.~\ref{fig:NDEs}(a).
We evaluated the quantum model with this quantum circuit using Eq.~\eqref{eq:RefxNDE}, and the result is consistent with the analytically exact solution, as shown in Fig.~\ref{fig:NDEs}(b).

We next tackle another example of a nonlinear differential equation with the quadratic nonlinearity on an unknown function, given by
\begin{equation}
\label{eq:NDE2}
\begin{aligned}
\frac{d^2f(x)}{dx^2} -2 f(x)^2 + x = 0 , \\ \mathcal{BC} = \{ f(-1)=-0.1, f(1)=0 .1\}.
\end{aligned}
\end{equation}
This NDE has no analytical solution, so instead a numerical solution to the NDE is found. The latent governing equation is $\mathbb{A} = \mathbb{N}_1 \Bigl( \mathbb{D}_n^{(0)}(x_s) \otimes {\mathbb{G}_n^T}^2 -2 \mathds{1}_{n} \otimes \mathds{1}_{n} \Bigr) +  \mathbb{N}_x \Bigl( \mathbb{D}_n^{(0)}(x_s) \otimes \mathbb{D}_n^{(0)}(x_s) \Bigr)$, where Eq.~\eqref{eq:Nx} is employed to create a linearly independent basis set for the linear $x$ term. Given $\mathcal{DC_I} = \{f(x_z=0.02615)\}$, the resultant effective Hamiltonian operator is $\mathbb{H}=\mathcal{T}(\mathbb{A})+ \mathcal{T} \Bigl(\mathbb{N}_1  \Bigl( \mathbb{D}_n^{(0)}(x_s) \otimes \mathbb{B}_{n}(x_z) \Bigr) \Bigr)$. $\mathcal{DC_R} = \{f(x_{s}=0.5)\}$ is utilized and the choice of $\mathcal{DC_R}$ only affects the accuracy of the solution. After performing the evaluations, the result is in excellent agreement with the numerical solution $f(x)$, as shown in Fig.~\ref{fig:NDEs}(c). 

Overall, the ability to represent quantum states in a higher-dimensional Hilbert space is the pivoting point for addressing NDEs in the degenerate eigenspace of $\mathbb{H}$ for the proposed framework.


\section*{Discussion}
Let us reflect on choices made when designing the algorithm, and consider potential avenues for improvement. The effective Hamiltonian approach is based on three parts, and here we shall discuss ideas for advancing physics-informed constraints, data-driven operation, and selected quantum subroutines. We will comment on scaling aspects, and describe steps towards hardware implementations. Finally, we will assess the utility of solving problems in latent spaces.

First, the implementation of physics-informed constraints largely depends on the choice of embedding (feature map). The latent-space basis plays a pivotal role in the formulation of effective Hamiltonians. Thanks to the purely-real Chebyshev differentiation matrix and linearly-independent Chebyshev basis set, we are able to translate a linear/nonlinear differential equation into the eigenvalue problem of a real-valued effective Hamiltonian in a finite-dimensional Hilbert space, described by a concise form of the total energy function [as shown in Eq.~\eqref{eq:Heff_LDE}, \eqref{eq:Heff_PDE} and \eqref{eq:Heff_NDE}]. By contrast, when the Fourier basis is utilized, the overlap model is formulated as a Fourier-series expansion, and the Fourier differentiation matrix has purely imaginary entries. Due to the complex-valued quantum state, the resulting total energy function is of a complicated form, rendering it difficult to find the ground state in this scenario. Similarly, if the Hartley basis~\cite{HYWu2024} is chosen, the top-left entry of Hartley differentiation matrix is not a constant, being a differential operator, and thus it does not allow us to easily access the derivative of the overlap model. Consequently, we focused on the quantum model built on the Chebyshev expansion. In addition, the $x$-dependent Chebyshev basis state can be readily prepared from the available Chebyshev feature map circuit. However, we do not exclude that other choices of basis can be adopted, especially when the map from constraints to effective Hamiltonians is modified. One guiding principle here can be the cost of differentiation, where the goal is minimizing the depth of quantum circuits to implement derivatives, multiplication etc.  

Second, the choice of data constraints is crucial for regularizing solutions and growing the gap between states that simply satisfy relations for derivatives, to that that solve the problem for specified conditions. So far we relied on the few data constraints, aiming to minimize the number of terms in effective Hamiltonians and show that the approach is applicable in a data-scarce regime. However, there is a growing value in working with more data, specifically when there is no easy way to perform calculations \cite{Brunton2020rev}. One idea here is to develop strategies with iterative improvement of solutions, similar in spirit to the model discovery \cite{Both2021,heim2021quantum}. Starting from some initial model and invariant constraint we can prepare first-approximation for our solution (low temperature state of $\mathbb{H}^{(0)}$). With the scaling factor estimated, we can further adjust physics-informed terms and add more data-based terms as operator constraints, and prepare the ground state for $\mathbb{H}^{(1)}$ etc. The gradual refinement leads to the ground state preparation with good initial states (large overlaps), and can be seen as a depth-frugal iterative improvement of the quantum solution \cite{williams2024iterative} with improved efficiency and convergence.

Another qualitative improvement for implementing boundary and data constraints can come from the use of \emph{quantum data} \cite{Williams2024readout}. So far we have followed the approach of data-loading for classically specified constraints (functions), as for instance performed in Refs. \cite{Marin2023,GonzalezConde2024efficientquantum,Rosenkranz2025quantumstate,bohun2025mps}. Alternatively, we can prepare states for $\mathcal{DC}$ directly using the tools from quantum simulation and ground state feature maps \cite{Umeano2024qcnn,umeano2024groundstate}, and using quantum signal processing to implement operators based on these states \cite{umeano2024community}. We believe that making quantum differential equation solvers more data-driven can lead to improvements in regimes where brute-force solvers may fail (deep turbulence), and consider this an important topic for future research. 


Third, the success of the effective Hamiltonian solvers relies on the efficiency of the ground state preparation. This in turn is defined by the efficiency of quantum subroutines that are employed in the state preparation process. The total budget for running of the algorithm depends on: 1) the cost of block-encoding $\mathbb{H}$; 2) the depth and success probability of GSP; 3) additional processing subroutines for basis transformations and overlap measurements (e.g. Chebyshev transform and Hadamard test). The cost of block-encoding $\mathbb{H}$ depends on the number of terms in the Hamiltonian, and can be approached in several ways. Since many DEs can be based on few terms, linear combination of unitaries is a favorable option \cite{Low2019}. We show that necessary terms have $\mathcal{O}\left(\text{poly}(n)\right)$ scaling and can be implemented efficiently. Next, the scaling of GSP depends on the gap of effective Hamiltonians. These are problem-dependent and in general are not easy to bound. So far our analysis shows a sizable gap for the studied problems, and this enabled QITE-based preparation. Generally, we expect the optimal preparation time $T$ to scale inversely proportional to the gap $\Delta$, being $\mathcal{O}\bigl(\tfrac{1}{\Delta}\,\mathrm{polylog}(1/\varepsilon)\bigr)$ for error $\varepsilon$ \cite{Lin2020nearoptimalground}. For the algorithmic thermalization one can prepare states with improved mixing time scaling and reaching $\mathcal{O}(1/\sqrt{\Delta})$ \cite{chen2023quantum,Scali2024_2}. Studying the  spectral gap dependence is an important question for the future work.

Next, let us consider steps towards possible quantum hardware implementations of the algorithm. In general, once we know the rules for composing effective Hamiltonians, the ground state preparation can proceed in a few different ways as outlined before. Given that differential operators for Chebyshev feature maps have non-trivial structure (see examples at the beginning of Supplementary Materials), the sensible approach is to block-encode them and proceed with algorithmic thermalization or imaginary-time GSP. However, this requires depth and ancillary register size that are available in the early fault-tolerant regime. One near-term alternative is to decompose $\mathbb{H}$ into Pauli strings and test the protocol by using variational state preparation. We demonstrate this for the key step in solving the Legendre DE example (Eq.~\ref{eq:VDE}), using a realistic emulation of neutral atom QPUs~\cite{henriet2020quantum} to prepare the relevant ground states. To do so, we adopt the library \texttt{pulser}~\cite{silverio2022pulser}, compatible with cloud-accessible devices, to design a pulse sequence preparing the targeted states with high-fidelity, while incorporating hardware constrains like the modulation bandwidth of the pulse. The details of the procedures are discussed in Sec.~\ref{sec:PulserGS} of Methods, pointing out that small-scale noisy operation can be tested.

Finally, we elucidate on the overall utility of solving DE problems in latent spaces. Given that readout of solutions is challenging for the full solution space, one options is to use the models to learn the necessary information \cite{Williams2024readout}. This process is highly sensitive to the chosen basis, and the effective Hamiltonian approach is beneficial for providing a small-scale but high-quality quantum data. For instance, this can be very beneficial for models that learn on sensitivities, which naturally benefit from automatic differentiation used here, as compared to real space grid-based approaches. Also, our algorithm makes sure that high-quality solutions (functions and their derivatives) can be prepared for all values of $x \in \mathcal{X}$ of a relevant domain. At the same time, the obtained Chebyshev-space model can be mapped into a real-space model via the quantum Chebyshev transform acting on an extended register \cite{delejarza2025QCPM}, $n \rightarrow n_{\mathrm{ext}}$, leading to fine grid solutions with $2^{n_{\mathrm{ext}}}$ points. This provides an advantage over finite differencing-based methods where the fine grid is needed at all points of solving DE-based for representing derivatives. 

%

\section*{Conclusions}
In this work, we have proposed a distinct paradigm for solving differential equations on quantum computers based on effective Hamiltonians. Our approach is motivated by embedding techniques from quantum machine learning, yet it relies on ground state preparation and bypasses the variational search. Specifically, we showed that physics-informed and data constraints can be designed in a latent space of Chebyshev polynomials, and solution of differential equations can be obtained by thermalizing states of the associated effective Hamiltonian. We have tested the approach on various problems, including multidimensional and nonlinear differential equations. We also presented quantum circuits for subroutines and measurement schedules to implement the algorithm efficiently. 

Whilst targeting early fault-tolerant QC operation, we also exemplified how to run the ground state preparation step on current hardware, allowing to experimentally test the key ideas of our described effective Hamiltonian approach. 

Summarizing, our work shows that there are more options to solving differential equations on quantum computers than finite differencing or variational search.
%
%


\appendix

\section*{Methods}
%
\subsection{Block encoding of structured and sparse matrices}
\label{sec:Dn(xs)andGnT}
\begin{figure}[b!]
\begin{center}
\includegraphics[width=0.6\linewidth]{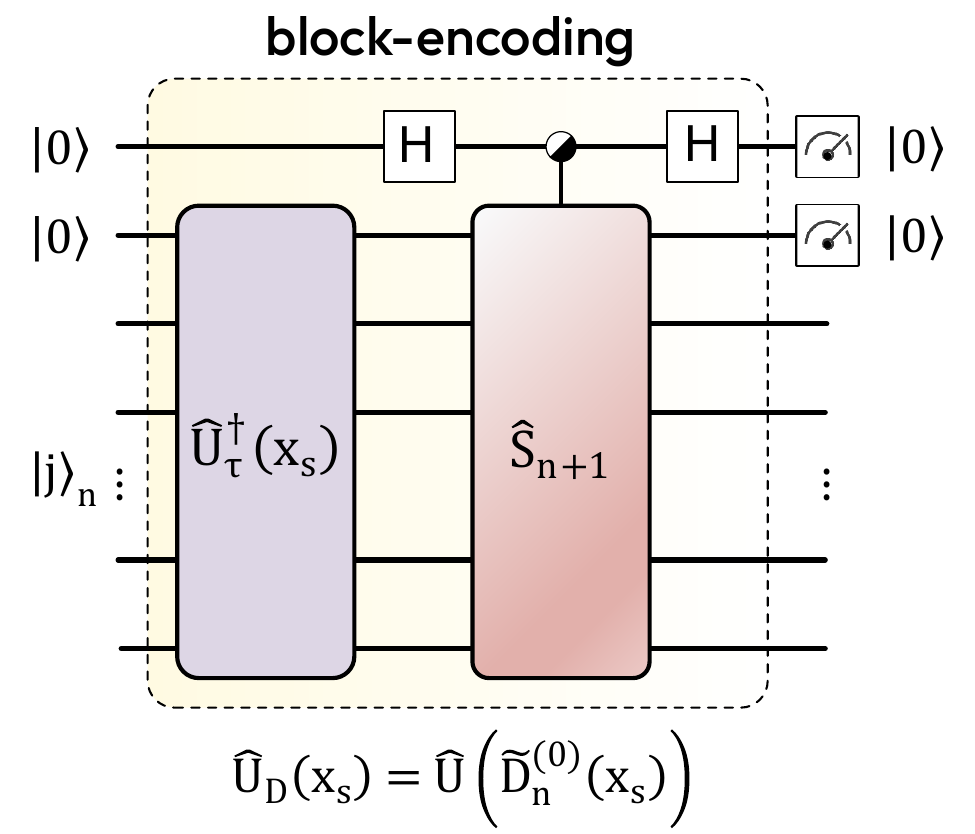}
\end{center}
\caption{
\textbf{Data (regular) constraint circuit.}
The implementation of $\hat{\text{U}}(\tilde{\mathbb{D}}_n^{(0)}(x_s))$ via linear combinations of unitary operations, where $\tilde{\mathbb{D}}_n^{(0)}(x_s) = \mathbb{B}_n(x_s)/\sqrt{2^{n+1}}$. $\hat{\mathcal{U}}_{\tau}^\dagger(x_s)$ is an inverse Chebyshev feature map (Fig.~\ref{fig:ChebFM}) and $\hat{\text{S}}_{n+1}$ is a Grover-type reflection about the all zeros state. 
Note that the global phase gate contributing the value of $-1$ in $\hat{\text{S}}_{n+1}$ circuit matters and cannot be ignored. Black-and-white circle on ancilla means that the unitary can be controlled by either 0 or 1.
}
\label{fig:U_D}
\end{figure}

\begin{figure*}[t!]
\begin{center}
\includegraphics[width=0.8\linewidth]{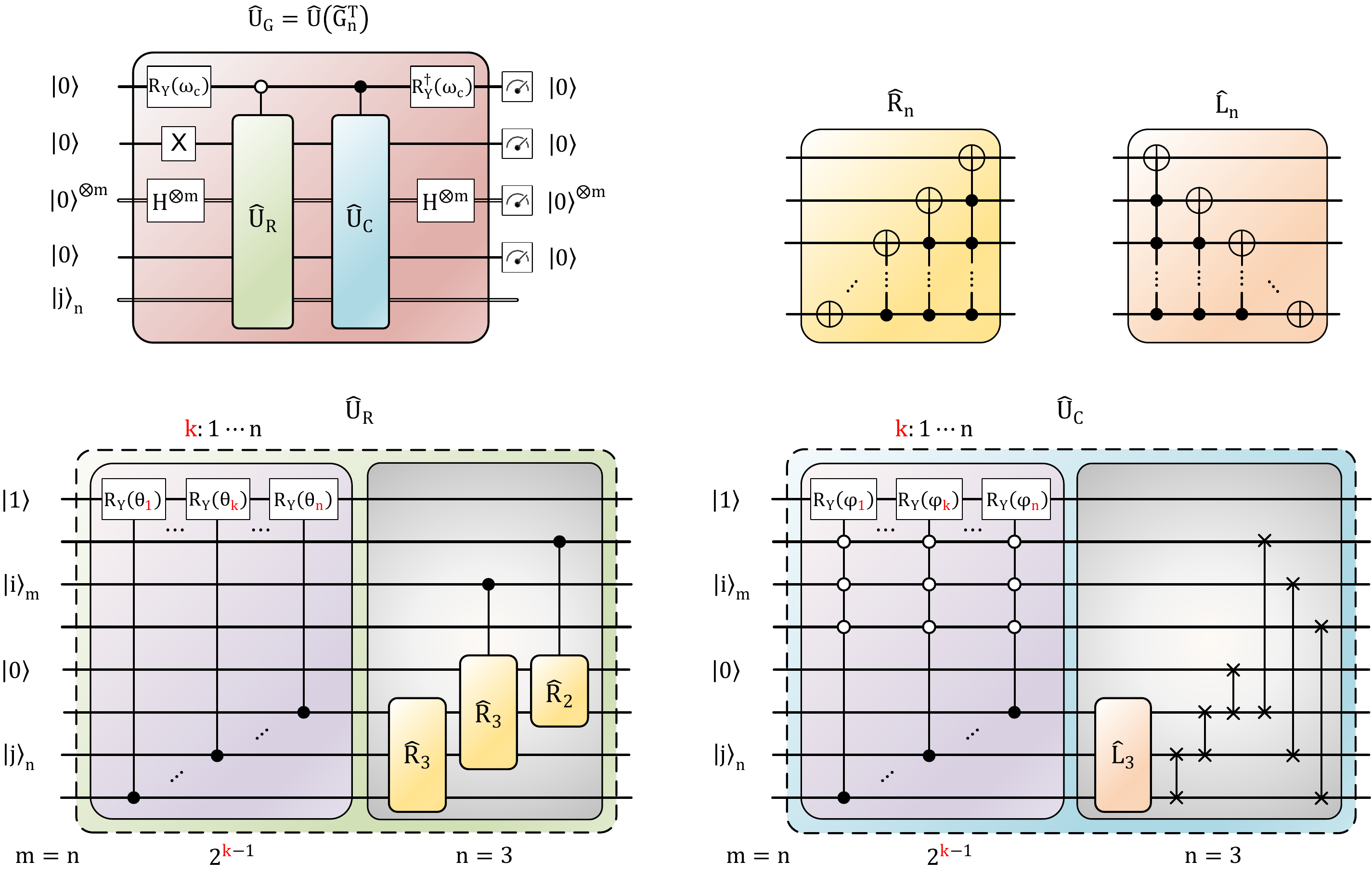}
\end{center}
\caption{
\textbf{Chebyshev differentiation circuit.}
The implementation of $\hat{\text{U}}(\tilde{\mathbb{G}}_n^T)$ via linear combinations of unitary operations, where $\tilde{\mathbb{G}}_n^T = \mathbb{G}_n^T/ \norm{\,\mathbb{G}_n^T\,}_{\text{S}}$. The $m$ register is equipped with $m$ Hadamard gates at the beginning and end of circuit $(m=n)$. Here, $\text{R}_\text{Y}$ gates acting on the top ancilla are controlled by the bit string $|j\rangle_n = |2^{k-1}\rangle$ in the $n$ register of $\hat{\text{U}}_{\text{R}}$, and bit strings $|i\rangle_m =|0\rangle^{\otimes m}$ and $|j\rangle_n = |2^{k-1}\rangle$ in $m$ and $n$ registers of $\hat{\text{U}}_{\text{C}}$, respectively. Argument correspond to $\omega_c=2\arccos{\sqrt{1/3}}$, $\alpha = 2/ \norm{\,\mathbb{G}_n^T\,}_{\text{S}}$, $\theta_k=2\arcsin{(-2^{k-1}\alpha)}$,
$\varphi_k=2\arcsin{[-2^{k-1}\alpha(1/\sqrt{2}-1)]}$. The last part of each unitary subroutine is composed of $n$ R-shift ($\hat{\text{R}}_{\text{n}}$) circuits for $\hat{\text{U}}_{\text{R}}$ and one L-shift ($\hat{\text{L}}_{\text{n}}$) circuit plus $2n$ SWAP gates for $\hat{\text{U}}_{\text{C}}$. The case for $n=3$ is shown here.
}
\label{fig:GnT}
\end{figure*}
We construct both data constraint and Chebyshev differentiation circuits via block encoding approaches that often require linear combination of unitaries~\cite{Childs2012,Low2019}. Since $\mathbb{D}_n^{(0)}(x_s)=\mathbb{B}_n(x_s)/f(x_s)$ is a structured and sparse matrix dependent on $n$ and $x_s$, the implementation o $\hat{\text{U}}(\tilde{\mathbb{D}}_n^{(0)}(x_s))$ is same as $\hat{\text{U}}(\tilde{\mathbb{B}}_n(x_i))$ described in the main text. 
As a result, $\mathbb{D}_n^{(0)}(x_s)$ can be obtained from $\tilde{\mathbb{D}}_n^{(0)}(x_s)$ embedded in the top-left $2^n \times 2^n$ block of $\hat{\text{U}}_{\text{D}}(x_s)$ multiplied by a prefactor of $\sqrt{2^{n+1}}/f(x_s)$.

The similar approach holds for $\mathbb{G}_n^T$ matrix exclusively dependent on $n$. The nature of this upper triangular matrix allows us to systematically construct its quantum circuit based on LCU operations and block encoding strategies proposed by Camps et al.~\cite{Daan_FABLE,Daan2022}. The block encoding of a properly scaled matrix $\tilde{\mathbb{G}}_n^T = \mathbb{G}_n^T/ \norm{\,\mathbb{G}_n^T\,}_{\text{S}}$ in a larger unitary matrix is denoted by  $\hat{\text{U}}(\tilde{\mathbb{G}}_n^T)$, where the subnormalization factor $\norm{\,\mathbb{G}_n(x)\,}_{\text{S}}$ is defined as the maximum of $\norm{\,\mathbb{G}_n(x)\mathbb{G}_n^T(x)\,}_{\text{1}}$ and $\norm{\,\mathbb{G}_n^T(x)\mathbb{G}_n(x)\,}_{\text{1}}$. The implementation of  $\hat{\text{U}}_{\text{G}}=\hat{\text{U}}(\tilde{\mathbb{G}}_n^T)$ circuit requires $m+n+3=2n+3$ qubits and two conditioned subroutines that assign the matrix elements on different bases. An exemplary $\hat{\text{U}}(\tilde{\mathbb{G}}_n^T)$ for $n=3$ is schematically illustrated in Fig.~\ref{fig:GnT}. For each subroutine $\hat{\text{U}}_{\text{R}}$ and $\hat{\text{U}}_{\text{C}}$,
the bit string $|i\rangle$ of the $m$ register specifies the $(2i+1)^{\text{th}}$ superdiagonal and $(i+1)^{\text{th}}$ row of $\tilde{\mathbb{G}}_n^T$, respectively, while the bit string $|j\rangle$ of the $n$ (system) register determines the $(j+1)^{\text{th}}$ column for both subcircuits. Specifically, $\hat{\text{U}}_{\text{R}}$ consists of $n$ single controlled $\text{R}_\text{Y}$ gates and $n$ R-shift ($\hat{\text{R}}_{\text{n}}$) circuits, while $\hat{\text{U}}_{\text{D}}$ comprises $n$ multi-controlled $\text{R}_\text{Y}$ gates, one L-shift ($\hat{\text{L}}_{\text{n}}$) and $2n$ SWAP gates. Eventually, $\mathbb{G}_n^T$ can be obtained from $\tilde{\mathbb{G}}_n^T$ block-encoded in the top-left $2^n \times 2^n$ block of $\hat{\text{U}}_{\text{G}}$ multiplied by a prefactor of $(2^{n-1}+2^n) \norm{\,\mathbb{G}_n^T\,}_{\text{S}}$ , with negligible errors in matrix entries that exponentially decrease as $n$ increases.

Overall, the number of gates scales linearly as $n$ for both $\hat{\text{U}}_{\text{D}}(x_s)$ and $\hat{\text{U}}_{\text{G}}$ circuits. For multi-controlled gates implemented with $\mathcal{O}(n^2)$ decomposition, the total circuit complexity is $\mathcal{O}(n^3)$, being polynomial in $n$ and thus efficient.
\begin{figure*}[ht]
\begin{center}
\includegraphics[width=0.8\linewidth]{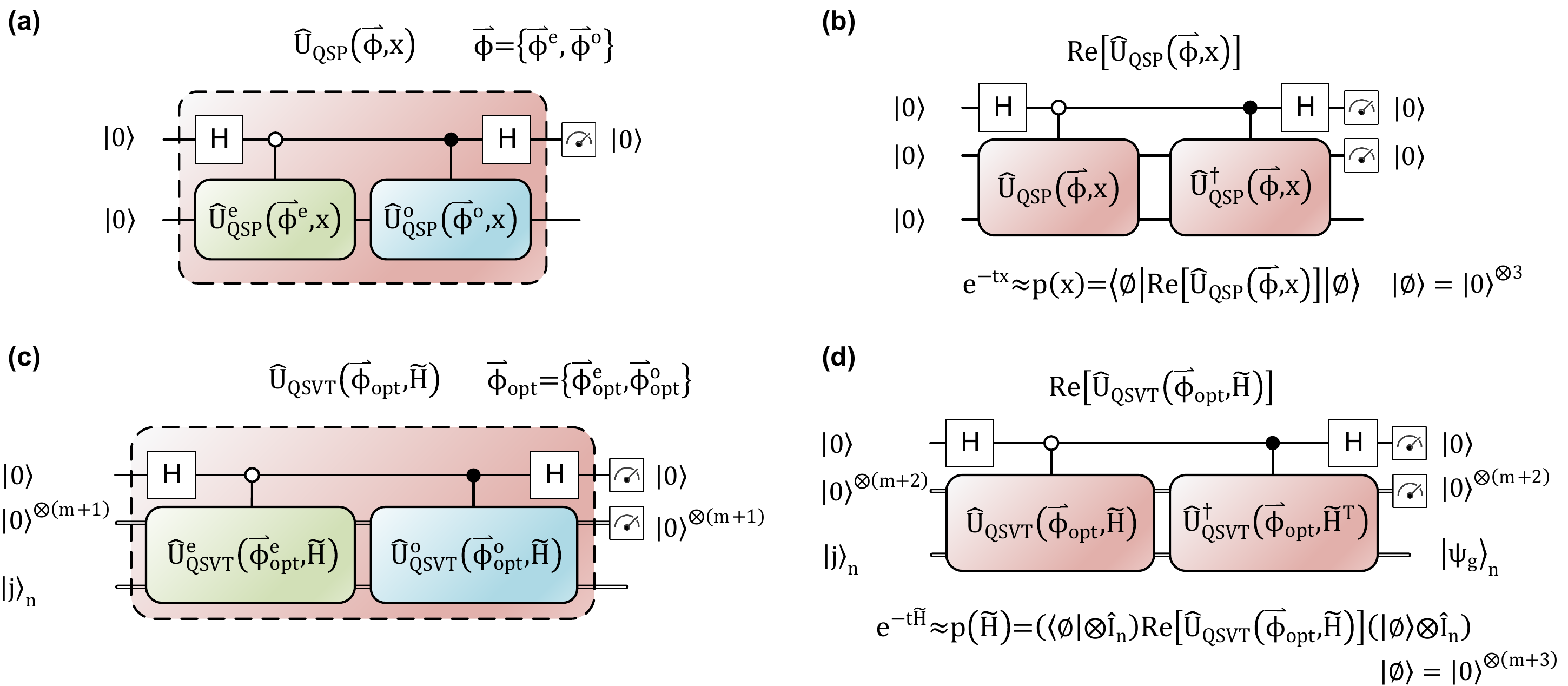}
\end{center}
\caption{
\textbf{QSVT-based quantum imaginary time evolution.} 
\textbf{(a)} $\hat{\mathcal{U}}_{\text{QSP}}(\vec{\phi},x)$ circuit used to generate a polynomial approximation of a mixed parity function based on the sum of even and odd polynomials via LCU. Other relevant QSP subroutine circuits are shown in Supplemental Materials, Fig.~\ref{fig:QSP}. \textbf{(b)} $\text{Re}[\hat{\mathcal{U}}_{\text{QSP}}(\vec{\phi},x)]$ circuit used to find the phase angles generating a polynomial approximation to $e^{-t x}$, where $t$ is the evolution time. \textbf{(c)} $\hat{\mathcal{U}}_{\text{QSVT}}(\vec{\phi}_{\text{opt}},\tilde{\mathbb{H}})$ circuit used to generate a mixed parity polynomial transformation of a normalized effective Hamiltonian operator $\tilde{\mathbb{H}}$ given a set of optimal phase angles obtained from \textbf{(b)}. Relevant QSVT subroutine circuits are shown in SM, Fig.~\ref{fig:QSVT}. \textbf{(d)} $\text{Re}[\hat{\mathcal{U}}_{\text{QSVT}}(\vec{\phi}_{\text{opt}},\tilde{\mathbb{H}})]$ circuit producing $e^{-t \tilde{\mathbb{H}}}$ in top left block used to find the ground state of $\tilde{\mathbb{H}}$ in the $n$-qubit system register after measuring the $(m+3)$-qubit register and post-selecting on the $|0\rangle^{\otimes (m+3)}$ outcome. Note that $\tilde{\mathbb{H}}$ is a real symmetric matrix.
}
\label{fig:QSPQSVT}
\end{figure*}
\subsection{Quantum imaginary-time evolution}
\label{sec:QITE}
As soon as the effective Hamiltonian operator $\mathbb{H}$ is ready, the lowest/zero energy eigenstate (ground state) can be prepared by the quantum imaginary-time evolution (QITE)~\cite{Motta}. We use quantum signal processing (QSP)~\cite{Low2017} and quantum singular value transformation (QSVT)~\cite{Gilyen,Martyn} to realize QITE. We first search for the phase angles of generating a polynomial approximation to a target function $p(x)\approx e^{-t x}$. Since the function $e^{-t x}$ has indefinite parity, $p(x)$ can be synthesized by the sum of the even and odd polynomials. This sum operation is accomplished by a mixed parity polynomial approximation circuit $\hat{\mathcal{U}}_{\text{QSP}}(\vec{\phi},x)$ consisting of opposite parity QSP circuits $\bigl( \hat{\mathcal{U}}_{\text{QSP}}^{e}(\vec{\phi}^{e},x)$ and $\hat{\mathcal{U}}_{\text{QSP}}^{o}(\vec{\phi}^{o},x) \bigr)$ via LCU, as shown in Fig.~\ref{fig:QSPQSVT}(a). The phase angles $\vec{\phi}=\{\vec{\phi}^e,\vec{\phi}^o\}$ are split into even and odd parity groups, $\vec{\phi}^e=\{\phi^e_1,\dots,\phi^e_{d^e+1}\}$ and $\vec{\phi}^o=\{\phi^o_1,\dots,\phi^o_{d^o+1}\}$ for parity-dependent QSP circuits (Fig.~\ref{fig:QSP}), with $d^e$ and $d^o$ being the maximum degrees of even and odd polynomial components of $p(x)$, respectively. Note that $p(x)$ is a purely real polynomial corresponding to the top-left entry of $\text{Re}[\hat{\mathcal{U}}_{\text{QSP}}(\vec{\phi},x)]$ circuit which can be similarly implemented using LCU, as shown in Fig.~\ref{fig:QSPQSVT}(b). The angles are adjusted until $p(x)$ gets closer to $e^{-t x}$ and then a set of optimal phase angles $\vec{\phi}_{\text{opt}} = \{\vec{\phi}^e_{\text{opt}},\vec{\phi}^o_{\text{opt}}\}$ is recorded.

Next, we upload the optimal phase angles to a mixed-parity polynomial transformation circuit $\hat{\mathcal{U}}_{\text{QSVT}}(\vec{\phi}_{\text{opt}},\tilde{\mathbb{H}})$ composed of opposite-parity QSVT circuits $ \bigl(\hat{\mathcal{U}}_{\text{QSVT}}^{e}(\vec{\phi}_{\text{opt}}^{e},\tilde{\mathbb{H}})$ and $\hat{\mathcal{U}}_{\text{QSVT}}^{o}(\vec{\phi}_{\text{opt}}^{o},\tilde{\mathbb{H}}) \bigr)$ via LCU, as shown in Fig.~\ref{fig:QSPQSVT}(c). Inside each QSVT circuit (Fig.~\ref{fig:QSVT}), the unitary $\hat{\text{U}}(\tilde{\mathbb{H}})$ is a block encoding of a non-unitary $\tilde{\mathbb{H}}$ embedded in the top-left block,
\begin{equation}
\label{eq:BE}
\begin{aligned}
\hat{\text{U}}(\tilde{\mathbb{H}}) = 
\begin{bmatrix}
\tilde{\mathbb{H}} & \sqrt{\mathds{1} - \tilde{\mathbb{H}}\tilde{\mathbb{H}}^{\dagger}} \\
\sqrt{\mathds{1} - \tilde{\mathbb{H}}^{\dagger}\tilde{\mathbb{H}}} & -\tilde{\mathbb{H}}^{\dagger}
\end{bmatrix},
\end{aligned}
\end{equation}
where $ \tilde{\mathbb{H}} = \mathbb{H}/ \norm{\,\mathbb{H}\,}_{\text{F}}$ is a normalized effective Hamiltonian operator with $\norm{\,\mathbb{H}\,}_{\text{F}}$ denoting the Frobenius norm of $\mathbb{H}$. In general, $\hat{\text{U}}(\tilde{\mathbb{H}})$ can be constructed using PREPARE and SELECT subroutines based on unitary decomposition of $\tilde{\mathbb{H}}$, with the number of ancillary qubits $m$ being the ceiling function of $\text{log}_2(L)$, where $L$ is the number of unitary operators. 

Finally, $p(\tilde{\mathbb{H}})\approx e^{-t \tilde{\mathbb{H}}}$ is a purely real matrix polynomial, corresponding to the top-left $2^n \times 2^n$ matrix block of $\text{Re}[\hat{\mathcal{U}}_{\text{QSVT}}(\vec{\phi}_{\text{opt}},\tilde{\mathbb{H}})]$ which can be similarly implemented via LCU, as shown in Fig.~\ref{fig:QSPQSVT}(d). The evolution time $t$ has an approximate lower bound, 
$ t \geq \lambda_{max}/(\lambda_{2} (n-1))$, where $\lambda_{max}$ and $\lambda_{2}$ are the largest and second smallest eigenvalues of $\mathbb{H}$ or $\tilde{\mathbb{H}}$, respectively. For the DEs shown in SM, Figs.~\ref{fig:VDESF2}(c,d), $t=15$ and $t=8$ are enough to recover the $n$-qubit ground state in the system register with $(d^e,d^o)=(6,7)$. Recently, generalized quantum signal processing (GQSP)~\cite{Motlagh2024}, generalized quantum eigenvalue and singular-value transformation (GQET and GQSVT)~\cite{GuangHaoLow2024,Christoph2023} have been developed, which could help to further improve the quantum circuit for the ground state preparation.

\subsection{Quantum overlap measurement for ODE}
\label{sec:M_ODE}

We describe a procedure for measuring overlaps and evaluating quantum models based on effective Hamiltonians. We use Eqs.~\eqref{eq:Refx} or~\eqref{eq:Refx_positive} for quantum models that contain a single variable. The real part reads
%
\begin{equation}
\label{eq:Refx}
\begin{aligned}
  \text{Re}(f_Q(x)) = \sqrt{\eta_{e} \, N} \left( \mathcal{N}_C \left| \hat{O}_C(x) \right|^2 - \left| \hat{O}_G(x) \right|^2 - \frac{1}{2N} \right),
\end{aligned}
\end{equation}
where $ |\hat{O}_G(x)|^2 = |\langle 0_a \mathrm{\o}|\hat{\mathcal{U}}_{\tau}^\dagger(x) |\mathrm{\psi}_G \rangle_{n+1} |^2$ is the probability of being in the Chebyshev basis state $|0_a \rangle |\tau(x) \rangle_n$ given $(n+1)$-qubit ground state $ |\mathrm{\psi}_G \rangle_{n+1} = |0_a\rangle |\mathrm{\psi}_g \rangle_n $, with $|\mathrm{\psi}_g \rangle_n$ prepared by the imaginary-time evolution based on the obtained $\mathbb{H}$. $|\hat{O}_C(x)|^2 = |\langle 0_a \mathrm{\o}|\hat{\mathcal{U}}_{\tau}^\dagger(x)  |\mathrm{\Psi}_C\rangle_{n+1}|^2$ is a probability of being in the Chebyshev basis state $|0_a \rangle |\tau(x) \rangle_n$ given $(n+1)$-qubit combined state $|\mathrm{\Psi}_C \rangle_{n+1} = \left( |\psi_G \rangle_{n+1} + |\psi_{r} \rangle_{n+1} \right)/\mathcal{N}_C$ with $ \mathcal{N}_C = | \left( |\psi_G \rangle_{n+1} + |\psi_r \rangle_{n+1} \right)|$ denoting the Frobenius norm. The reference state $|\psi_r \rangle_{n+1} $ 
is set as $|0\rangle^{\otimes (n+1)}$ so as to easily decouple the reference function when computing the real part of the product term. Here, $\eta_e$ is the scaling factor and $N=2^n$ is used.

In the special cases of either $f(x) \geq 0$ [for instance Figs.~\ref{fig:CDEs}(a),~\ref{fig:NDEs}(b,c),~\ref{fig:CDESF1}(b) and~\ref{fig:VDESF2}(a)] or $f(x) \leq 0$ [Fig.~\ref{fig:VDESF3}(a) for example] $\forall x \in [-1,1]$, we only need to evaluate the $|\hat{O}_G(x)|^2$ term,
\begin{equation}
\label{eq:Refx_positive}
\begin{aligned}
  f_Q(x) = \begin{cases} \sqrt{2\,\eta_{e}}   \sqrt{\left| \hat{O}_G(x) \right|^2 }, ~ &f(x) \geq 0, \\
   - \sqrt{2\,\eta_{e}}   \sqrt{\left| \hat{O}_G(x) \right|^2 } ~ &f(x) \leq 0
            \end{cases}
\end{aligned}
\end{equation}
\subsection{Quantum overlap measurement for PDE}
\label{sec:M_PDE}

Continuing to the multivariate case, we make use of Eq.~\eqref{eq:Refxy} to evaluate the quantum models containing two independent variables. The overlap reads as
\begin{equation}
\label{eq:Refxy}
\begin{aligned}
  \text{Re}(f_Q(x,y)) = \sqrt{\eta_{e}} \,2N \left( \mathcal{N}_C \left| \hat{O}_C(x,y) \right|^2 - \left| \hat{O}_G(x,y) \right|^2 - \frac{1}{4 N^2} \right),
\end{aligned}
\end{equation}
where $ |\hat{O}_G(x,y)|^2 = |\langle 0_a \mathrm{\o}0_a \mathrm{\o}| \left(\hat{\mathcal{U}}_{\tau}^\dagger(x) \otimes \hat{\mathcal{U}}_{\tau}^\dagger(y) \right) |\mathrm{\psi}_G \rangle_{2n+2}|^2 $ is a probability of being in the two-dimensional Chebyshev basis state $|0_a \rangle |\tau(x) \rangle_n |0_a \rangle |\tau(y) \rangle_n$ given $2(n+1)$-qubit ground state $|\mathrm{\psi}_G \rangle_{2n+2} = |0_a\rangle|\mathrm{\psi}_{gx} \rangle_n|0_a\rangle|\mathrm{\psi}_{gy} \rangle_n = \left( \mathds{1}_1 \otimes \hat{\text{P}} \right) |0_a 0_a\rangle |\mathrm{\psi}_g \rangle_{2n}$, with $|\mathrm{\psi}_g \rangle_{2n}$  prepared by the imaginary-time evolution based on the obtained $\mathbb{H}$. Here $\hat{\text{P}}$ is a permutation circuit responsible for reshuffling prepared states in a specific order. 
Next, $|\hat{O}_C(x,y)|^2 = |\langle 0_a \mathrm{\o}0_a \mathrm{\o}| \left( \hat{\mathcal{U}}_{\tau}^\dagger(x) \otimes \hat{\mathcal{U}}_{\tau}^\dagger(y) \right) |\mathrm{\Psi}_C \rangle_{2n+2}|^2 $ is a probability of being in the two-dimensional Chebyshev basis state $|0_a \rangle |\tau(x) \rangle_n |0_a \rangle |\tau(y) \rangle_n$ given $2(n+1)$-qubit combined state $|\mathrm{\Psi}_C \rangle_{2n+2} = \left( |\psi_{G} \rangle_{2n+2} + |\psi_r \rangle_{2n+2} \right) /\mathcal{N}_C $ with $ \mathcal{N}_C = | \left( |\psi_{G} \rangle_{2n+2} + |\psi_r \rangle_{2n+2} \right) |$ denoting the Frobenius norm. The reference state $|\psi_r \rangle_{2n+2}$ 
is set as $|0\rangle^{\otimes (2n+2)}$ so as to easily decouple the reference function when computing the real part of the product term. Again, $\eta_e$ is the scaling factor and $N=2^n$ is used.

\subsection{Quantum overlap measurement for NDE}
\label{sec:M_NDE}

To evaluate quantum models containing quadratic nonlinearities we use 
\begin{equation}
\label{eq:RefxNDE}
\begin{aligned}
  \text{Re}(f_Q(x)) = \eta_{e} \frac{\sqrt{8N^3}}{f(x_s)} \left( \mathcal{N}_C \left| \hat{O}_C(x) \right|^2 - \left| \hat{O}_G(x) \right|^2 - \frac{1}{4 N^2} \right),
\end{aligned}
\end{equation}
where $|\hat{O}_G(x)|^2 = |\langle 0_a 0_a \mathrm{\o} 0_a \mathrm{\o}| \left (\mathds{1}_1 \otimes \hat{\text{H}}^{\otimes(n+1)} \otimes \hat{\mathcal{U}}_{\tau}^\dagger(x) \right) |\mathrm{\Psi}_G \rangle_{2n+3}|^2 $ is a probability of being in the Hadamard-Chebyshev basis state $|0_a \rangle|+_a \rangle |+ \rangle^{\otimes n} |0_a \rangle |\tau(x) \rangle_n$ given $(2n+3)$-qubit ground state $|\mathrm{\Psi}_G \rangle_{2n+3} = \left( \hat{\text{U}}_{\text{D}}(x_s) \otimes \mathds{1}_{1+n} \right) \left( \mathds{1}_2 \otimes \hat{\text{P}} \right) |0_a0_a0_a\rangle |\mathrm{\Psi}_g \rangle_{2n}$, with $|\mathrm{\Psi}_g \rangle_{2n}$ prepared by imaginary-time evolution based on the obtained $\mathbb{H}$.
$\hat{\text{P}}$ is a permutation circuit and $\hat{\text{U}}_{\text{D}}(x_s)$ is a data constraint circuit [Fig.~\ref{fig:U_D}] accompanied with a regular constraint $f(x_s)$. Next, $|\hat{O}_C(x)|^2 = |\langle 0_a 0_a \mathrm{\o} 0_a \mathrm{\o}| \left( \mathds{1}_1 \otimes \hat{\text{H}}^{\otimes(n+1)} \otimes \hat{\mathcal{U}}_{\tau}^\dagger(x) \right) |\mathrm{\Psi}_C \rangle_{2n+3}|^2 $ is a probability of being in the Hadamard-Chebyshev basis state  $|0_a \rangle |+_a \rangle |+ \rangle^{\otimes n} |0_a \rangle |\tau(x) \rangle_n$ given $(2n+3)$-qubit combined state $|\mathrm{\Psi}_C \rangle_{2n+3} = \left( |\Psi_{G} \rangle_{2n+3} + |\Psi_r \rangle_{2n+3} \right) /\mathcal{N}_C $ with $ \mathcal{N}_C = | \left( |\Psi_{G} \rangle_{2n+3} + |\Psi_r \rangle_{2n+3} \right) |$ denoting the Frobenius norm. The reference state $|\Psi_r \rangle_{2n+3}$ is set as $|0\rangle^{\otimes (2n+3)}$ so as to easily decouple the reference function when computing the real part of the product term. Once more, $\eta_e$ is the scaling factor and $N=2^n$ is used.

\subsection{Alternative effective Hamiltonian workflow}
\label{sec:AEHW}

The beauty of the proposed physics-informed effective Hamiltonian consists in the capability of systemically building effective Hamiltonian and expressing the ground state of a multidimensional system. Moreover, our algorithm offers flexibility in adapting to the input of a multivariate system and a means to reduce the complexity of the circuit by incorporating complicated elements into the Hamiltonian. 
\begin{figure}[thb]
\begin{center}
\includegraphics[width=1.0\linewidth]{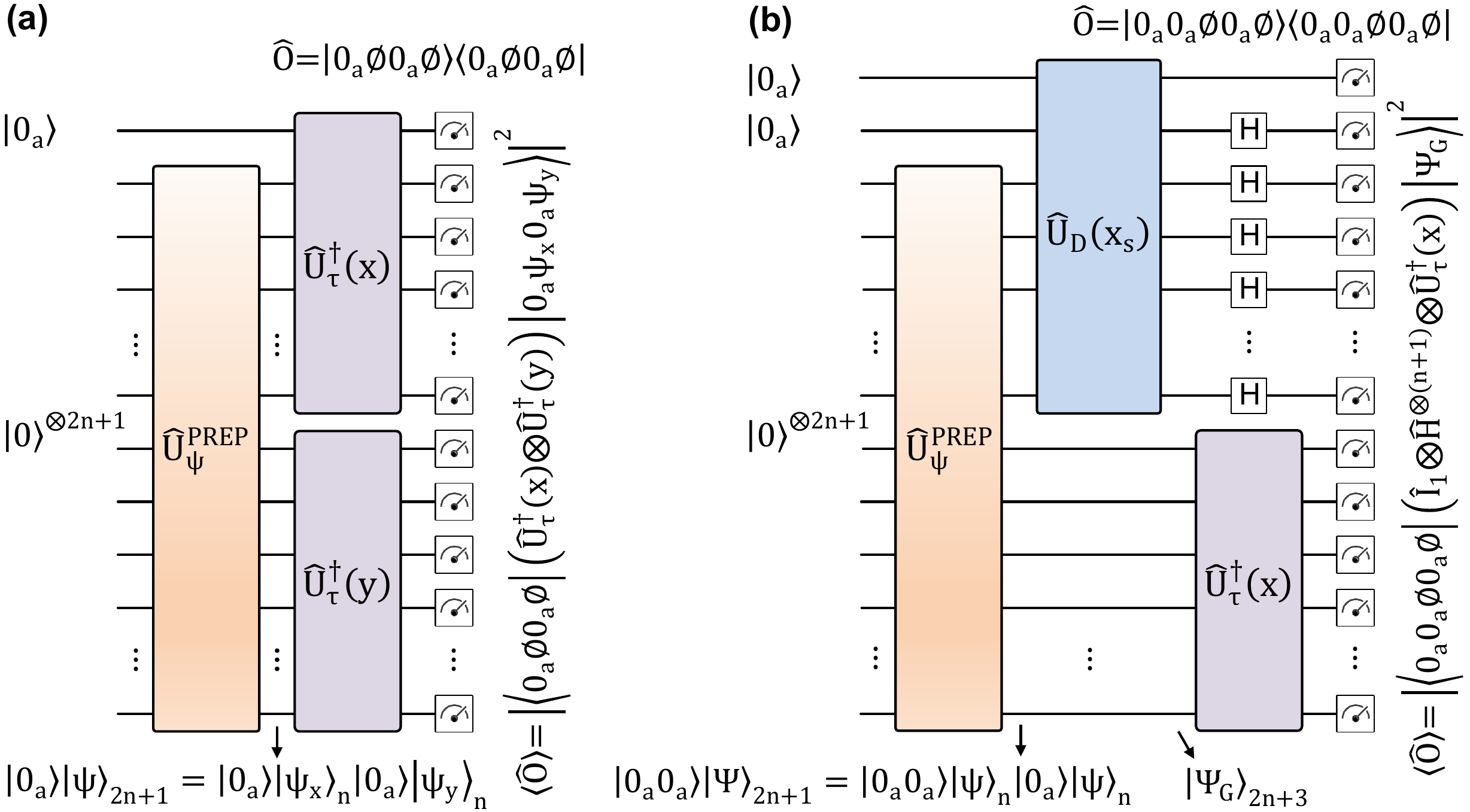}
\end{center}
\caption{\textbf{Alternative circuits for evaluating PDEs and NDEs models.} 
Quantum circuits used to evaluate the two-dimensional state overlaps $ \langle 0_a \tau(x) 0_a \tau(y) | 0_a \psi_x 0_a \psi_y \rangle$ for Eqs.~\eqref{eq:Laplce},~\eqref{eq:Heat} and~\eqref{eq:Wave} \textbf{(a)}, and the state overlaps $ \Bigl(\langle0_a| \langle+_a| \langle+|^{\otimes n} \langle0_a| \langle\tau(x)| \Bigr)|\Psi_G \rangle$ for Eqs.~\eqref{eq:NDE1},~\eqref{eq:NDE2} and~\eqref{eq:NDE4} \textbf{(b)}. The effect of the permutation circuit has been absorbed to the state preparation circuit. The measurement is still carried out through the interferometric measurement where a single overlap measurement is obtained from two separate measurements of overlap probabilities [Eqs.~\eqref{eq:Refxy} and~\eqref{eq:RefxNDE}]. 
}
\label{fig:new2Dmodel}
\end{figure}
For example, the previously required permutation circuit $\hat{\text{P}} $ in PDEs can be removed if Eq.~\eqref{eq:2D quantum model} is replaced with the equivalent quantum model given by
\begin{equation}
\label{eq:2D quantum model_new}
\begin{aligned}
f_Q(x,y) =& \sqrt{\eta} \bigl(\langle \tau(x)|_n \otimes \langle 0| \otimes \langle \tau(y)|_n \bigr) |\psi \rangle_{2n+1} \\ =& \sqrt{\eta} \bigl( \langle \tau(x)|_n \otimes \langle \tau(y)|_n \bigr) \mathbb{P}_a  |\psi \rangle_{2n+1}, \\ 
& \text{s.t.} \; \langle \psi |_{2n+1} | \psi \rangle_{2n+1} = 1,
\end{aligned}
\end{equation}
where $\mathbb{P}_a$ is a non-square constant matrix that has only one 1 in each row (see SM). For this new quantum model, the unknown quantum state $|\psi \rangle_{2n+1}$ is implicitly engineered to directly follow the desired input pattern of the inverse Chebyshev feature map, $|0_a\rangle |\psi \rangle_{2n+1} =|0_a\rangle |\psi_x \rangle_{n} |0_a \rangle |\psi_y \rangle_{n}$, and therefore the $\hat{\text{P}} $ circuit is no longer needed. Dirichlet and Neumann invariant constraints (Eqs.~\eqref{eq:bc1xy} and~\eqref{eq:bc2xy}) accordingly become as
\begin{equation}
\label{eq:bc1xy_new}
\begin{aligned}
f(x_z,y) = 0 \rightarrow
\sqrt{\eta} \langle \tau(x,y)|_{2n} \left( \mathbb{B}_n(x_z) \otimes \mathds{1}_n \right) \mathbb{P}_a |\psi \rangle_{2n+1} = 0, \\ 
f(x,y_z) = 0 \rightarrow 
\sqrt{\eta} \langle \tau(x,y)|_{2n} \left( \mathds{1}_n \otimes \mathbb{B}_n(y_z)  \right) \mathbb{P}_a |\psi \rangle_{2n+1} = 0, 
\end{aligned}
\end{equation}
\begin{equation}
\label{eq:bc2xy_new}
\begin{aligned}
\pdv{f(x_m, y)}{x} = 0  \rightarrow
\sqrt{\eta} \langle \tau(x,y)|_{2n} \left( \mathbb{B}_n(x_m) \mathbb{G}_n^T \otimes \mathds{1}_n \right) \mathbb{P}_a |\psi \rangle_{2n+1} = 0, \\ 
\pdv{f(x, y_m)}{y} = 0 \rightarrow
\sqrt{\eta} \langle \tau(x,y)|_{2n} \left( \mathds{1}_n \otimes \mathbb{B}_n(y_m) \mathbb{G}_n^T \right) \mathbb{P}_a |\psi \rangle_{2n+1} = 0 
\end{aligned}
\end{equation}
%
For the case of Laplace's equation (Eq.~\eqref{eq:Laplce}), the corresponding quantum model is $\langle \tau(x,y)|_{2n} \sqrt{\eta} \mathbb{A} |\psi \rangle_{2n+1} = 0$ with $\mathbb{A} = ( {\mathbb{G}_n^T}^2 \oplus {\mathbb{G}_n^T}^2 ) \mathbb{P}_a $. The $(2n+1)$-qubit ground state can be found by minimizing the total energy function $\mathcal{E}=  \eta \langle \psi|_{2n+1} \mathbb{H} |\psi \rangle_{2n+1} = 0$, with $\mathbb{H} = \mathcal{T}( \mathbb{A} ) + \mathcal{T} \left( (\mathbb{B}_n(\pm 1) \otimes \mathds{1}_n )\mathbb{P}_a \right) + \mathcal{T} \left( (\mathds{1}_n \otimes \mathbb{B}_n(-1)) \mathbb{P}_a \right) $ using the same approach as described previously. After discarding the $2^{2n}$ degenerate zero-energy states, we can find the ground state $|\psi_g \rangle_{2n+1}$ associated with the lowest eigenvalue. Both approaches return the same value of $\eta_e$ under the same data constraints for different PDE problems, but the new model provides the ground state exactly matching the inverse Chebyshev feature maps without the need for an extra permutation circuit. 
Therefore, the resulting quantum circuit [Fig.~\ref{fig:new2Dmodel}(a)] used to evaluate the two-dimensional state overlaps is similar to that in ODEs [Fig.~\ref{fig:CDEs}(a)]. 
\begin{figure*}[t!]
\begin{center}
\includegraphics[width=\linewidth]{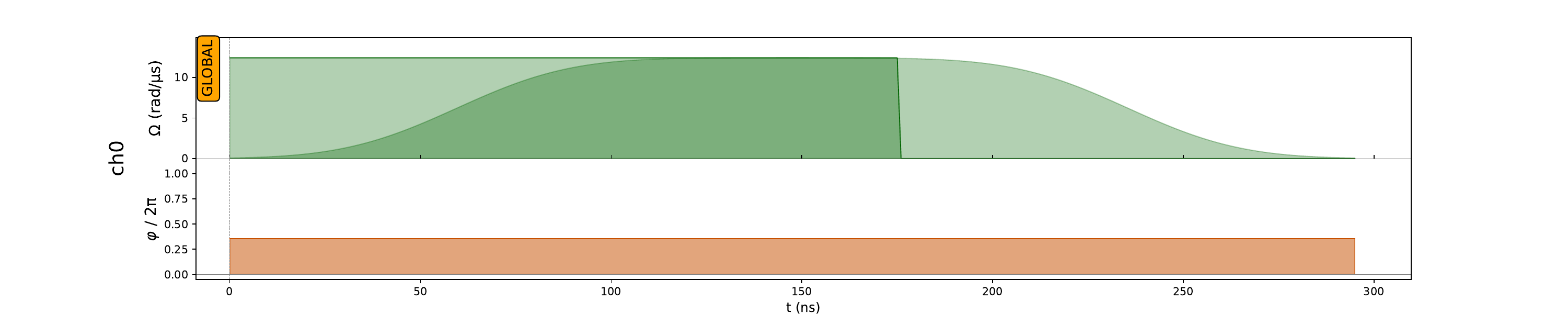}
\end{center}
\caption{Pulse sequence to prepare the state Eq.~\eqref{eq:pulserstate1}, generated using Pulser~\cite{silverio2022pulser}. $\Omega$ represents the amplitude of pulse and $\phi$ represents its phase. The state is obtained with fidelity $99.999\%$, discarding $12.95\%$ of the samples as a part of post selection. The modulation of the square pulse to account for the finite modulation bandwidth of the hardware is shown.}
\label{fig:PulserSequence}
\end{figure*}

To verify whether the same procedure is applicable to NDEs, we rewrite Eqs.~\eqref{eq:nonlinear_quantum model},~\eqref{eq:nonlinear_bc1n+1} and ~\eqref{eq:nonlinear_bc2n+1} as
\begin{equation}
\label{eq:nonlinear_quantum model_new}
\begin{aligned}
f_Q(x) =& \eta \bigl(\langle \tau(x)|_n \otimes \langle 0| \otimes \langle \tau(x)|_n \bigr) \Bigl(\mathbb{D}_n^{(0)} \otimes \mathds{1}_1 \otimes \mathds{1}_n \Bigr)  |\Psi \rangle_{2n+1} \\ =& \eta \langle \tau(x)|_{n+1} \mathbb{Q}_a \Bigl(\mathbb{D}_n^{(0)} \otimes \mathds{1}_1 \otimes \mathds{1}_n \Bigr) |\Psi \rangle_{2n+1}, \\
& \text{s.t.} \; \langle \Psi |_{2n+1} | \Psi \rangle_{2n+1} = 1,
\end{aligned}
\end{equation}
\begin{equation}
\label{eq:nonlinear_bc1n+1_new}
\begin{aligned}
&f(x_z) = 0  \rightarrow \\ 
\eta \langle \tau(x)|_{n+1} \mathbb{Q}_a & \Bigl( \mathbb{D}_n^{(0)}(x_s) \otimes \mathds{1}_1 \otimes \mathbb{B}_n (x_z) \Bigr) |\Psi \rangle_{2n+1} = 0,
\end{aligned}
\end{equation}
\begin{equation}
\label{eq:nonlinear_bc2n+1_new} 
\begin{aligned}   
&f'(x_m) = 0 \rightarrow \\ 
\eta \langle \tau(x)|_{n+1} \mathbb{Q}_a & \Bigl( \mathbb{D}_n^{(0)}(x_s) \otimes \mathds{1}_1 \otimes \mathbb{B}_{n}(x_m) \mathbb{G}_n^T  \Bigr) |\Psi \rangle_{2n+1} = 0,
\end{aligned}
\end{equation}
where $\mathbb{Q}_a=\mathbb{N}_1\mathbb{P}_a$ is an isometry (corresponding to a non-square constant matrix). For the case of Eq.~\eqref{eq:NDE3}, the corresponding quantum model is $ \langle \tau(x)|_{n+1} \eta \mathbb{A} |\Psi \rangle_{2n+1} = 0$, with $\mathbb{A} = \mathbb{Q}_a \Bigl[ \mathbb{D}_n^{(0)}(x_s) \otimes \mathds{1}_1 \otimes \Bigl(4{\mathbb{G}_n^T}^2 + \mathds{1}_n \Bigr) +2(\mathbb{G}_n^T \otimes \mathds{1}_1 \otimes \mathbb{G}_n^T) \Bigr]$, and the total energy function is $\mathcal{E} = \eta^2 \langle \Psi|_{2n+1} \mathbb{H} |\Psi \rangle_{2n+1}= 0$, with $\mathbb{H}=\mathcal{T}(\mathbb{A})+ \mathcal{T} \Bigl(\mathbb{Q}_a  \Bigl( \mathbb{D}_n^{(0)}(x_s) \otimes \mathds{1}_1 \otimes \mathbb{B}_{n}(x_m) \mathbb{G}_n^T \Bigr) \Bigr)$. The $(2n+1)$-qubit ground state can be found from the degenerate zero-energy space of $\mathbb{H}$ with the number of degeneracy increased by $2^{2n}$. Again, both approaches return the same value of $\eta_e$. The resulting quantum circuit used to evaluate the state overlap related to the quadratic nonlinearity is depicted in Fig.~\ref{fig:new2Dmodel}(b).
\subsection{Handle a function without zero crossings and slopes}
\label{sec:IDCnotavailable}
What if the solution function has neither zero crossings nor zero slopes in the Chebyshev domain? In such a case, invariant constraints do not exist. For example, let us consider this situation with a NDE whose analytical solution is a cubic function $-(x-3)^3/27$, written as
\begin{equation}
\label{eq:NDE3}
\begin{aligned}
3f(x)\frac{df^2(x)}{dx^2}-2\left( \frac{df(x)}{dx} \right)^2 = 0, \\ \mathcal{BC} = \{ f(0) = 1, f'(0) = -1 \},
\end{aligned}
\end{equation}
Apparently, invariant constraints are not available for this analytical solution. We define a new dependent variable as $\bar{f}(x)=f(x)-f(0)=f(x)-1$ such that Eq.~\eqref{eq:NDE3} becomes as
\begin{equation}
\label{eq:NDE4}
\begin{aligned}
3\bar{f}(x)\frac{d\bar{f}\,^2(x)}{dx^2} &+ 3\frac{d\bar{f}\,^2(x)}{dx^2} -2\left( \frac{d\bar{f}(x)}{dx} \right)^2 = 0, \\ \mathcal{BC} =& \{ \bar{f}(0) = 0, \bar{f}\,'(0) = -1 \},
\end{aligned}
\end{equation}
Now, Eq.~\eqref{eq:NDE4} has $\mathcal{DC_I} = \{\bar{f}(x_z=0)=0\}$ and thus our algorithms are applicable. The corresponding effective
Hamiltonian operator is $\mathbb{H}=\mathcal{T}(\mathbb{A})+ \mathcal{T} \Bigl(\mathbb{N}_1  \Bigl( \mathbb{D}_n^{(0)}(x_s) \otimes \mathbb{B}_{n}(x_z) \Bigr) \Bigr)$ with $\mathbb{A} = \mathbb{N}_1 \Bigl[3 \Bigl( \mathds{1}_n + \mathbb{D}_n^{(0)}(x_s)\Bigr)  \otimes {\mathbb{G}_n^T}^2 -2(\mathbb{G}_n^T \otimes \mathbb{G}_n^T) \Bigr]$, where $\mathcal{DC_R} = \{\bar{f}(x_{s}=-1/16)=f(x_{s}=-1/16)-1\}$ is utilized. As soon as $\bar{f}_Q^{\star}(x)$ is obtained, the original quantum model solution can be recovered by $f_Q^{\star}(x) = \bar{f}_Q^{\star}(x) + 1 $, as shown in SM, Fig.~\ref{fig:zerocrossing}.

\subsection{Prepare the ground states with Pulser} \label{sec:PulserGS}

Given that solutions to differential equations require ground state preparation, here we consider a first task of compiling such preparation to be run on neutral atom-based hardware. Specifically, two states are selected for preparation that can solve Legendre's DEs with $m=0$ [see SM and Fig.~\ref{fig:VDESF2}~(c,d) therein]. These simple states correspond to
\begin{align}
    |\psi_g\rangle_n &= \left(0.514496 \ket{01} + 0.857493 \ket{11}\right), \label{eq:pulserstate1} \\
    |\psi_g\rangle_n &= \left(0.426401 \ket{00} + 0.904534 \ket{10}\right), \label{eq:pulserstate2}
\end{align}
%
and can be readily prepared. As hardware, the cloud-accessible industrial QPU from Pasqal, Fresnel, is targeted. Fresnel is a neutral atom QPU made of individual $\prescript{87}{}{\text{Rb}}$ atoms trapped in an array of optical tweezers that operates in the ground-Rydberg qubit basis with global analog control~\cite{henriet2020quantum}. The results are simulated using Pulser~\cite{silverio2022pulser}, maintaining compatibility with hardware capabilities, and can be experimentally performed on the QPU. With only global pulses available a post-selection method is used to obtain the desired state. 

For the simulation, a register consisting of two qubits $7\,\mu m$ apart is prepared, with both qubits in the ground state. (As of the date of the simulation, Fresnel operates at Rydberg level 60 and supports a maximum possible amplitude of $4\pi \,\text{rad}/\mu$s, which sets the Rydberg blockade radius at $6.40\,\mu$m. This puts an interatomic distance of $7\,\mu$m in a regime of significant interaction.) We subject the register to a global pulse of amplitude $4\pi \,\text{rad}/\mu$s. Pulser applies suitable modulation to the pulses to account for the finite modulation bandwidth of the hardware, as shown in Fig.~\ref{fig:PulserSequence}. The duration and phase of this pulse is optimized to obtain the desired state after post section, by maximizing the fidelity, as described above. 

The duration and phase of the pulse was optimized using black box optimization to maximize fidelity while minimizing the discarded states. In the absence of interaction, this is simply accomplished with a duration of $2\sin^{-1}(b)/4\pi \,\mu$s and phase $\pi/2$ (here $b$ denotes the amplitude of excited component). In our setup, a sequence was obtained to prepare the state Eq.~\eqref{eq:pulserstate1} with fidelity $99.999\%$ by discarding $12.95\%$ of the samples as a part of post selection. For state Eq.~\eqref{eq:pulserstate2}, a fidelity of $99.995\%$ was obtained, while discarding $49.94\%$ of the samples as a part of post-selection.
%
%
\section*{Acknowledgements}
We thank Stefano Scali for useful discussion on the question of overlap measurements, Kaonan Micadei \& Raja Selvarajan for insightful suggestions on the hardware implementation.  O.K. acknowledges support from the QCi3 Hub (grant number EP/Z53318X/1).
\section*{Competing interests}
The authors have no conflict of interest. A patent application for the method described in this manuscript has been submitted by Pasqal with H.-Y. W. and O.K. as inventors.
%

%

\clearpage 
\onecolumngrid

\makeatletter
\setcounter{figure}{0}
\renewcommand \thesection{S\@arabic\c@section}
\renewcommand\thetable{S\@arabic\c@table}
\renewcommand \thefigure{S\@arabic\c@figure}
\makeatother

\section*{Supplementary Materials}
\subsection*{A. Relevant non-unitary matrices used in this work}
The non-unitary matrices symbolized with double-struck capital letters for $n=2$ are listed here. These $x$-independent matrices depend exclusively on $n$ and can therefore be reused, except for $\mathbb{B}_n(x)$ and $\mathbb{B}_{n+1}(x)$ whose elements uniquely determined by both $n$ and $x$.

\doublespacing

$\mathbb{G}_n^T =
\begin{bmatrix}
\begin{smallmatrix}
0 & \sqrt{2}  & 0 & 3\sqrt{2} \\
0 & 0 & 4 & 0 \\
0 & 0 & 0 & 6 \\
0 & 0 & 0 & 0
\end{smallmatrix}
\end{bmatrix}$,
\,
$\mathbb{M}_1 =
\begin{bmatrix}
\begin{smallmatrix}
\sqrt{2} & 0 & 0 & 0\\
0 & \sqrt{2} & 0 & 0 \\
0 & 0 & \sqrt{2} & 0 \\
0 & 0 & 0 & \sqrt{2} \\
0 & 0 & 0 & 0 \\
0 & 0 & 0 & 0 \\
0 & 0 & 0 & 0 \\
0 & 0 & 0 & 0
\end{smallmatrix}
\end{bmatrix}$,
\,
$\mathbb{M}_x =
\begin{bmatrix}
\begin{smallmatrix}
0 & 1 & 0 & 0 \\
1 & 0 & 1\over \sqrt{2} & 0 \\
0 & 1\over \sqrt{2} & 0 & 1\over \sqrt{2} \\
0 & 0 & 1\over \sqrt{2} & 0 \\
0 & 0 & 0 & 1\over \sqrt{2} \\
0 & 0 & 0 & 0 \\
0 & 0 & 0 & 0 \\
0 & 0 & 0 & 0
\end{smallmatrix}
\end{bmatrix}$,
\,
$\mathbb{M}_{x^2} =
\begin{bmatrix}
\begin{smallmatrix}
1\over \sqrt{2} & 0 & 1\over 2 & 0 \\
0 & 3\over 2\sqrt{2} & 0 & 1\over 2 \sqrt{2} \\
1\over 2 & 0 & 1\over \sqrt{2} & 0 \\
0 & 1\over 2\sqrt{2} & 0 & 1\over \sqrt{2} \\
0 & 0 & 1\over 2\sqrt{2} & 0 \\
0 & 0 & 0 & 1\over 2\sqrt{2} \\
0 & 0 & 0 & 0 \\
0 & 0 & 0 & 0
\end{smallmatrix}
\end{bmatrix}$,
\,
$\mathbb{M}_{x^3} =
\begin{bmatrix}
\begin{smallmatrix}
0 & 3\over 4 & 0 & 1\over 4 \\
3\over 4 & 0 & 1\over \sqrt{2} & 0 \\
0 & 1\over \sqrt{2} & 0 & 3\over 4\sqrt{2} \\
1\over 4 & 0 & 3\over 4\sqrt{2} & 0 \\
0 & 1\over 4\sqrt{2} & 0 & 3\over 4\sqrt{2} \\
0 & 0 & 1\over 4\sqrt{2} & 0 \\
0 & 0 & 0 & 1\over 4\sqrt{2} \\
0 & 0 & 0 & 0
\end{smallmatrix}
\end{bmatrix}$,
\,
$\mathbb{M}_{x^4} =
\begin{bmatrix}
\begin{smallmatrix}
3\over 4\sqrt{2} & 0 & 1\over 2 & 0 \\
0 & 5\over 4\sqrt{2} & 0 & 5\over 8 \sqrt{2} \\
1\over 2 & 0 & 7\over 8\sqrt{2} & 0 \\
0 & 5\over 8\sqrt{2} & 0 & 3\over 4\sqrt{2} \\
1\over 8 & 0 & 1\over 2\sqrt{2} & 0 \\
0 & 1\over 8\sqrt{2} & 0 & 1\over 2\sqrt{2} \\
0 & 0 & 1\over 8\sqrt{2} & 0 \\
0 & 0 & 0 & 1\over 8\sqrt{2}
\end{smallmatrix}
\end{bmatrix}$,
\,
$\mathbb{N}_1 =
\begin{bmatrix}
\begin{smallmatrix}
1\over \sqrt{2} & 0 & 0 & 0 & 0 & 1\over \sqrt{2} & 0 & 0 & 0 & 0 & 1\over \sqrt{2} & 0 & 0 & 0 & 0 & 1\over \sqrt{2} \\
0 & 1\over \sqrt{2} & 0 & 0 & 1\over \sqrt{2} & 0 & 1\over 2 & 0 & 0 & 1\over 2 & 0 & 1\over 2 & 0 & 0 & 1\over 2 & 0 \\
0 & 0 & 1\over \sqrt{2} & 0 & 0 & 1\over 2 & 0 & 1\over 2 & 1\over \sqrt{2} & 0 & 0 & 0 & 0 & 1\over 2 & 0 & 0 \\
0 & 0 & 0 & 1\over \sqrt{2} & 0 & 0 & 1\over 2 & 0 & 0 & 1\over 2 & 0 & 0 & 1\over \sqrt{2} & 0 & 0 & 0 \\
0 & 0 & 0 & 0 & 0 & 0 & 0 & 1\over 2 & 0 & 0 & 1\over 2 & 0 & 0 & 1\over 2 & 0 & 0 \\
0 & 0 & 0 & 0 & 0 & 0 & 0 & 0 & 0 & 0 & 0 & 1\over 2 & 0 & 0 & 1\over 2 & 0 \\
0 & 0 & 0 & 0 & 0 & 0 & 0 & 0 & 0 & 0 & 0 & 0 & 0 & 0 & 0 & 1\over 2 \\
0 & 0 & 0 & 0 & 0 & 0 & 0 & 0 & 0 & 0 & 0 & 0 & 0 & 0 & 0 & 0
\end{smallmatrix}
\end{bmatrix}$,
\,
$\mathbb{B}_n(x) = \sqrt{2}
\begin{bmatrix}
\begin{smallmatrix}
T_0(x)\over\sqrt{2} & T_1(x) & T_2(x) & T_3(x) \\
0 & 0 & 0 & 0 \\
0 & 0 & 0 & 0 \\
0 & 0 & 0 & 0 
\end{smallmatrix}
\end{bmatrix}$, 

\doublespacing

$\mathbb{N}_x =
\begin{bmatrix}
\begin{smallmatrix}
0 & 1\over 2 & 0 & 0 & 1\over 2 & 0 & 1\over 2\sqrt{2} & 0 & 0 & 1\over 2\sqrt{2} & 0 & 1\over 2\sqrt{2} & 0 & 0 & 1\over 2\sqrt{2} & 0 \\
1\over 2 & 0 & 1\over 2\sqrt{2} & 0 & 0 & 3\over 4 & 0 & 1\over 4 & 1\over 2\sqrt{2} & 0 & 1\over 2 & 0 & 0 & 1\over 4 & 0 & 1\over 2  \\
0 & 1\over 2\sqrt{2} & 0 & 1\over 2\sqrt{2} & 1\over 2\sqrt{2} & 0 & 1\over 2 & 0 & 0 & 1\over 2 & 0 & 1\over 4 & 1\over 2\sqrt{2} & 0 & 1\over 4 & 0 \\
0 & 0 & 1\over 2\sqrt{2} & 0 & 0 & 1\over 4 & 0 & 1\over 2 & 1\over 2\sqrt{2} & 0 & 1\over 4 & 0 & 0 & 1\over 2 & 0 & 0 \\
0 & 0 & 0 & 1\over 2\sqrt{2} & 0 & 0 & 1\over 4  & 0 & 0 & 1\over 4  & 0 & 1\over 4  & 1\over 2 \sqrt{2}  & 0 & 1\over 4  & 0 \\
0 & 0 & 0 & 0 & 0 & 0 & 0 & 1\over 4  & 0 & 0 & 1\over 4  & 0 & 0 & 1\over 4  & 0 & 1\over 4  \\
0 & 0 & 0 & 0 & 0 & 0 & 0 & 0 & 0 & 0 & 0 & 1\over 4  & 0 & 0 & 1\over 4  & 0 \\
0 & 0 & 0 & 0 & 0 & 0 & 0 & 0 & 0 & 0 & 0 & 0 & 0 & 0 & 0 & 1\over 4
\end{smallmatrix}
\end{bmatrix}$,
\,
$\mathbb{B}_{n+1}(x) = \sqrt{2}
\begin{bmatrix}
\begin{smallmatrix}
T_0(x)\over\sqrt{2} & T_1(x) & T_2(x) & T_3(x) & T_4(x) & T_5(x) & T_6(x) & T_7(x) \\
0 & 0 & 0 & 0 & 0 & 0 & 0 & 0 \\
0 & 0 & 0 & 0 & 0 & 0 & 0 & 0 \\
0 & 0 & 0 & 0 & 0 & 0 & 0 & 0 \\
0 & 0 & 0 & 0 & 0 & 0 & 0 & 0 \\
0 & 0 & 0 & 0 & 0 & 0 & 0 & 0 \\
0 & 0 & 0 & 0 & 0 & 0 & 0 & 0 \\
0 & 0 & 0 & 0 & 0 & 0 & 0 & 0
\end{smallmatrix}
\end{bmatrix}$,

\doublespacing

$\mathbb{P}_a =
\begin{bmatrix}
\begin{smallmatrix}
1 & 0 & 0 & 0 & 0 & 0 & 0 & 0 & 0 & 0 & 0 & 0 & 0 & 0 & 0 & 0 & 0 & 0 & 0 & 0 & 0 & 0 & 0 & 0 & 0 & 0 & 0 & 0 & 0 & 0 & 0 & 0 \\
0 & 1 & 0 & 0 & 0 & 0 & 0 & 0 & 0 & 0 & 0 & 0 & 0 & 0 & 0 & 0 & 0 & 0 & 0 & 0 & 0 & 0 & 0 & 0 & 0 & 0 & 0 & 0 & 0 & 0 & 0 & 0 \\
0 & 0 & 1 & 0 & 0 & 0 & 0 & 0 & 0 & 0 & 0 & 0 & 0 & 0 & 0 & 0 & 0 & 0 & 0 & 0 & 0 & 0 & 0 & 0 & 0 & 0 & 0 & 0 & 0 & 0 & 0 & 0 \\
0 & 0 & 0 & 1 & 0 & 0 & 0 & 0 & 0 & 0 & 0 & 0 & 0 & 0 & 0 & 0 & 0 & 0 & 0 & 0 & 0 & 0 & 0 & 0 & 0 & 0 & 0 & 0 & 0 & 0 & 0 & 0 \\
0 & 0 & 0 & 0 & 0 & 0 & 0 & 0 & 1 & 0 & 0 & 0 & 0 & 0 & 0 & 0 & 0 & 0 & 0 & 0 & 0 & 0 & 0 & 0 & 0 & 0 & 0 & 0 & 0 & 0 & 0 & 0 \\
0 & 0 & 0 & 0 & 0 & 0 & 0 & 0 & 0 & 1 & 0 & 0 & 0 & 0 & 0 & 0 & 0 & 0 & 0 & 0 & 0 & 0 & 0 & 0 & 0 & 0 & 0 & 0 & 0 & 0 & 0 & 0 \\
0 & 0 & 0 & 0 & 0 & 0 & 0 & 0 & 0 & 0 & 1 & 0 & 0 & 0 & 0 & 0 & 0 & 0 & 0 & 0 & 0 & 0 & 0 & 0 & 0 & 0 & 0 & 0 & 0 & 0 & 0 & 0 \\
0 & 0 & 0 & 0 & 0 & 0 & 0 & 0 & 0 & 0 & 0 & 1 & 0 & 0 & 0 & 0 & 0 & 0 & 0 & 0 & 0 & 0 & 0 & 0 & 0 & 0 & 0 & 0 & 0 & 0 & 0 & 0 \\
0 & 0 & 0 & 0 & 0 & 0 & 0 & 0 & 0 & 0 & 0 & 0 & 0 & 0 & 0 & 0 & 1 & 0 & 0 & 0 & 0 & 0 & 0 & 0 & 0 & 0 & 0 & 0 & 0 & 0 & 0 & 0 \\
0 & 0 & 0 & 0 & 0 & 0 & 0 & 0 & 0 & 0 & 0 & 0 & 0 & 0 & 0 & 0 & 0 & 1 & 0 & 0 & 0 & 0 & 0 & 0 & 0 & 0 & 0 & 0 & 0 & 0 & 0 & 0 \\
0 & 0 & 0 & 0 & 0 & 0 & 0 & 0 & 0 & 0 & 0 & 0 & 0 & 0 & 0 & 0 & 0 & 0 & 1 & 0 & 0 & 0 & 0 & 0 & 0 & 0 & 0 & 0 & 0 & 0 & 0 & 0 \\
0 & 0 & 0 & 0 & 0 & 0 & 0 & 0 & 0 & 0 & 0 & 0 & 0 & 0 & 0 & 0 & 0 & 0 & 0 & 1 & 0 & 0 & 0 & 0 & 0 & 0 & 0 & 0 & 0 & 0 & 0 & 0 \\
0 & 0 & 0 & 0 & 0 & 0 & 0 & 0 & 0 & 0 & 0 & 0 & 0 & 0 & 0 & 0 & 0 & 0 & 0 & 0 & 0 & 0 & 0 & 0 & 1 & 0 & 0 & 0 & 0 & 0 & 0 & 0 \\
0 & 0 & 0 & 0 & 0 & 0 & 0 & 0 & 0 & 0 & 0 & 0 & 0 & 0 & 0 & 0 & 0 & 0 & 0 & 0 & 0 & 0 & 0 & 0 & 0 & 1 & 0 & 0 & 0 & 0 & 0 & 0 \\
0 & 0 & 0 & 0 & 0 & 0 & 0 & 0 & 0 & 0 & 0 & 0 & 0 & 0 & 0 & 0 & 0 & 0 & 0 & 0 & 0 & 0 & 0 & 0 & 0 & 0 & 1 & 0 & 0 & 0 & 0 & 0 \\
0 & 0 & 0 & 0 & 0 & 0 & 0 & 0 & 0 & 0 & 0 & 0 & 0 & 0 & 0 & 0 & 0 & 0 & 0 & 0 & 0 & 0 & 0 & 0 & 0 & 0 & 0 & 1 & 0 & 0 & 0 & 0 \\
\end{smallmatrix}
\end{bmatrix}$, $\mathbb{Q}_a = \mathbb{N}_1 \mathbb{P}_a$

%

\begin{figure*}[ht]
\begin{center}
\includegraphics[width=0.6\linewidth]{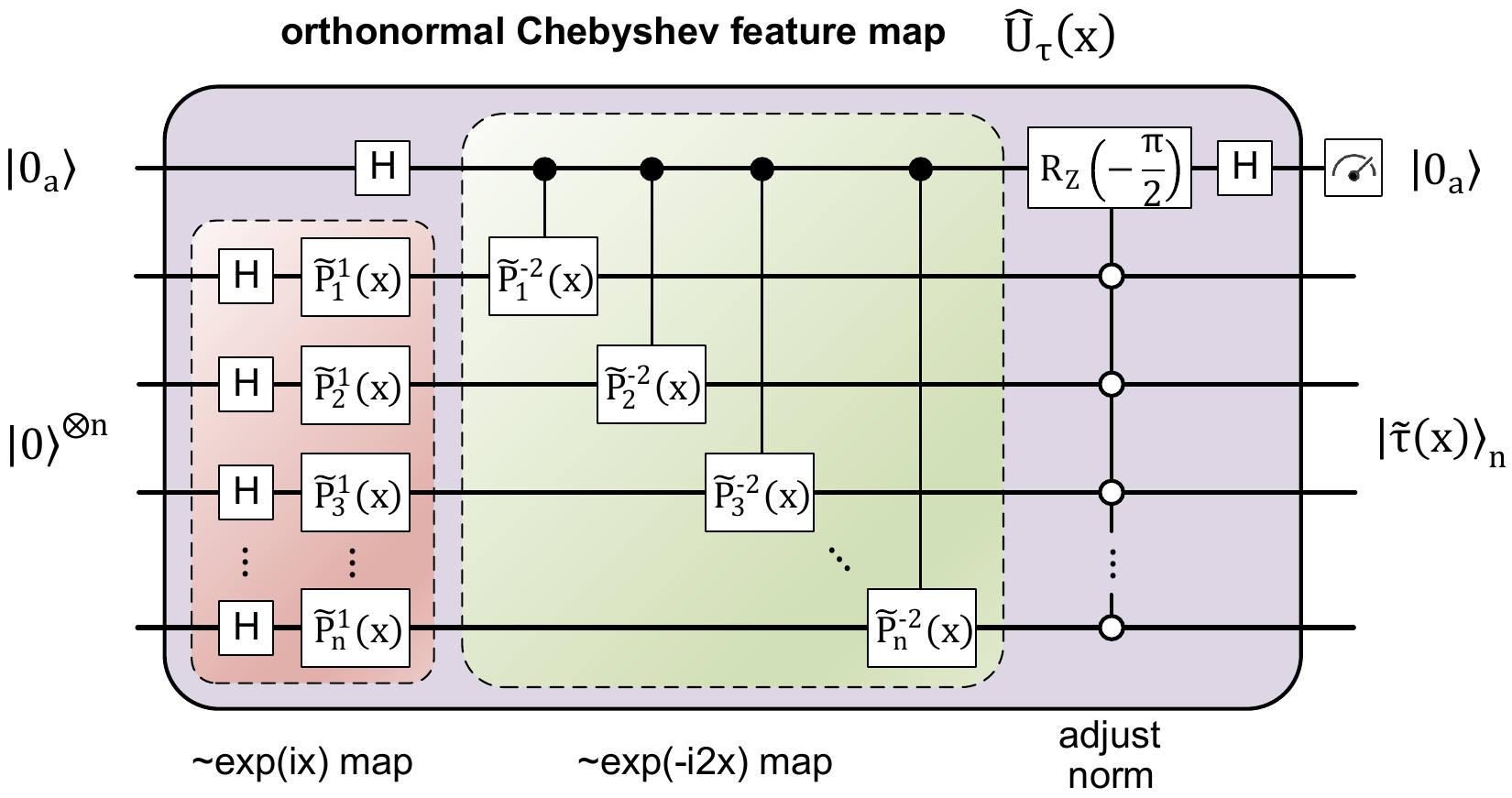}
\end{center}
\caption{
\textbf{Quantum Chebyshev feature map $\hat{\mathcal{U}}_{\tau}(x)$ circuit.} The $(n+1)$-qubit $\hat{\mathcal{U}}_{\tau}(x)$ prepares a normalized Chebyshev state with the real amplitude upon the ancillary measurement yields $|0\rangle$ outcome~\cite{Williams2023}. Here, scaled single-qubit phase shift gate is defined as $\tilde{\mathrm{P}}^{m}_l(x)=\mathrm{diag}\{1,\exp(i\,m\, 2^n \arccos(x) /2^l)\}$, where $l \in [1,\cdots,n]$ is the qubit index and $m$ takes values of $1$ and $-2$, for any continuous variable $x \in [-1,1]$.
}    
\label{fig:ChebFM}
\end{figure*}

\begin{figure*}[ht]
\begin{center}
\includegraphics[width=0.5\linewidth]{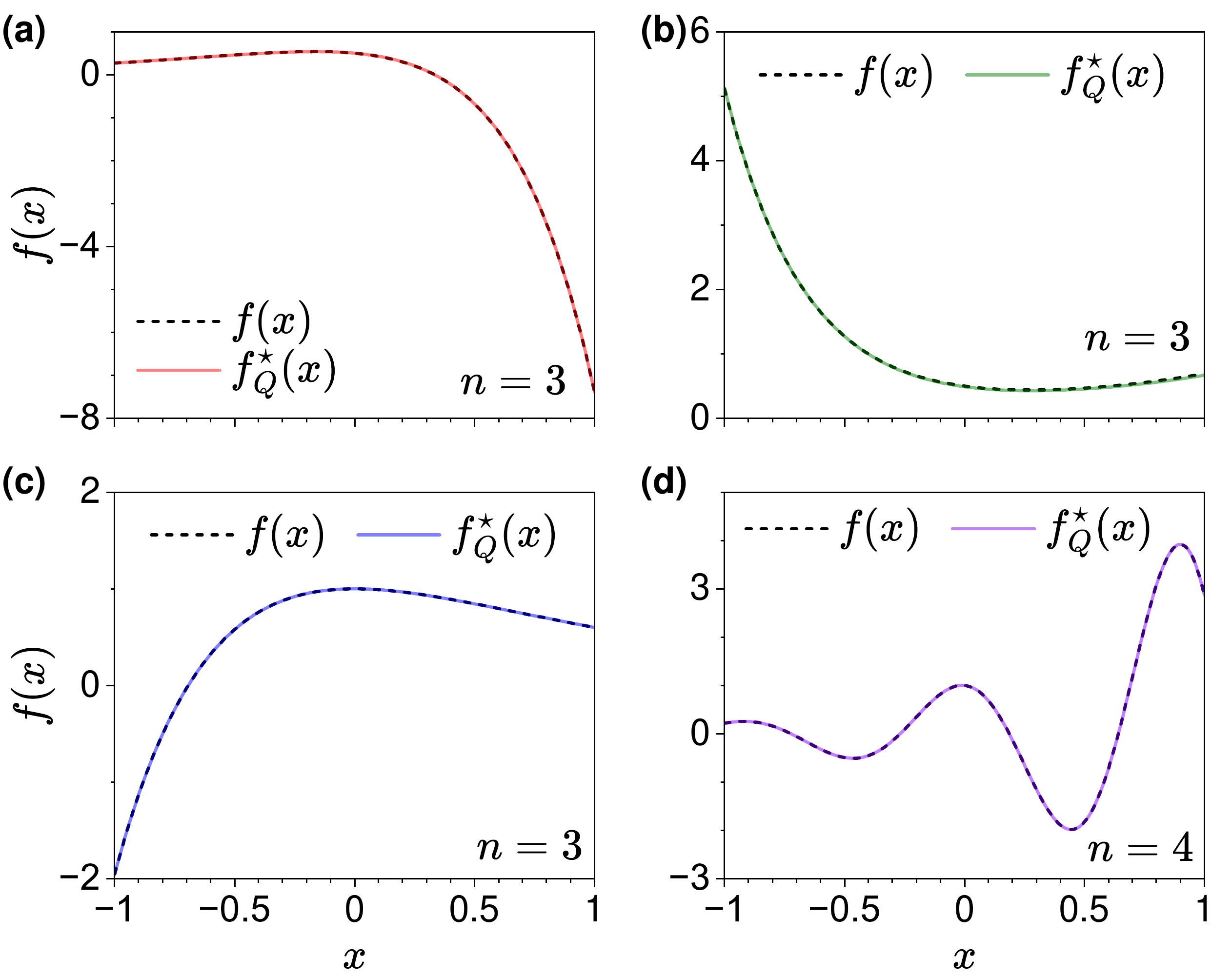}
\end{center}
\caption{
\textbf{Results of solving second-order DEs with constant coefficients.} In all cases, Eq.~\eqref{eq:CDE} with coefficients $\mathcal{C} = \{a,b,c\}$ and boundary conditions $\mathcal{BC} = \{f(0),f'(0)\}$ is used to determine a unique analytical solution $f(x)$. 
Plots of $f(x)$ and $f_Q^{\star}(x)$ for
\textbf{(a)} $\mathcal{C} = \{1,-4,4\}, \mathcal{BC} = \{0.5,-0.5\}$ and $\mathcal{DC_I} = \{f(x_z=0.333)\}, \eta_{e} = 65$,
\textbf{(b)} $\mathcal{C} = \{1,2,-3\}, \mathcal{BC} = \{0.5,-0.5\}$ and $\mathcal{DC_I} = \{f'(x_m=0.275)\}, \eta_{e} = 37.03$, \textbf{(c)} $\mathcal{C} = \{1,3,2\}, \mathcal{BC} = \{1,0\}$ and $\mathcal{DC_I} = \{f'(x_m=0)\}, \eta_{e} = 7.18$, and
\textbf{(d)} $\mathcal{C} = \{1,-3,50\}, \mathcal{BC} = \{1,-0.5\}$ and $\mathcal{DC_I} = \{f(x_z=0.187)\}, \eta_{e} = 50.25$.
In each panel, $n$ and $\eta_{e}$ represent the number of qubits and the scaling factor of the quantum model, respectively. 
The corresponding analytical solutions are \textbf{(a)} $f(x)=0.5 \,\text{exp}(2x) - 1.5 \, x\, \text{exp}(2x)$, \textbf{(b)} $f(x)=0.25 \left(\text{exp}(-3x) + \text{exp}(x) \right)$, \textbf{(c)} $f(x)=2 \, \text{exp}(-x) - \text{exp}(-2x)$ and \textbf{(d)} $f(x)=\text{exp}(1.5x) \left( \text{cos}(\sqrt{191}x/2) - 4/\sqrt{191} \, \text{sin}(\sqrt{191}x/2) \right)$, respectively.
}    
\label{fig:CDESF1}
\end{figure*}
\begin{figure*}[ht]
\begin{center}
\includegraphics[width=0.5\linewidth]{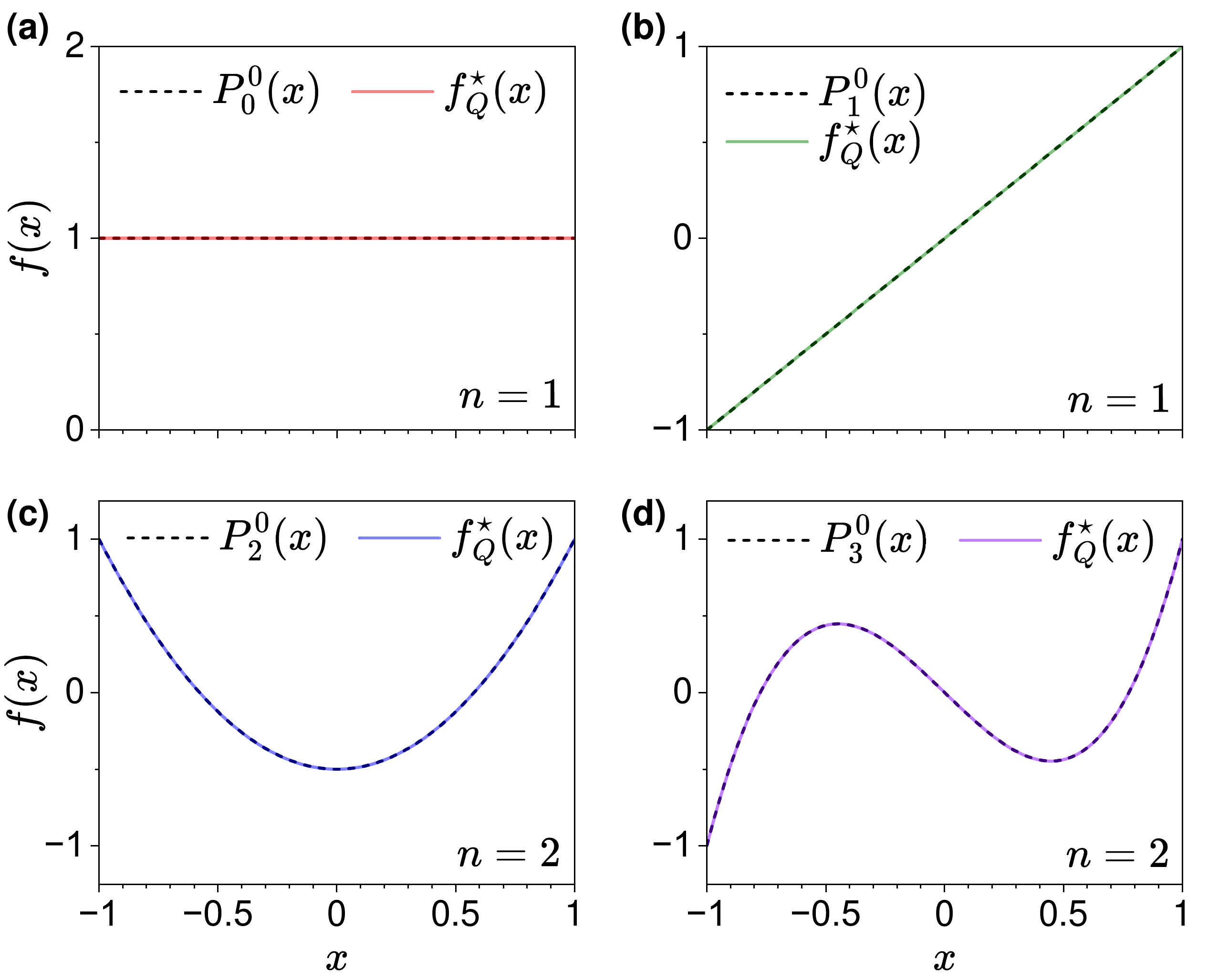}
\end{center}
\caption{
\textbf{Results of solving Legendre’s differential equations (m=0).}
In all cases, Eq.~\eqref{eq:VDE} with $l$-dependent boundary conditions $\mathcal{BC} = \{f(-1)=(-1)^l,f(1)=1\}$ are used to determine a unique analytical solution $f(x) = P^0_l(x)$. 
Plots of $P^0_l(x)$ and $f_Q^{\star}(x)$ for \textbf{(a)} $l=0$ and $\mathcal{DC_I} = \{f'(x_m=0)\}, \eta_{e}=2$, \textbf{(b)} $l=1$ and $\mathcal{DC_I} = \{f(x_z=0)\}, \eta_{e}=1$, \textbf{(c)} $l=2$ and $\mathcal{DC_I} = \{f'(x_m=0)\}, \eta_{e}=1.38$, and \textbf{(d)} $l=3$ and $\mathcal{DC_I} = \{f(x_z=0)\}, \eta_{e}=1.06$. In each panel, $n$ and $\eta_{e}$ represent the number of qubits and the scaling factor of the quantum model, respectively.
}
\label{fig:VDESF2}
\end{figure*}
\begin{figure*}[ht]
\begin{center}
\includegraphics[width=0.5\linewidth]{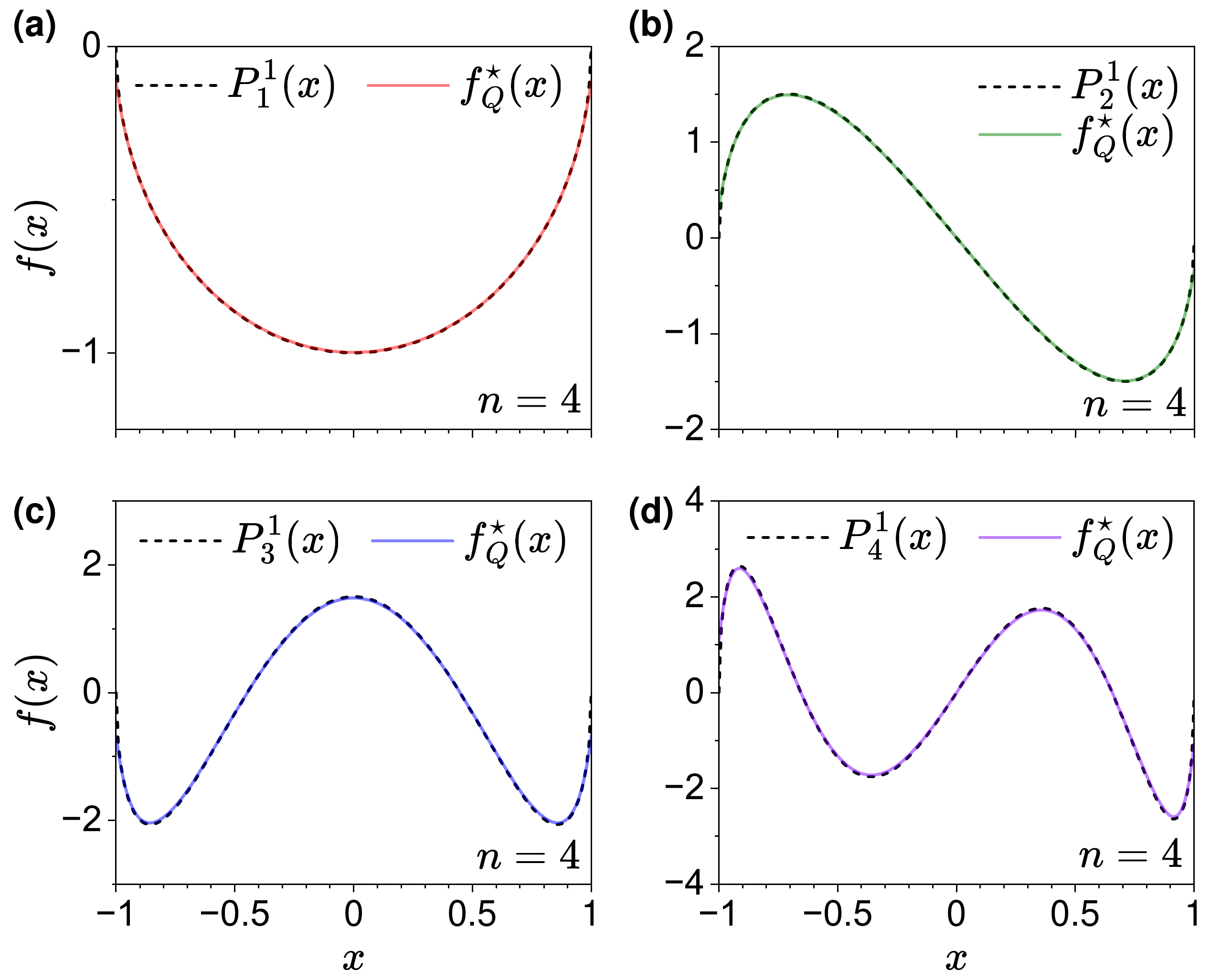}
\end{center}
\caption{\textbf{Results of solving associated Legendre’s differential equations (m=1).}
In all cases, Eq.~\eqref{eq:VDE} with fixed boundary conditions $\mathcal{BC} = \{f(-1)=0,f(1)=0\}$ are used to determine a unique analytical solution $f(x) = P^1_l(x)$. 
Plots of $P^1_l(x)$ and $f_Q^{\star}(x)$ for 
\textbf{(a)} $l=1$ and $\mathcal{DC_I} = \{f'(x_m=0)\}, \eta_{e}=8$, 
\textbf{(b)} $l=2$ and $\mathcal{DC_I} = \{f(x_z=0)\}, \eta_{e}=18$, 
\textbf{(c)} $l=3$ and $\mathcal{DC_I} = \{f'(x_m=0)\}, \eta_{e}=29.21$, and 
\textbf{(d)} $l=4$ and $\mathcal{DC_I} = \{f(x_z=0)\}, \eta_{e}=41.31$. 
In each panel, $n$ and $\eta_{e}$ represent the number of qubits and the scaling factor of the quantum model, respectively.
}
\label{fig:VDESF3}
\end{figure*}
\begin{figure*}[ht]
\begin{center}
\includegraphics[width=0.7\linewidth]{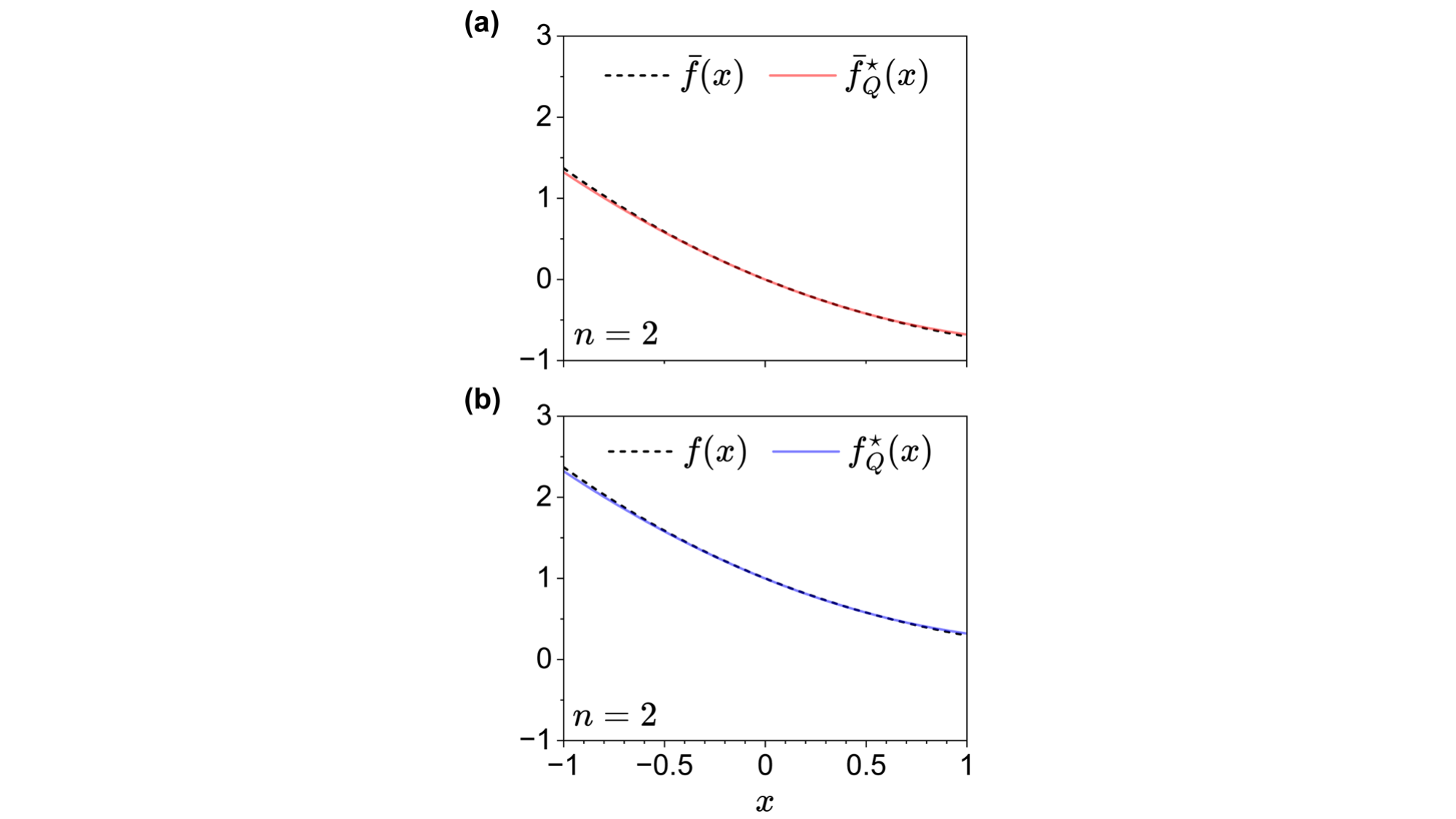}
\end{center}
\caption{
\textbf{Example of solving Eq.~\eqref{eq:NDE3} whose solution $f(x)$ has neither zero crossings nor zero slopes.} 
\textbf{(a)} Instead of addressing Eq.~\eqref{eq:NDE3} directly, we solve Eq.~\eqref{eq:NDE4} with a dependent variable defined as $\bar{f}(x)=f(x)-f(0)$, where $f(0)$ is one of its boundary conditions for Eq.~\eqref{eq:NDE3}, and then follow the same workflow to get $\bar{f}_Q^{\star}(x)$. \textbf{(b)} $f_Q^{\star}(x)$ is accessible based on $f(x)=\bar{f}(x)+f(0)$. 
}
\label{fig:zerocrossing}
\end{figure*}
\begin{figure*}[ht]
\begin{center}
\includegraphics[width=0.8\linewidth]{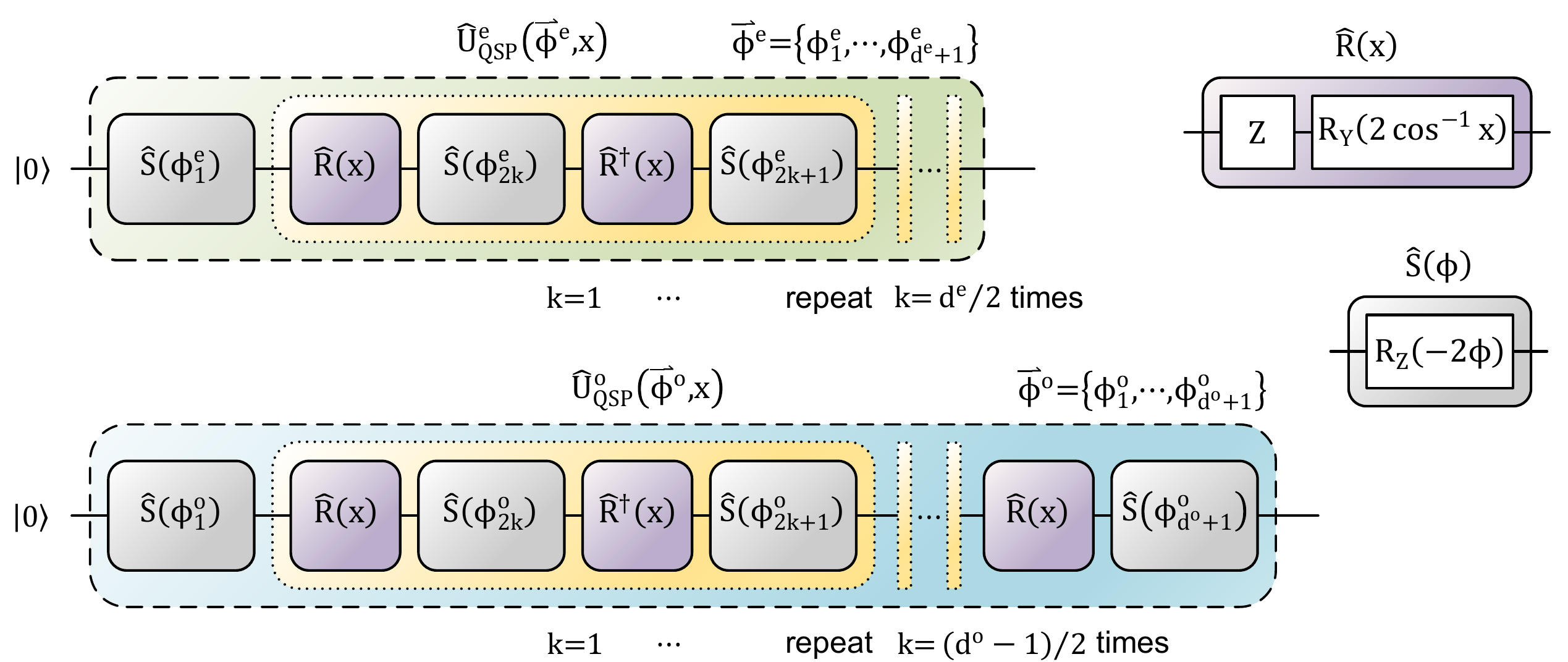}
\end{center}
\caption{
\textbf{Even and odd parity QSP circuits.} 
 Quantum signal processing (QSP) is performed by the repeated applications of the single-qubit reflection and parameterized $Z$-rotation gates, $\hat{R}(x)$ and $\hat{S}(\phi)$. The phase angles $\vec{\phi}=\{\vec{\phi}^e,\vec{\phi}^o\}$ comprise even and odd parity groups, $\vec{\phi}^e=\{\phi^e_1,\dots,\phi^e_{d^e+1}\}$ and $\vec{\phi}^o=\{\phi^o_1,\dots,\phi^o_{d^o+1}\}$, where $d^e$ and $d^o$ are the maximum degrees of even and odd polynomial components of the target function $p(x)$, respectively. Note that $\hat{R}(x)$ is a symmetric operator.
}
\label{fig:QSP}
\end{figure*}
\begin{figure*}[ht]
\begin{center}
\includegraphics[width=0.9\linewidth]{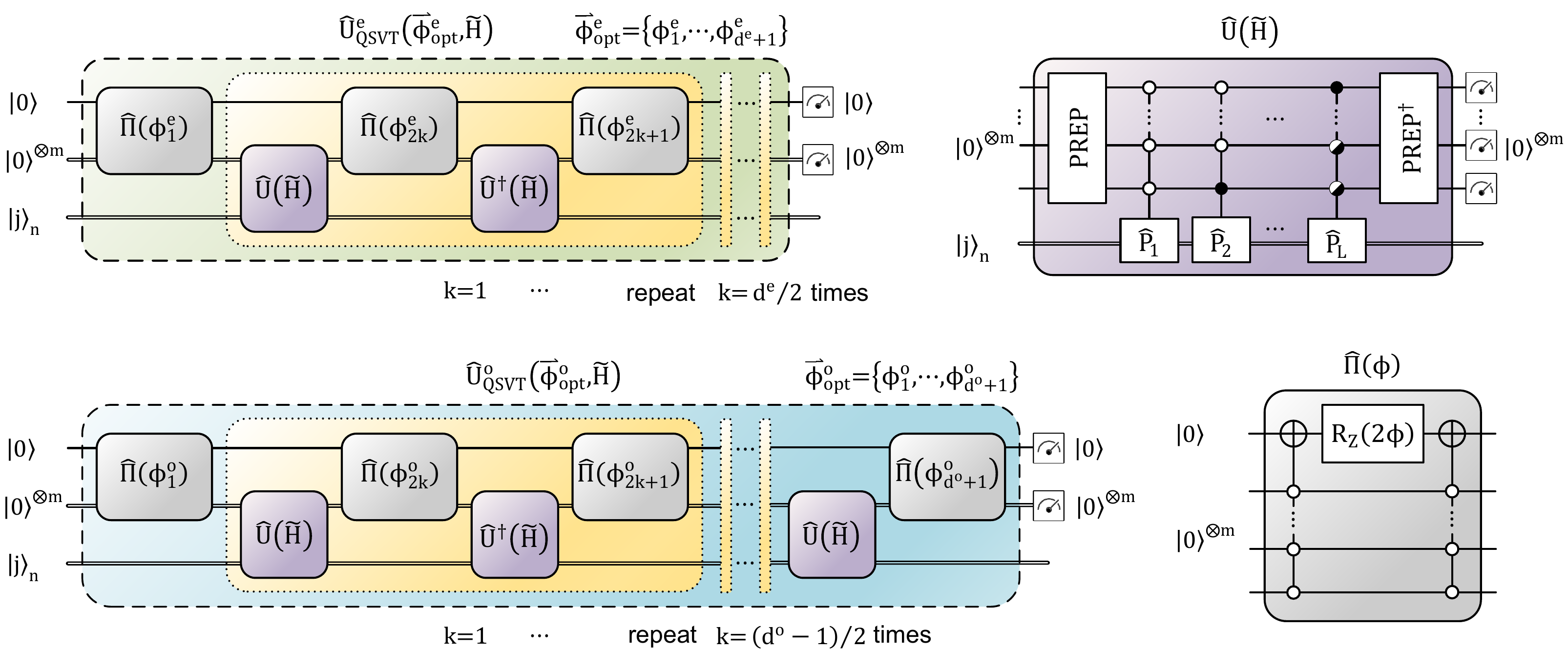}
\end{center}
\caption{
\textbf{Even and odd parity QSVT circuits.} 
Quantum singular value transformation (QSVT) operation is composed of repeated applications of block-encoding unitary and projector-controlled phase gates, $\hat{\text{U}}(\tilde{\mathbb{H}})$ and $\hat{\Pi}(\phi)$, where $\tilde{\mathbb{H}} = \mathbb{H}/ \norm{\,\mathbb{H}\,}_{\text{F}}$ is a normalized effective Hamiltonian (real symmetric) operator.  $\vec{\phi}_{\text{opt}} = \{\vec{\phi}^e_{\text{opt}},\vec{\phi}^o_{\text{opt}}\}$ is a set of optimal phase angles. $\hat{\text{U}}(\tilde{\mathbb{H}})$ is a block encoding of $\tilde{\mathbb{H}}$, which can be constructed using the combination of PREPARE and SELECT subroutines based on unitary decomposition of $\tilde{\mathbb{H}}$ with the number of ancillary qubits $m = \ceil{\text{log}_2L}$, where $L$ is the number of decomposed unitaries. Black-and-white circles mean that the unitary can be controlled by either 0 or 1.
}
\label{fig:QSVT}
\end{figure*}

\end{document}